\documentclass[aps,reprint,nofootinbib,nobibnotes,showpacs,superscriptaddress]{revtex4-1}

\usepackage[english]{babel}
	\addto\captionsenglish{%
	}

\usepackage{graphicx}
\usepackage{amssymb}
\usepackage{color}
\usepackage{epsfig}
\usepackage[absolute,overlay]{textpos}

\usepackage{amsmath}
\usepackage{amssymb}
\usepackage[version=3]{mhchem}	
\usepackage{bm}	
\usepackage{nicefrac}	
\usepackage{mathrsfs}

\usepackage[hidelinks]{hyperref}
\hypersetup{
    colorlinks,
    linkcolor={blue!70!black},
    citecolor={blue!70!black},
    urlcolor={blue!70!black}
}

\usepackage[usenames,dvipsnames]{xcolor}	
\usepackage{lineno}
\usepackage{verbatim}
\usepackage[version=3]{mhchem}

\newcommand{\xrs}{X-rays}
\newcommand{\gr}{$\gamma$-ray}
\newcommand{\grs}{$\gamma$-rays}
\newcommand{\ber}{\ce{^7Be}}
\newcommand{\bor}{\ce{^8B}}
\newcommand{\C}{\ce{^{14}C}}
\newcommand{\Coseven}{\ce{^{57}Co}}
\newcommand{\Sr}{\ce{^{85}Sr}}
\newcommand{\Cenine}{\ce{^{139}Ce}}
\newcommand{\Hg}{\ce{^{203}Hg}}
\newcommand{\Tl}{\ce{^{208}Tl}}
\newcommand{\Po}{\ce{^{210}Po}}
\newcommand{\Bi}{\ce{^{210}Bi}}
\newcommand{\Pofour}{\ce{^{214}Po}}
\newcommand{\Rn}{\ce{^{222}Rn}}
\newcommand{\Theight}{\ce{^{228}Th}}
\newcommand{\Th}{\ce{^{232}Th}}
\newcommand{\U}{\ce{^{238}U}}

\newcommand{\pep}{\emph{pep}}
\newcommand{\pp}{\emph{pp}}
\newcommand{\birks}{$k_B$}

\newcommand{\Bipo}{\ce{^{214}Bi} - \ce{^{214}Po}}

\newcommand{\AmBe}{\ce{^{241}Am}-\ce{^{9}Be}}
\newcommand{\noflambda}{n($\lambda$)}

\newcommand{\APC}{AstroParticule et Cosmologie, Universit\'e Paris Diderot, CNRS/IN2P3, CEA/IRFU, Observatoire de Paris, Sorbonne Paris Cit\'e, 75205 Paris Cedex 13, France}
\newcommand{\Dubna}{Joint Institute for Nuclear Research, 141980 Dubna, Russia}
\newcommand{\Genova}{Dipartimento di Fisica, Universit\`a degli Studi e INFN, 16146 Genova, Italy}
\newcommand{\Hamburg}{Institut f\"ur Experimentalphysik, Universit\"at Hamburg, 22761 Hamburg, Germany}

\newcommand{\Krakow}{M.~Smoluchowski Institute of Physics, Jagiellonian University, 30059 Krakow, Poland}
\newcommand{\Kiev}{Kiev Institute for Nuclear Research, 03680 Kiev, Ukraine}
\newcommand{\Kurchatov}{National Research Centre Kurchatov Institute, 123182 Moscow, Russia}
\newcommand{\Kurchatovb}{ National Research Nuclear University MEPhI (Moscow Engineering Physics Institute), 115409 Moscow, Russia}
\newcommand{\LNGS}{INFN Laboratori Nazionali del Gran Sasso, 67010 Assergi (AQ), Italy}
\newcommand{\Milano}{Dipartimento di Fisica, Universit\`a degli Studi e INFN, 20133 Milano, Italy}
\newcommand{\Perugia}{Dipartimento di Chimica, Biologia e Biotecnologie, Universit\`a degli Studi e INFN, 06123 Perugia, Italy}
\newcommand{\Peters}{St. Petersburg Nuclear Physics Institute NRC Kurchatov Institute, 188350 Gatchina, Russia}
\newcommand{\Princeton}{Physics Department, Princeton University, Princeton, NJ 08544, USA}
\newcommand{\PrincetonChemEng}{Chemical Engineering Department, Princeton University, Princeton, NJ 08544, USA}
\newcommand{\UMass}{Amherst Center for Fundamental Interactions and Physics Department, University of Massachusetts, Amherst, MA 01003, USA}
\newcommand{\Virginia}{Physics Department, Virginia Polytechnic Institute and State University, Blacksburg, VA 24061, USA}
\newcommand{\Munchen}{Physik-Department and Excellence Cluster Universe, Technische Universit\"at  M\"unchen, 85748 Garching, Germany}
\newcommand{\Lomonosov}{ Lomonosov Moscow State University Skobeltsyn Institute of Nuclear Physics, 119234 Moscow, Russia}
\newcommand{\GSSI}{ Gran Sasso Science Institute (INFN), 67100 L'Aquila, Italy}
\newcommand{\Huston}{Department of Physics, University of Houston, Houston, TX 77204, USA}
\newcommand{\Dresda}{Department of Physics, Technische Universit\"at Dresden, 01062 Dresden, Germany}
\newcommand{\UCLA}{Physics and Astronomy Department, University of California Los Angeles (UCLA), Los Angeles, California 90095, USA}
\newcommand{\Mainz}{Institute of Physics and Excellence Cluster PRISMA, Johannes Gutenberg-Universit\"at Mainz, 55099 Mainz, Germany}
\newcommand{\Honolulu}{Department of Physics and Astronomy, University of Hawaii, Honolulu, HI 96822, USA}
\newcommand{\Juelich}{IKP-2 Forschungzentrum J\"ulich, 52428 J\"ulich, Germany}
\newcommand{\RWTH}{RWTH Aachen University, 52062 Aachen, Germany}
\newcommand{\Canfranc}{Also at: Laboratorio Subterr\'aneo de Canfranc, Paseo de los Ayerbe S/N, 22880 Canfranc Estacion Huesca, Spain}
\newcommand{\DIEGO}{Present address: Physics Department, University of California, San Diego, CA 92093, USA}
\newcommand\blfootnote[1]{%
  \begingroup
  \renewcommand\thefootnote{}\footnote{#1}%
  \addtocounter{footnote}{-1}%
  \endgroup
}
\newcommand{\spokes}{Corresponding author: spokeperson-borex@lngs.infn.it}

\begin{document}

\title{The Monte Carlo simulation of the Borexino detector}


\author{M.~Agostini}
\affiliation{\GSSI}
\author{K.~Altenm\"{u}ller}
\affiliation{\Munchen}
\author{S.~Appel}
\affiliation{\Munchen}
\author{V.~Atroshchenko}
\affiliation{\Kurchatov}
\author{Z.~Bagdasarian}
\affiliation{\Juelich}
\author{D.~Basilico}
\affiliation{\Milano}
\author{G.~Bellini}
\affiliation{\Milano}
\author{J.~Benziger}
\affiliation{\PrincetonChemEng}
\author{D.~Bick}
\affiliation{\Hamburg}
\author{G.~Bonfini}
\affiliation{\LNGS}
\author{L.~Borodikhina}
\affiliation{\Kurchatov}
\author{D.~Bravo}
\affiliation{\Virginia}
\affiliation{\Milano}
\author{B.~Caccianiga}
\affiliation{\Milano}
\author{F.~Calaprice}
\affiliation{\Princeton}
\author{A.~Caminata}
\affiliation{\Genova}
\author{S.~Caprioli}
\affiliation{\Milano}
\author{M.~Carlini}
\affiliation{\LNGS}
\author{P.~Cavalcante}
\affiliation{\LNGS}
\affiliation{\Virginia}
\author{A.~Chepurnov}
\affiliation{\Lomonosov}
\author{K.~Choi}
\affiliation{\Honolulu}
\author{D.~D'Angelo}
\affiliation{\Milano}
\author{S.~Davini}
\affiliation{\Genova}
\author{A.~Derbin}
\affiliation{\Peters}
\author{X.F.~Ding}
\affiliation{\GSSI}
\author{L.~Di Noto}
\affiliation{\Genova}
\author{I.~Drachnev}
\affiliation{\GSSI}
\affiliation{\Peters}
\author{K.~Fomenko}
\affiliation{\Dubna}
\author{A.~Formozov}
\affiliation{\Milano}
\author{D.~Franco}
\affiliation{\APC}
\author{F.~Froborg}
\affiliation{\Princeton}
\author{F.~Gabriele}
\affiliation{\LNGS}
\author{C.~Galbiati}
\affiliation{\Princeton}
\author{C.~Ghiano}
\affiliation{\Genova}
\author{M.~Giammarchi}
\affiliation{\Milano}
\author{M.~Goeger-Neff}
\affiliation{\Munchen}
\author{A.~Goretti}
\affiliation{\Princeton}
\author{M.~Gromov}
\affiliation{\Lomonosov}
\author{C.~Hagner}
\affiliation{\Hamburg}
\author{T.~Houdy}
\affiliation{\APC}
\author{E.~Hungerford}
\affiliation{\Huston}
\author{Aldo~Ianni$^{\dagger}$}\blfootnote{$^{\dagger}$ \Canfranc}
\affiliation{\LNGS}
\author{Andrea~Ianni}
\affiliation{\Princeton}
\author{A.~Jany}
\affiliation{\Krakow}
\author{D.~Jeschke}
\affiliation{\Munchen}
\author{V.~Kobychev}
\affiliation{\Kiev}
\author{D.~Korablev}
\affiliation{\Dubna}
\author{G.~Korga}
\affiliation{\Huston}
\author{D.~Kryn}
\affiliation{\APC}
\author{M.~Laubenstein}
\affiliation{\LNGS}
\author{E.~Litvinovich}
\affiliation{\Kurchatov}
\affiliation{\Kurchatovb}
\author{F.~Lombardi$^{\ddagger}$}\blfootnote{$^{\ddagger}$ \DIEGO}
\affiliation{\LNGS}
\author{P.~Lombardi}
\affiliation{\Milano}
\author{L.~Ludhova}
\affiliation{\Juelich}
\affiliation{\RWTH}
\author{G.~Lukyanchenko}
\affiliation{\Kurchatov}
\author{I.~Machulin}
\affiliation{\Kurchatov}
\affiliation{\Kurchatovb}
\author{G.~Manuzio}
\affiliation{\Genova}
\author{S.~Marcocci}
\affiliation{\GSSI}
\affiliation{\Genova}
\author{J.~Martyn}
\affiliation{\Mainz}
\author{E.~Meroni}
\affiliation{\Milano}
\author{M.~Meyer}
\affiliation{\Hamburg}
\author{L.~Miramonti}
\affiliation{\Milano}
\author{M.~Misiaszek}
\affiliation{\Krakow}
\author{V.~Muratova}
\affiliation{\Peters}
\author{B.~Neumair}
\affiliation{\Munchen}
\author{L.~Oberauer}
\affiliation{\Munchen}
\author{B.~Opitz}
\affiliation{\Hamburg}
\author{F.~Ortica}
\affiliation{\Perugia}
\author{M.~Pallavicini}
\affiliation{\Genova}
\author{L.~Papp}
\affiliation{\Munchen}
\author{A.~Pocar}
\affiliation{\UMass}
\author{G.~Ranucci}
\affiliation{\Milano}
\author{A.~Re}
\affiliation{\Milano}
\author{A.~Romani}
\affiliation{\Perugia}
\author{R.~Roncin}
\affiliation{\LNGS}
\affiliation{\APC}
\author{N.~Rossi}
\affiliation{\LNGS}
\author{S.~Sch\"onert}
\affiliation{\Munchen}
\author{D.~Semenov}
\affiliation{\Peters}
\author{P.~Shakina}
\affiliation{\Peters}
\author{M.~Skorokhvatov}
\affiliation{\Kurchatov}
\affiliation{\Kurchatovb}
\author{O.~Smirnov}
\affiliation{\Dubna}
\author{A.~Sotnikov}
\affiliation{\Dubna}
\author{L.F.F.~Stokes}
\affiliation{\LNGS}
\author{Y.~Suvorov}
\affiliation{\UCLA}
\affiliation{\Kurchatov}
\author{R.~Tartaglia}
\affiliation{\LNGS}
\author{G.~Testera}
\affiliation{\Genova}
\author{J.~Thurn}
\affiliation{\Dresda}
\author{M.~Toropova}
\affiliation{\Kurchatov}
\author{E.~Unzhakov}
\affiliation{\Peters}
\author{A.~Vishneva}
\affiliation{\Dubna}
\author{R.B.~Vogelaar}
\affiliation{\Virginia}
\author{F.~von~Feilitzsch}
\affiliation{\Munchen}
\author{H.~Wang}
\affiliation{\UCLA}
\author{S.~Weinz}
\affiliation{\Mainz}
\author{M.~Wojcik}
\affiliation{\Krakow}
\author{M.~Wurm}
\affiliation{\Mainz}
\author{Z.~Yokley}
\affiliation{\Virginia}
\author{O.~Zaimidoroga}
\affiliation{\Dubna}
\author{S.~Zavatarelli}
\affiliation{\Genova}
\author{K.~Zuber}
\affiliation{\Dresda}
\author{G.~Zuzel} 
\affiliation{\Krakow}

\collaboration{Borexino Collaboration$^{*}$}\blfootnote{$^{*}$ \spokes}
\noaffiliation{}

\begin{abstract}
We describe the Monte Carlo (MC) simulation package of the Borexino detector and discuss the agreement of its output with data. 
The Borexino MC ``ab initio" simulates the energy loss of particles in all detector components and generates the resulting scintillation photons 
and their propagation within the liquid scintillator volume. The simulation accounts for absorption, reemission, and scattering of the optical photons and tracks them until they either are absorbed or reach the photocathode of one of the photomultiplier tubes. 
Photon detection is followed by a comprehensive simulation of the readout electronics response.
The algorithm proceeds with a detailed simulation of the electronics chain.
The MC is tuned using data collected with radioactive calibration sources deployed inside and around the scintillator volume.
The simulation reproduces the energy response of the detector, its uniformity within the fiducial scintillator volume relevant to neutrino physics, and the time distribution of 
detected photons to better than 1\% between 100\,keV and several MeV. 
The techniques developed to simulate the Borexino detector and their level of refinement are of possible interest to the neutrino community, especially for current and 
future large-volume liquid scintillator experiments such as Kamland-Zen, SNO+, and Juno.
\end{abstract}

	\pacs{02.70.Uu,\, 
	14.60.Lm,\,  
	26.65.+t,\,   
	29.40.Mc,\,  
	78.70.Ps,\,  
	96.60.Jw 
	}
		   
	\maketitle



\section{Introduction}
Large volume, liquid scintillator detectors have contributed greatly to neutrino physics in recent years. 
Starting twenty-five years ago, Borexino pioneered the development of ultra-low radioactive background detectors that could push the energy threshold down to ~100\,keV allowing the measurement 
of most of the solar neutrino spectrum~\cite{bib:BxOldPaper}. 
A wide variety of liquid scintillator detectors, listed in~Table~\ref{tab:list_exp}, are either running or being designed for measurements of neutrino oscillations, 
for neutrino-less double beta decay searches, or for use as active $\gamma$-ray and neutron veto devices for direct WIMP dark matter experiments~\cite{Agnes:2015ftt}.

The accurate modeling by means of Monte Carlo (MC) methods of the entire detection chain is essential to extract quality data from running detectors and to inform the design of future experiments. 
The simulation needs to include the energy lost by primary particles, the generation and transport of scintillation light in the large volume of liquid scintillator and its detection 
by photomultipliers, and the detector response.

In this paper we describe the MC modeling of the Borexino experiment and we compare its output to data.
The methods and the results presented here were developed and specifically optimized for the Borexino detector and scintillator mixture, but 
they can be easily adapted to other liquid scintillators and detector geometries.

A collection of experiments employing large liquid scintillator detectors is shown in Table~\ref{tab:list_exp}. A feature common to all of them is the focus on events of relatively low energy, up to a few tens of MeV.
In this energy range it is possible to perform detailed simulations that generate and tracke individual optical photons emitted produced in scintillation events, with typical photon yield of $\simeq$$10^4$ photons/MeV.
The MC model can in turn very accurately reproduce physical events with minimal use of effective parameters, 
provided that all the physics processes are included and realistically described (e.\,g. light emission, absorption, reemission, scattering, reflection, ...).

\begin{figure*}[tb]
	\centering
	\includegraphics[width=0.85\columnwidth]{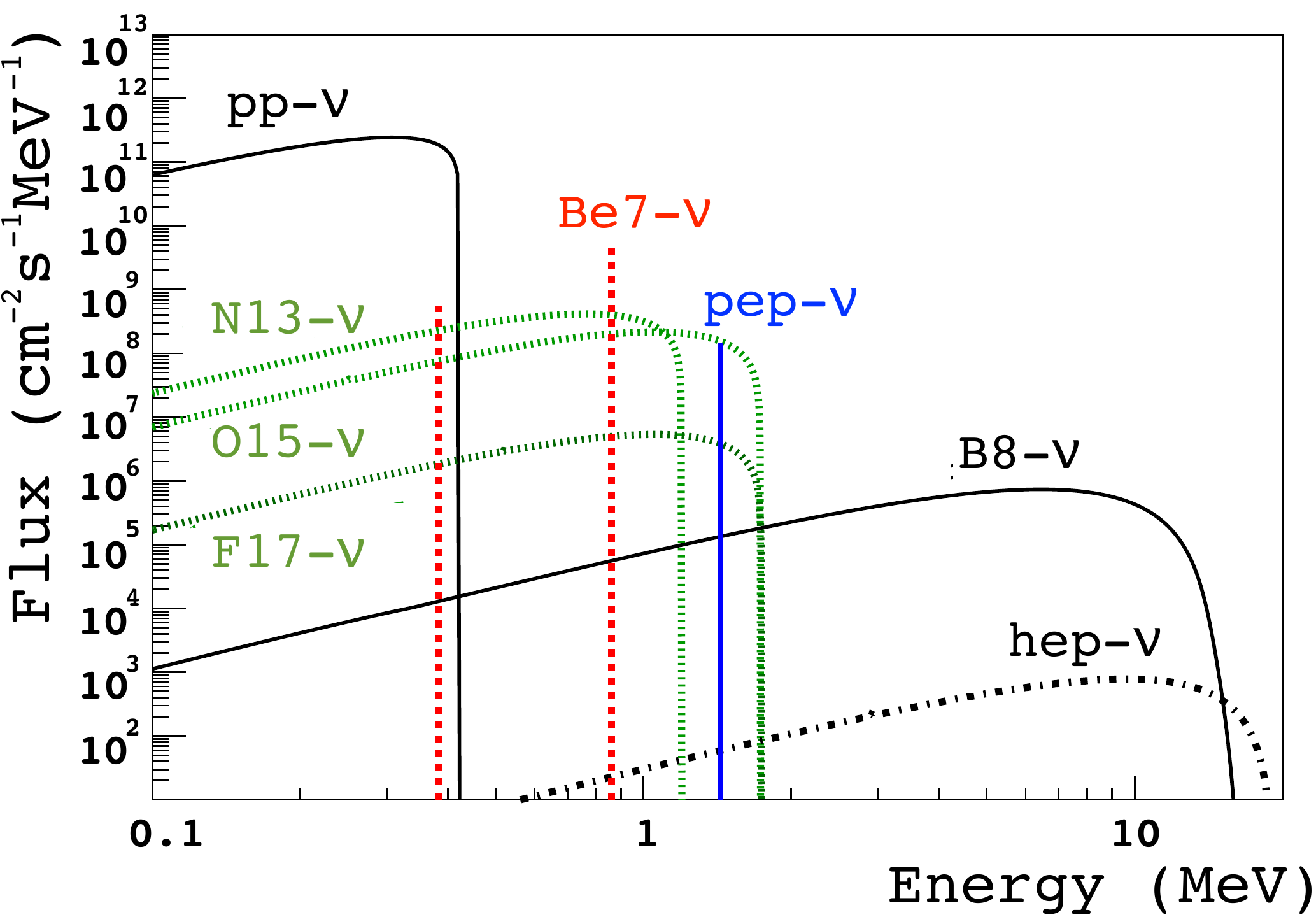}
	\includegraphics[width=\columnwidth]{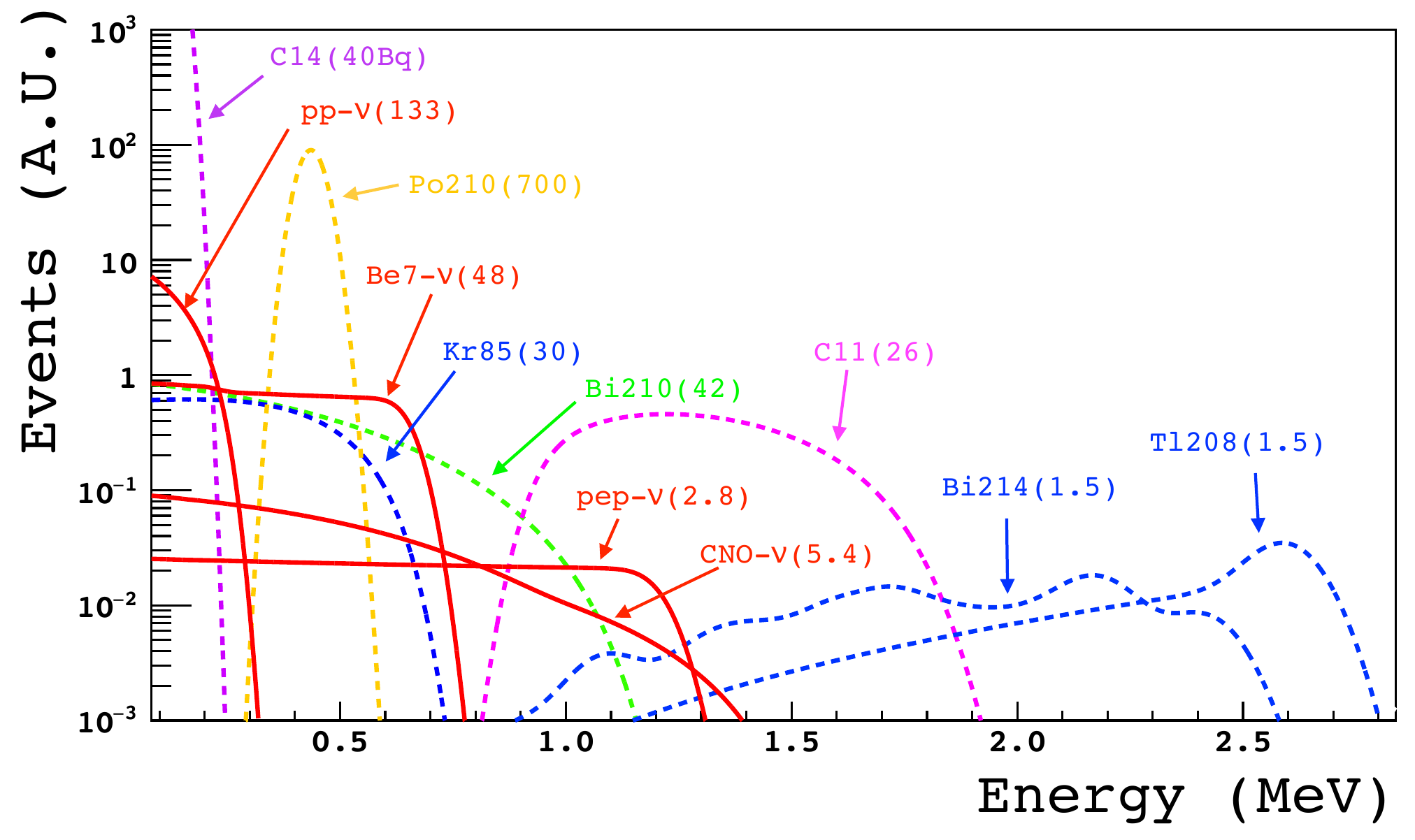}
	\\[0.3cm]
	\includegraphics[width=0.85\columnwidth]{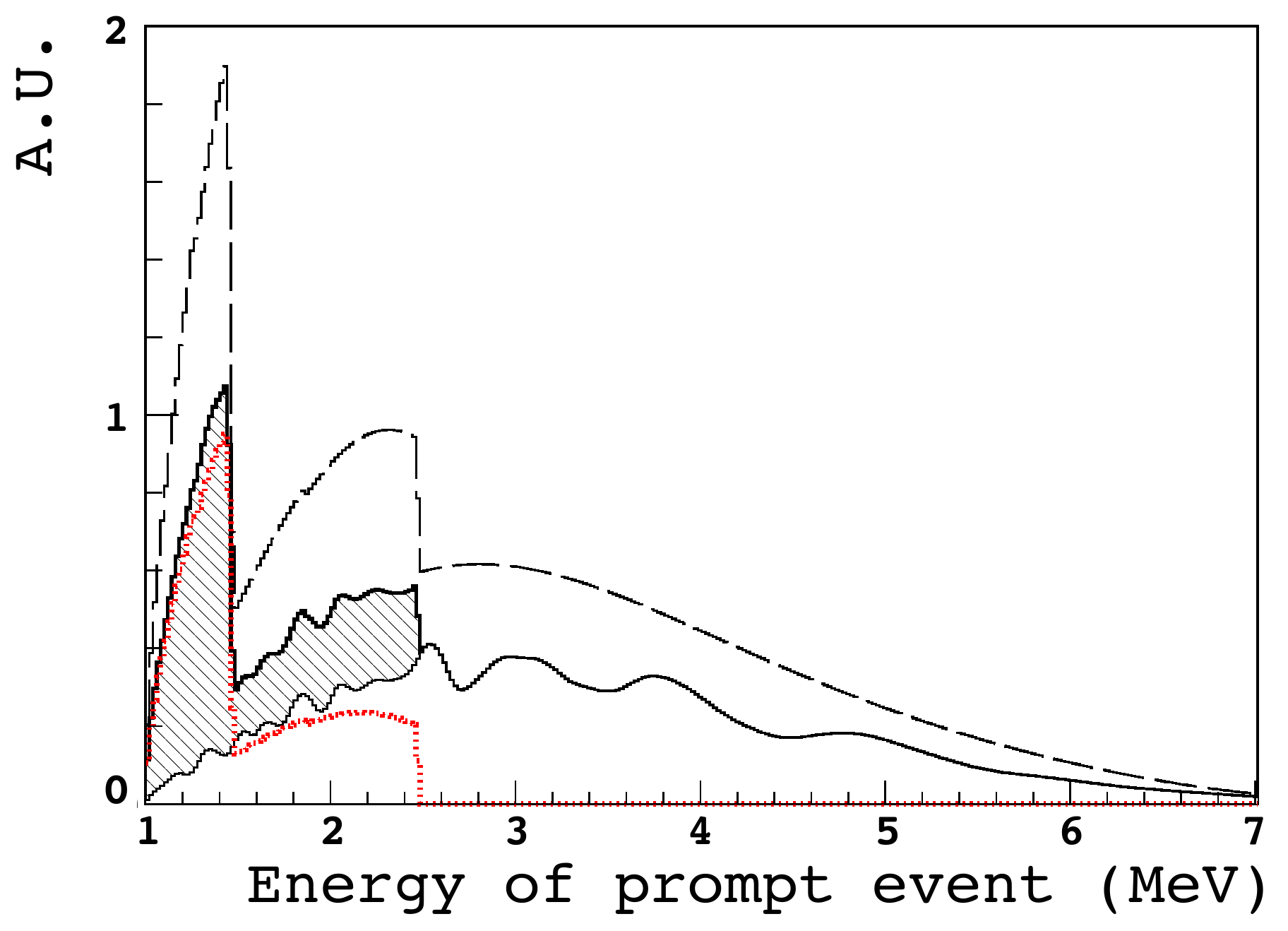}
	\includegraphics[width=1.02\columnwidth]{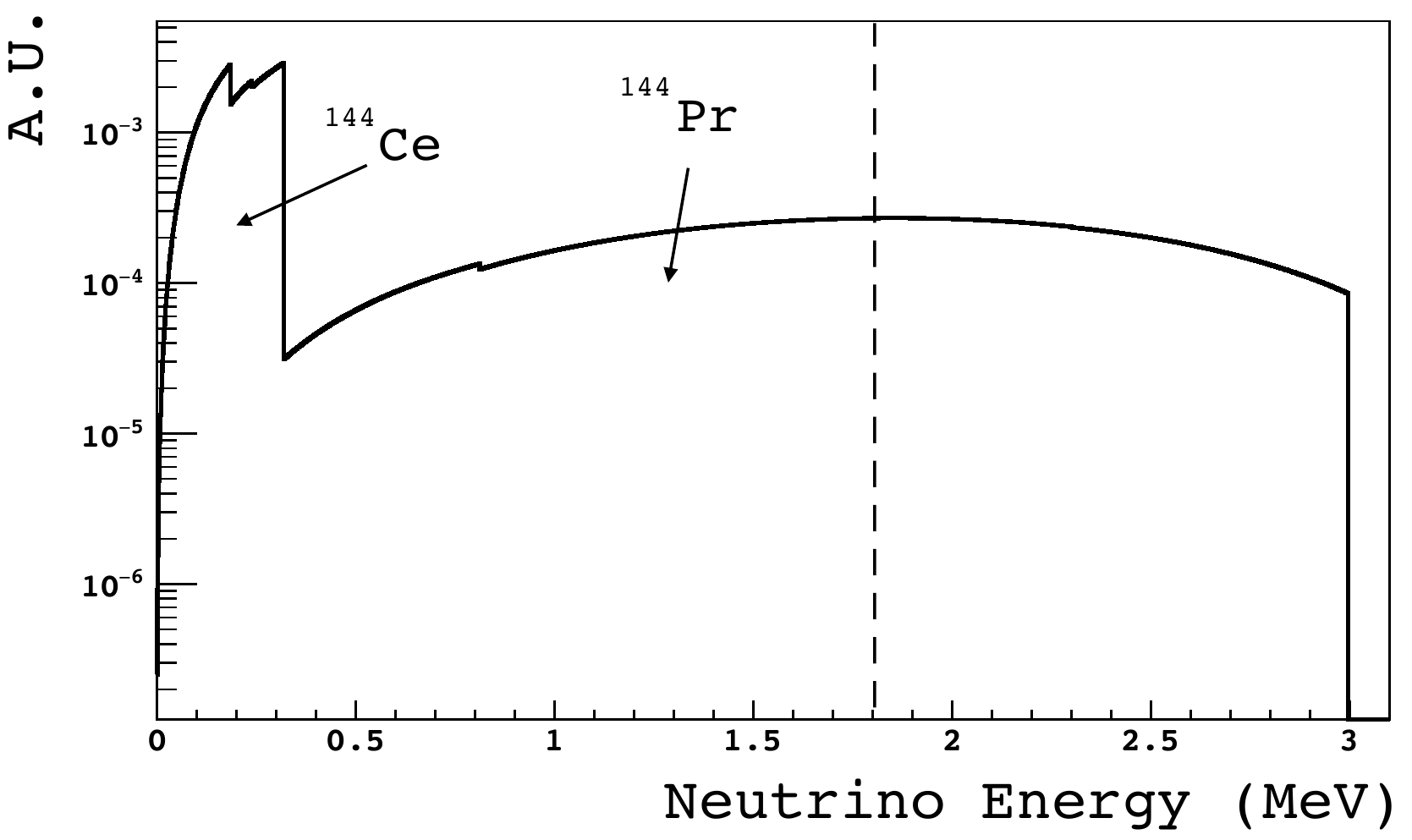}
	\caption{Sources of signals in Borexino.
		\emph{Top left}: expected spectrum of solar neutrinos~\cite{bahcall}. The flux (vertical scale) is given in cm$^{-2}$ s$^{-1}$ for monochromatic lines.
		\emph{Top right}: energy spectrum of electron recoils induced by the interaction of solar neutrinos in Borexino together with the spectral shapes of some known radioactive contaminants in the scintillator. The numbers shown in parenthesis are in counts per day per 100 ton for all species apart from \ce{^{14}C}, for which Bq per 100 ton is used.
		\emph{Bottom left}: expected spectrum of electron anti-neutrinos in Borexino.  The horizontal axis shows the kinetic plus annihilation (1.022\,MeV) energy of the prompt positron event. Dashed line: total geo--$\bar{\nu}_e$ plus reactor--$\bar{\nu}_e$ spectrum without oscillations. Solid thick lines: geo--$\bar{\nu}_e$ and reactor--$\bar{\nu}_e$ with oscillations.  Dotted line (red): geo--$\bar{\nu}_e$ with the high (low) energy peak due to decays in the \U\ chain (\U\ and \Th\ chains).  Solid thin line: reactor--$\bar{\nu}_e$.
		\emph{Bottom right}: energy spectrum of the emitted $\bar{\nu}_e$ in the decay of the \ce{^{144}Ce}-\ce{^{144}Pr} for the SOX experiment~\cite{bib:SOX}.
		Only the portion of the $\bar{\nu}_e$ spectrum above 1.8\,MeV  (dashed vertical line) can be detected via inverse $\beta$ decay on protons.}
	\label{fig:Signals}
\end{figure*}

Borexino is an un-segmented calorimeter consisting of $\sim$278 t of organic liquid scintillator~\ref{fig:detector}. 
The detector has been in continuous operation since May 2007 at the underground Laboratori Nazionali del Gran Sasso (LNGS) in Italy. 
It is designed to measure the low energy portion of the solar neutrino spectrum (see the top left panel of Fig.~\ref{fig:Signals}) and, in particular, the $0.862$\,MeV \ber\ solar neutrinos, 
whose interaction rate was measured with 5\% precision~\cite{bib:BxBe3} with no observed day-night rate asymmetry~\cite{bib:BxDayNight}.

The extremely low levels of radioactivity achieved within the scintillator allowed Borexino to broaden its science reach beyond the original design goal. 
Borexino reported first evidence for the observation of the $1.44$\,MeV \pep\ neutrinos and holds the current best upper limit on the solar CNO flux~\cite{bib:BxPep}. 
At higher end of the solar neutrino spectrum, the electron recoil induced by \bor\ neutrinos was measured with a record-low threshold of 3\,MeV~\cite{bib:BxB8}.
A scintillator purification campaign performed $\sim$3 years into data taking, which further reduced radioactive contamination, allowed Borexino to measure the spectrum 
of \pp\ solar neutrinos (with end-point energy of $0.42$ MeV)~\cite{bib:BxPp}. 
Further physics results of Borexino include the detection of geo-neutrinos in the range between $1.8$ and $~3$\,MeV~\cite{bib:BxGeo, bib:BxGeo2013, bib:BxGeo2015}, 
searches for anti-neutrinos from astrophysical sources up to 15 MeV~\cite{bib:BxAntiNu, bib:BXGRB}, 
searches for solar axions at $5.5$\,MeV~\cite{bib:BxAxion}, as well as other exotic searches~\cite{bib:BxEDec, bib:BxPauli, bib:BxHeavy}. 
The investigation of short baseline anti-neutrino oscillations into light sterile neutrinos using an artificial $^{144}$Ce-$^{144}Pr$ source is planned for the near future within the SOX project~\cite{bib:SOX}.

Figure~\ref{fig:Signals} provides an overview of the signals of interest in Borexino, their energy ranges, and most relevant backgrounds.
The Borexino data analysis calls for an accurate and complete understanding of the detector response over these energy ranges. 
This is achieved through meticulous energy calibration and the understanding of the degree of uniformity of the energy response of the detector over its volume, 
of the trigger efficiency, of the detector stability over time, and of the spatial reconstruction and time response for events of different type occurring at different positions, 
including the scintillation pulse shape characteristics.
The Borexino MC simulation reproduces physics and calibrations data with 1\% precision or better for all quantities relevant for experiment's physics program. 

Short descriptions of the Borexino detector and of the calibration hardware are presented in Sec.~\ref{sec:detector} and Sec.~\ref{sec:calibration}, respectively. 
We describe the MC package and the physical models it implements in Sec.~\ref{sec:mc}, Sec.~\ref{sec:EvGen}, Sec.~\ref{sec:MCbxelec}, and Sec.~\ref{sec:reco}). 
In Sec.~\ref{sec:tuning} we show how we tuned the simulation parameters with the aid of laboratory measurements, source calibration data, and physics-mode data. 
Finally, we compare MC results and data in Sec.~\ref{sec:validation}.

\begin{table*}
	\begin{center}
		\begin{tabular}{c||c||c||c||c} 
			Experiment              & Mass & Physics investigation & Status & Reference         \\ \hline \hline
			Chooz & 5\,t LS$+$Gd (0.1\%) $+$ 107\,t LS & $\nu$ oscillations & past & \cite{bib:Chooz}\\
			KamLAND & 1\,kt LS & $\nu$ oscillations & past & \cite{bib:KamLAND}\\ 
			Karmen & 56\,t $+$Gd foils & $\nu$ oscillations & past & \cite{bib:Karmen}\\
			LSND & 167\,t LS & $\nu$ oscillations & past & \cite{bib:LSND}\\
			Palo Verde & 11\,t LS$+$Gd (0.1\%)& $\nu$ oscillations & past & \cite{bib:Palo_Verde}\\
			Borexino & 278\,t LS & $\nu$ oscillations & ongoing & \cite{bib:BxDet}\\
			Daya Bay & 20\,t LS$+$Gd (0.1\%) $+$ 20\,t LS & $\nu$ oscillations & ongoing & \cite{bib:daya_bay}\\
			Double Chooz & 8\,t LS$+$Gd (0.1\%) $+$ 18\,t$+$80\,t LS& $\nu$ oscillations & ongoing & \cite{bib:double_chooz}\\
			Reno &  16\,t LS$+$Gd (0.1\%) + 30\,t LS & $\nu$ oscillations & ongoing & \cite{bib:Reno}\\
			LENS &  125\,t LS$+$In (8\%) & $\nu$ oscillations & future & \cite{bib:LENS}\\
			Juno & 20\,kt LS & $\nu$ oscillations & future & \cite{An:2015jdp}\\
			Reno-50 & 18\,kt LS & $\nu$ oscillations & future & \cite{bib:RENO-50} \\
			\hline \hline
			KamLAND-Zen & 13\,t LS$+$Xe (2.9\%) $+$ 1\,kt LS&$0\nu\beta\beta$ decay & ongoing & \cite{KamLAND-Zen:2016pfg}\\
			SNO$+$ & 780\,t LS$+$Te (0.5\%) & $0\nu\beta\beta$ decay & commissioning & \cite{bib:SNO+} \\
			\hline \hline
			LVD & 1.8\,kt LS & SN $\nu$ & ongoing & \cite{bib:LVD}\\
			\hline \hline
			Nucifer & 0.75\,t LS$+$Gd (0.2\%) & reactor monitoring & ongoing & \cite{bib:Nucifer} \\
			\hline \hline
			Neos & 1\,t LS+Gd (0.5\%)& sterile $\nu$ & ongoing & \cite{bib:Neos}\\
			Neutrino-4 & 0.35\,t LS$+$Gd (0.1\%) & sterile $\nu$ & ongoing & \cite{bib:Neutrino4} \\
			Prospect & 3--13\,t LS$+^6$Li & sterile $\nu$ & ongoing & \cite{Ashenfelter:2015uxt} \\
			Stereo & 1.8\,t LS$+$Gd (0.2\%) & sterile $\nu$ & commissioning & \cite{bib:Stereo} \\
			SOX & 278\,t LS & sterile $\nu$ & future & \cite{bib:SOX} \\
			\hline \hline
			Dark Side-50 & 30\,t LS$+$TMB (5\%)& DM veto & ongoing & \cite{bib:DSVeto} \\
			Dark Side-20k & $\sim$250\,t LS$+$TMB (20\%) & DM veto & future & \cite{bib:DS20k} \\
			LZ & 20.8\,t LS$+$Gd (0.1\%)& DM veto & future & \cite{bib:LZ}\\
			SABRE & 2\,t LS & DM veto & future & \cite{Tomei:2017rkg}\\
		\end{tabular}
	\end{center}
	
	
	
	
	
	
	
	
	
	
	
	
	
	
	
	
	
	
	\caption{Compilation of past, present and future liquid scintillator based experiments excluding Borexino. 
		In the table, ``LS'' stands for liquid scintillator, ``Gd'' (gadolinium) and ``TMB'' (trimethilborate) are used for enhancing neutron captures,
		``In'' is the chemical element indium,
		``DM'' signifies dark matter, ``SN'' stands for supernova and $0\nu\beta\beta$ for neutrino-less double beta decay. }
	\label{tab:list_exp}
\end{table*}

\section{Short description of the Borexino detector}
\label{sec:detector}

\begin{figure}
\begin{center}
\includegraphics[width =\columnwidth]{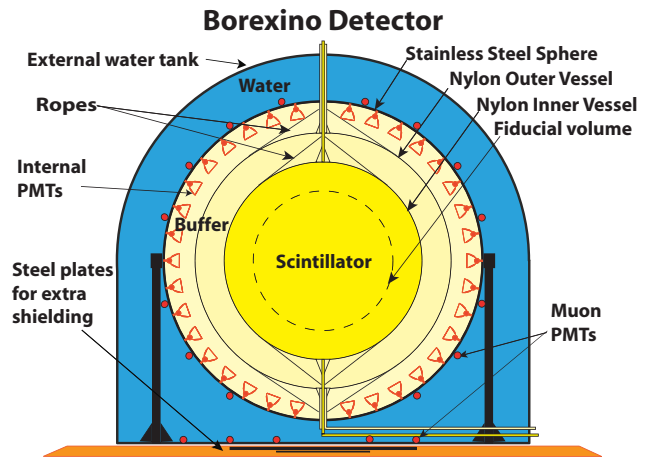}
\caption{Schematic drawing of the Borexino detector.}
\label{fig:detector}
\end{center}
\end{figure}


The Borexino detector is schematically depicted in Fig.~\ref{fig:detector} and widely described in Ref.~\cite{bib:BxDet}. 
The inner detector is enclosed by a stainless steel sphere (SSS) that serves both as the container of the scintillator and buffer liquids and as the mechanical support of the photomultipliers (PMTs). 
Within this sphere, two nylon vessels (0.125 mm thick) separate the volume in three shells of radii 4.25\,m, 5.50\,m and 6.85\,m, the latter being the radius of the SSS itself. 
The inner nylon vessel contains the liquid scintillator solution ($\sim$278 t or $\sim$300 m$^3$), 
namely PC (pseudocumene, 1,2,4-trimethylbenzene C$_6$H$_3$(CH$_3$)$_3$) as a solvent and the fluor PPO (2,5-diphenyloxazole, C$_{15}$H$_{11}$NO) 
as a solute at a concentration of 1.5~g/l (0.17~\% by weight). 
The second and the third shells (buffer regions) contain PC with a small amount (5~g/l)\footnote{The DMP concentration was 5~g/l at the beginning of data taking. 
After the discovery of a tiny leak in the inner vessel, the concentration was reduced to 2.8~g/l in order to decrease the pressure difference between the inner vessel and the buffer.} of DMP 
(dimethylphthalate, C$_6$H$_4$(COOCH$_3$)$_2$) as a light quencher added to further reduce the scintillation yield of pure PC~\cite{bib:DMP}.

The 2212 internal PMTs (8'' ETL 9351, formerly Thorn EMI) mounted on the inner side of the SSS detect the scintillation light.
In order to enhance the photon detection efficiency, 1828 PMTs out of the total 2212 were equipped with aluminum optical concentrators~\cite{bib:BXLC}.
Their field of view is designed to collect all scintillation photons incident under an angle of less than $\delta_{max}=44^\circ$, corresponding to scintillator signals from within the inner vessel. 
Photons hitting the concentrator surfaces with angles greater than $\delta_{max}$ are reflected back into the detector.

A 18\,m in diameter, 16.9\,m height domed cylinder filled by ultra-pure water contains the SSS and acts both as radiation shielding 
and as \v Cerenkov detector (outer detector) for identifying and vetoing cosmic muons. For this purpose, 208 additional PMTs are mounted on the outer side of the SSS and on the water tank floor.
A detailed description of the detector, of the electronics, and of the purification plants used to prepare the scintillator and fill the detector can be found in Ref.~\cite{bib:BxDet} and Ref.~\cite{bib:BxLiquid}.
The muon detector design and performances are detailed in Ref.~\cite{bib:BxMuon}.

Solar neutrinos are detected through their elastic scattering on electrons. The measurement of different solar neutrino components is possible through a  fit of the electron recoil 
energy spectrum (see the top right panel of Fig.~\ref{fig:Signals}), aiming at disentangling the contribution of solar neutrinos and that of background signals~\cite{bib:BxLong}.

Anti-neutrinos ($\bar{\nu}_{e}$) are detected via inverse $\beta$ decay:
\begin{equation}
\bar{\nu}_e + p \rightarrow e^+ + n,
\label{Eq:InvBeta}
\end{equation}
with a threshold of 1.806\,MeV. The positron promptly comes to rest in the liquid scintillator and annihilates emitting two 511\,keV \grs, yielding a prompt event 
with a visible energy of $E_{\rm prompt} = E_{\bar{\nu}_e} - 0.782\,{\rm MeV}$.  
The free emitted neutron is typically captured on protons within a mean time $\tau\,\sim$256\,$\mu$s~\cite{bib:BxMuon}, resulting then in the emission of a 2.22\,MeV de-excitation \gr, which provides a coincident delayed event. 
The characteristic time and spatial coincidences of prompt and delayed events offer a clean signature of the $\bar{\nu}_e$ detection.

\section{The source calibration campaigns}
\label{sec:calibration}

\begin{table}
\begin{center}
\begin{tabular}{l||c||c}
Isotope              & Type            & Energy (keV)    \\ \hline \hline
\Coseven\         & $\gamma$   &  122                 \\
\Cenine\            & $\gamma$   &  165                \\ 
\Hg\       		& $\gamma$   &  279                 \\ 
\Sr\          		& $\gamma$   &  514                   \\ 
\ce{^{54}Mn}        & $\gamma$   &  834                \\ 
\ce{^{65}Zn}         & $\gamma$   &  1115               \\
\ce{^{60}Co}         & $\gamma$   &  1173 - 1332    \\ 
\ce{^{40}K}           & $\gamma$   &  1460                \\ \hline \hline
\Rn\      		&  $\alpha$/$\beta$ & 0 $\div$ 3200          \\ \hline \hline
\C\                     &  $\beta$     &  0 $\div$ 156                           \\ \hline \hline
\ce{^{241}Am}-\ce{^9Be}   &  n    &  $\sim$0 $\div$ 10000       \\ \hline \hline
Ext. \Theight\ & $\gamma$ & 2615 \\
\end{tabular}
\end{center}
\caption{Radioactive isotopes used for Borexino calibration. The \AmBe\ source allows to study the thermalization of neutrons in the scintillator
and the neutron captures on \ce{H} and \ce{C}.}
\label{table:source}
\end{table}

A series of calibration campaigns based on various types of radioactive sources inserted directly into the detector volume were performed in November 2008, January 2009, and June-July 2009. 
In July 2010, an external \Theight\ \gr\ source was positioned in dedicated pipes close to the SSS in order to study the response to the external \grs. 
The hardware used for a safe, air-tight, clean, and accurate deployment of small radioactive sources in several locations within the scintillator target is described in Ref.~\cite{bib:BxCalib}, 
while the design, construction and performance of the \Theight\ source can be found in Ref.~\cite{Maneschg2012161}. 

Calibration data have been essential in validating the physics model adopted for the description of the scintillation light emission, propagation, and detection by PMTs. 
The results are in full agreement with observations in the Counting Test Facility, the small 4\,t Borexino prototype~\cite{bib:CTFLight}. 

The radioactive sources deployed inside the scintillator volume (see Table~\ref{table:source}) were selected to study the detector response to $\alpha$ and $\beta$ particles, \grs, and neutrons in a wide energy region from 
122\,keV and $\sim$10\,MeV. 

More specifically, the goals of the Borexino calibration campaigns included:
\begin{itemize}
\item the measurement of the position reconstruction accuracy and resolution for events distributed in the whole inner vessel and over the energy range.
\item the calibration of the absolute energy scale and resolution especially in the energy region of interest for studying \ber, \pep, and \pp\ solar neutrinos interactions (i.\,e. $\sim$100\,keV $\div$ $\sim$1.5\,MeV).
\item the measurement of the non-uniformity of the energy response as a function of the event position and energy.
\item the production of signals mimicking the external background.
\end{itemize}

Characterizing the  position reconstruction is of high importance for all Borexino analyses.
It permits to define a fiducial volume in which the background from the radioactive contaminants on the vessel surfaces or \grs\ 
originating from the outer parts of the detector are minimized. The optimal choice of the fiducial volume depends on the type of analysis performed, as explained in Ref.~\cite{bib:BxLong}.
In addition, the accuracy of the reconstruction of the scintillation vertex determines the precision by which the target volume is defined, thus directly affecting
the uncertainty of the measurement of absolute neutrino or anti-neutrino fluxes. 
In SOX, the search for the short-baseline oscillation pattern possibly induced by sterile neutrinos relies heavily on the accurate spatial reconstruction of the inverse $\beta$ decay events.

Spectral distortions in the energy response, due to effects like the ionization quenching or the dependence of the light collection efficiency on the vertex position, are relevant for the analysis.
A correct understanding of such effects is of primary importance for spectroscopic measurements, in particular
the precision measurement of the \ber\ solar neutrino interaction rate, which relies on an absolute energy calibration on a level better than 1\%.
Finally, an accurate understanding of the external background is fundamental for the detection of \pep, CNO, and \bor\ neutrinos in Borexino.



The source vials were carefully designed to mitigate the risk of introducing unwanted contaminations into the Borexino scintillator target. 
Spherical quartz vessels (1'' diameter) were either filled by \Rn\ loaded liquid scintillator (identical to the Borexino one) or \gr\ emitters in aqueous solution. 
The sources were deployed along the polar axis of the inner vessel and in several positions off axis. 
An optical system consisting of a LED mounted on the source support and of a series of cameras allowed to measure the position of the source within the inner vessel with $\sim$2\,cm accuracy~\cite{bib:BxCalib}. 

A dedicated procedure was developed for loading scintillator inside the vials with  \Rn, minimizing the quenching due to oxygen contamination.
The comparison of the \Pofour\ energy peak from within the source with the \Bipo\ tagged coincidences from contaminants dissolved in  the detector gave evidence of small but non-negligible light quenching.
Fast coincidence decays from the \Rn\ source were used to study the accuracy of the position reconstruction algorithm and to characterize the energy response uniformity in the detector volume.

Pure beta sources could not be introduced directly in the detector because of the high risk of contaminating the scintillator. 
Instead, beta emitters were dissolved in a non-scintillating solution, for suppressing the pure $\beta$ component and allowing, at the same time, \grs\ to escape the vial and to induce electrons  in the scintillator by Compton scattering or by photoelectric effect.
The \grs\ provided the absolute energy scale calibration over the whole region of interest. 

Moreover, we used a commercial \AmBe\ neutron source inserted in a properly designed shielding. 
The neutron scattering on protons during thermalization provided a calibration for proton ionization quenching.
The \grs\ emitted by the nuclei capturing neutrons (mostly hydrogen and $^{12}$C in the scintillator or in the materials of the source shield) allowed calibration signals in the highest energy range.

\section{The Monte Carlo simulation code of the Borexino detector}
\label{sec:mc}

Particles depositing energy in the inner vessel or in the buffer regions produce scintillation and \v Cerenkov photons, which propagate inside the detector and possibly reach the PMTs, yielding a detectable signal.  
The agreement between measured observables (energy estimators and PMT pulse times~\cite{bib:BxLong}) and the physical quantities (deposited energy, position, type of particles generating the signal) 
depends on the knowledge and understanding of all the physical processes governing the particle energy loss in the various materials, the scintillator light production, propagation, and detection. 
Besides, it depends on the characteristics of the electronics and of the triggering system. 
The Borexino MC simulation was designed and optimized to fully model and reproduce all these processes up to the signal detection.

The MC simulation chain consists of a set of numerical codes that: 
\begin{enumerate}
\item provide a wide range of event generators, from solar neutrino interactions, to radioactive decays, geo-neutrinos, and calibration source events.
\item simulate the energy loss of each specific particle in every material present in the detector, either active (the scintillator, buffer liquid, and water in the muon detector) or passive.
\item generate a number of scintillation or \v Cerenkov photons considering the particle energy loss in the media and the properties of the scintillator and/or the buffer.
\item track each single optical photon including its interactions with the scintillator and with the materials, until a PMT is reached or the photon is absorbed.
\item generate the PMT response for photons absorbed at the PMT cathode, considering the quantum efficiency of each individual PMT. 
\item generate the PMT pulse signal taking into account the specific design of the front end and of the digital electronics chain of Borexino.
\item simulate the trigger generation and save the final output for triggering events.
\item produce a raw data file with the same structure as the one produced by the Borexino data acquisition system.
\end{enumerate}


Technically, the code is structured in three separate programs, which have to be chained:
\begin{itemize}
\item the event generation and light tracking (described in Sec.~\ref{sec:EvGen}), accomplishing the tasks 1 to 5.
\item the simulation of the electronics response (discussed in Sec.~\ref{sec:MCbxelec}), realizing the items 6 to 8.
\item the event reconstruction, that converts the binary data generated by the DAQ into physical observables, such as number of photoelectrons, event position, pulse shape variables. 
This code is the same for both real and MC data. Its short description can be found in Sec.~\ref{sec:reco}.
\end{itemize}
The code developments were driven by the constant comparison of simulations with calibration and real data.

\section{Event generation and light tracking}
\label{sec:EvGen}
The event generation is implemented within the Geant4 package and uses the standard libraries therein~\cite{bib:Geant4}. 
Geant4 is an object oriented \texttt{C++} toolkit for the simulation of the passage of particles through matter. 

A high precision simulation depends on a careful description of the system geometry and construction materials.
The rich geometry tools of the package allow to obtain the necessary accurate description of the elements of the detector and of the physical properties of solid components, liquids, and light sensors.

\subsection{Outer detector geometry}

\begin{figure}[tb]
\centering
\includegraphics[width=0.53\columnwidth]{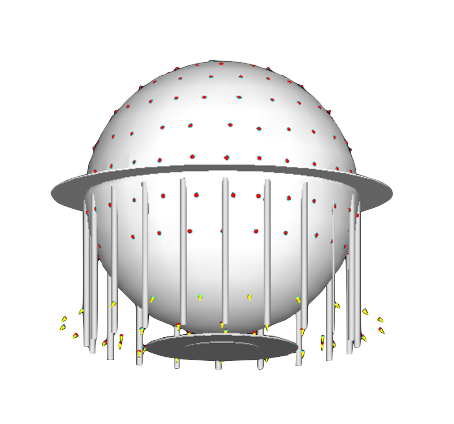} 
\includegraphics[width=0.43\columnwidth]{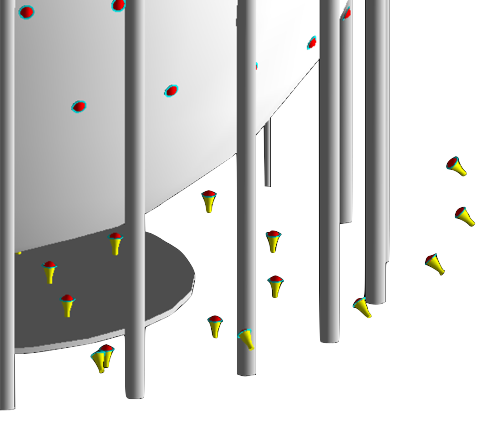} 
\caption{Simulated geometry for the Borexino outer detector in Geant4.}
\label{fig:simgeo1}
\end{figure}

\begin{figure}[bt]
\centering
\includegraphics[width=0.7\columnwidth]{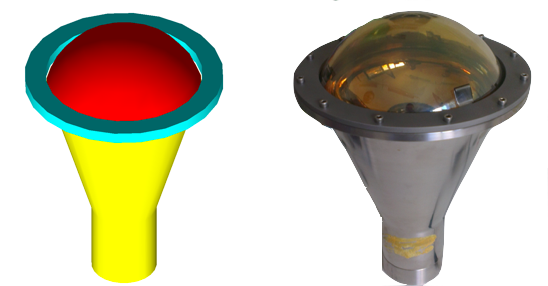} 
\caption{Comparison between the schematization of the outer detector PMT in the simulation (left) and the real object (right).
}
\label{fig:odpmt}
\end{figure}

The Geant4 implementation of the outer detector geometry is shown in Fig.~\ref{fig:simgeo1}.
Important elements are the legs supporting the stainless steel sphere (SSS) and the steel platforms at the bottom of the detector, which 
were inserted as shielding against the rock radioactivity~\cite{bib:BxDet}.
The simulation precisely follows the detector geometry.
PMTs are placed on the floor of the water tank and on the outer surface of the SSS.
The schematization of the outer detector PMT geometry directly follows the real design, as it is shown in Fig.~\ref{fig:odpmt}. 
PMTs are enclosed in an outer shielding against water and pressure, and this is reproduced in the simulation. 
For PMTs attached to the SSS, reflective tyvek foils cover theit surface
leaving free only the photocathodes facing the water. The tyvek foils placed on the inner surface of the water tank are also simulated.

Besides the geometrical details implemented in the simulation, the optical properties of all the materials involved are taken into account according to specific measurements or data available
in the literature. In the specific case of the outer detector simulation, the tyvek reflectivity dependence upon the wavelength, the PMT quantum efficiencies, and the water properties are of particular importance.

\subsection{Inner detector geometry}
\label{sec:IDgeometry}

\begin{figure}[tb]
\centering
\includegraphics[width=0.9\columnwidth]{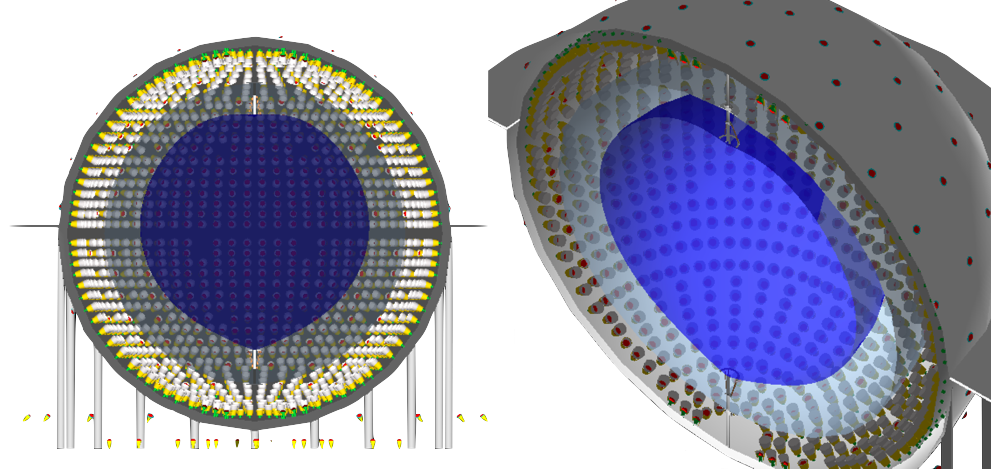} 
\caption{Simulated geometry for the Borexino inner detector in Geant4. The realistic inner vessel shape (dark blue) is updated on a weekly basis.}
\label{fig:simgeo2}
\end{figure}

Figure~\ref{fig:simgeo2} shows a cross section of the  Borexino inner detector as implemented in the simulation. All the most relevant geometrical features are included, such as the PMTs
with their real shapes and positions, the nylon vessels, and the holding endcaps. 
Special tools were added to consider the real shape of the inner vessel and its time dependent shape that deviates from a perfect sphere. As described in Ref.~\cite{bib:BxLong},
the inner vessel profile is determined on a weekly basis  by using background events from vessel contamination and external \grs. 
The $r- \theta$ distribution of these events is fitted and used as input of the MC simulation for the proper simulation of the vessel shape.
Therefore, when simulating data throughout a long period of time, the vessel evolution is taken into account. This is important, 
since the vessel shape can affect both the amount of external contaminations reaching the standard fiducial volume (FV) and the energy response of the detector 
(see e.\,g. Sec.\ \ref{sec:MCtuninglightcollection} and Sec.\ \ref{sec:extcomparison}). 
The granularity with which the real vessel shape is simulated can be adjusted by the user with optimal values usually 
around $\sim$5~cm (i.\,e. the shape is approximated with a polygonal shape in the $r-\theta$ plane with sides of $\sim$5~cm). 
This is a good tradeoff between the intrinsic uncertainty of the vessel shape determination (a few $\mbox{cm}$~\cite{bib:BxLong}) and an optimized code performance.

In Fig.\ \ref{fig:simgeo2} the nylon vessel endcaps are also visible. They support the vessels and provide the connection between the innermost volumes and the Borexino purification plants.
A careful simulation of the endcaps is important, since they induce light shadowing and reflections, which affect the reconstruction for events close to the north and the south poles of the inner vessel.

\begin{figure}
 \centering
   \includegraphics[width=0.58\columnwidth]{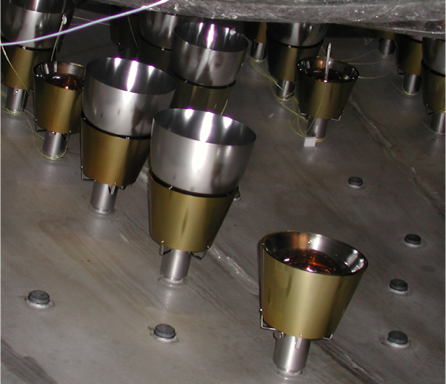}
 \begin{minipage}[b]{0.4\columnwidth}
  \includegraphics[width=0.9\columnwidth]{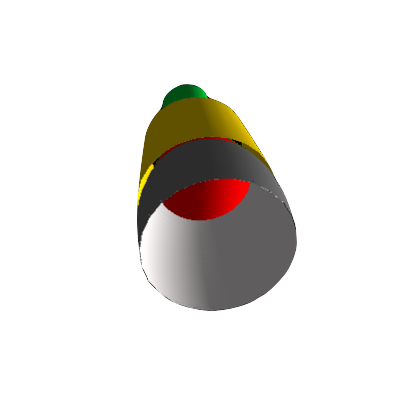} \\
\includegraphics[width=0.6\columnwidth]{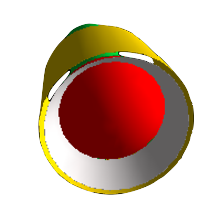} 
 \end{minipage}
 \caption{\emph{Left Panel:} Real Borexino inner detector PMTs mounted on the stainless steel sphere. \emph{Right Panel:} MC schematization of the PMTs with (top) and without (bottom) light concentrators.} 
\label{fig:simgeo3}
\end{figure}

The key components of the Borexino geometry simulation are the PMTs of the inner detector. Their geometry, materials, and optical properties highly impact the detector response.
In Fig.\ \ref{fig:simgeo3}, a comparison between the real PMTs mounted on the inner surface of the SSS and the MC modeling is shown. As anticipated in Sec.~\ref{sec:detector},
$1828$ PMTs are equipped with light concentrators, while $384$ are not.
The two different configurations are shown both in the left and right panels of Fig.\ \ref{fig:simgeo3}.
PMTs with concentrators are simulated as shown at the top of the right panel of Fig.\ \ref{fig:simgeo3}, while PMTs without concentrators are simulated as shown at the bottom
of the same panel.
The schematization of the inner detector PMTs is made of a common PMT base (almost invisible, but in green in the right panel of Fig.\ \ref{fig:simgeo3}) on top of which the photocathode is placed
(red in the picture). The photocathode is made of bialkali and it is a spherical cap. The whole PMT body is surrounded by a conical $\mu$-metal shielding (yellow in the picture),
which reduces the effect of the Earth magnetic field on the PMT response. 

\begin{figure}[b]
\begin{center}
\includegraphics[width = \columnwidth]{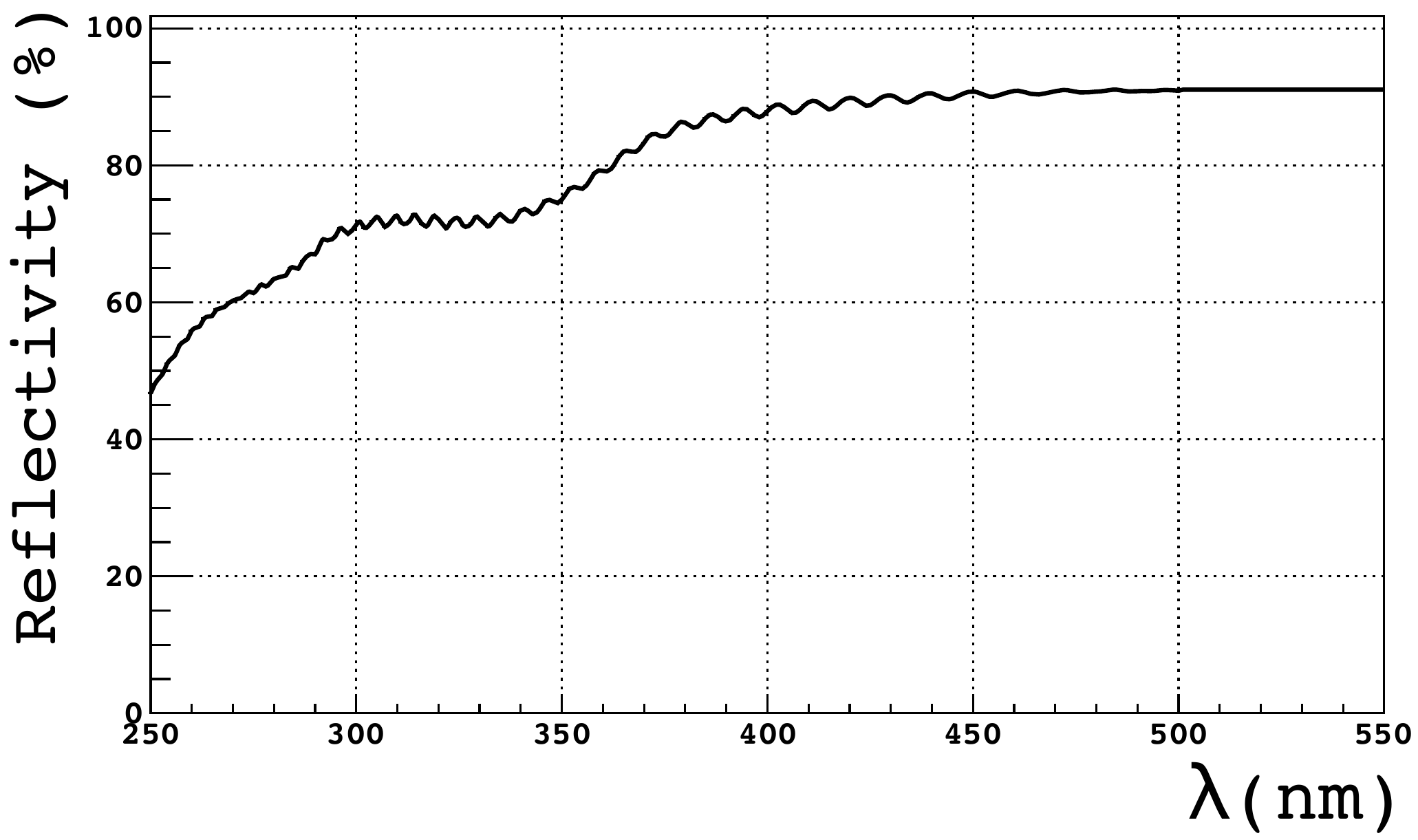}
\caption{Reflectivity of polished aluminum light guides as a function of photon wavelength~\cite{bib:BXLC}.}
\label{fig:LGRefl}
\end{center}
\end{figure}

The light concentrator shape is reproduced in the simulation with sub-$\mbox{cm}$ precision, 
since it significantly affects the amount of light collected as a function of the event position. 
PMTs without concentrators are equipped with a small steel ring which surrounds the photocathode and supports it. 
Figure~\ref{fig:LGRefl} shows the wavelength dependence of the aluminum reflectivity of light concentrators as implemented in the simulation. 

Particular care was devoted to the implementation of the optical properties of the simulated materials in the inner detector, since they are the most relevant for solar neutrino analyses.
Apart from the scintillator and buffer properties, which are discussed in more detail in Sec.\ \ref{sec:lighttracking}, the nylon vessel absorption length as a function of wavelength
and its index of refraction are also included in the simulation. The vessel transparency was measured with a dedicated  campaign carried out at the time of the vessel 
construction. 
The PMT quantum efficiency spectral dependence was implemented according to the specifics provided by the manufacturer~\cite{bib:BxDet} and it is shown in Fig.\ \ref{fig:ppoemission}.
Furthermore, the light concentrators and the SSS inner surface are modeled with a finite probability that the incoming light is reflected or absorbed. In case of reflection,
the photon can be reflected specularly or diffusively (Lambertian diffusion). 
The same holds for the steel rings of the PMTs not equipped with cones, since they act similarly to light concentrators albeit at a reduced efficiency.
The precise determination of all these coefficients is performed through the tuning procedure outlined in Sec.\ \ref{sec:tuning}.

\subsection{The particle energy loss}
\label{sec:energy_loss}

The Geant4 libraries allow to simulate radioactive decay chains and energy losses of every type of particle (electrons, \grs\, positrons, neutrons, $\alpha$-particles...) 
down to 100\,eV in each material of the detector. 
For all charged particles, the physics of interactions and propagation is completely handled through Geant4 libraries.
Special modules developed within the collaboration generate scintillation and \v Cerenkov photons according to the particle type and to the models of light generation and propagation 
described in Sec.~\ref{sec:lighttracking}.

\subsubsection{Positron annihilation}
Positrons slow down in the scintillator and a fraction of them are captured by electrons before annihilation, forming para- or ortho-positronium.
In the former case, the emission of two 511\,keV \grs\ is almost immediate, while in the latter it is delayed by a few ns. 
The ortho-positronium lifetime in the PC+PPO mixture was measured in Ref.~\cite{PhysRevC.83.015504} and is $\sim$3\,ns, while the ortho-positronium formation probability was found to be close to 50\%. 
The ortho-positronium lifetime is comparable to the fastest scintillation time constant ($\simeq$4\,ns, see Sec.~\ref{sec:tuning_timeresp}) 
and therefore its formation has a significant impact on the light pulse shape. This phenomenon can be used to identify and subtract $\beta^+$ radioactive backgrounds~\cite{bib:BxLong}. 
We developed a custom physics module that handles this effect. 
According to the measured formation probability, we model the positron energy loss and either the para- or ortho-positronium formation and decay. 
The two processes are simulated as three-body vertexes, composed by the positron track and two delayed annihilation $\gamma$ decay. The difference between the former and the latter
is that the annihilation happens immediately after the positron is stopped. 
Ortho-positronium in vacuum decays emitting three \grs\, while in the liquid scintillator it mainly annihilates emitting only two \grs.
This is due to the \emph{pick-off} effect, for which the positron of an ortho-positronium state annihilates with an electron from the surrounding scintillator. 
The delay of the \gr\ generation is  described by a simple exponential law depending on the ortho-positronium mean life in the scintillator.

\subsection{Primary event generators}
\label{sec:primary_generators}
Several generators were developed to properly simulate radioactive decays inside the scintillator, solar neutrinos, radioactive sources encapsulated as in the calibration campaign, 
the \ce{^{241}Am}-\ce{^{9}Be} source, and anti-neutrino interactions.
 
\subsubsection{Radioactive decays}
Standard Geant4 classes manage radioactive decays providing the correct daughter spectra and branching ratios for most of the radionuclides. 
The radioactive decays which we treat differently from the Geant4 standard approach are those of \C\ and \Bi.

The beta decay of \C\ into \ce{^{14}N} is an allowed ground-state-to-ground-state transition. This decay was investigated both theoretically and experimentally 
by many groups, but some unsatisfactory features remain. For instance, its anomalously long half-life ($\sim$5730 years) with respect to ``standard'' beta decays has been subject of considerable interest.
In addition, there have been different experimental investigations aiming to assess the deviations from the expected allowed decay spectrum~\cite{bib:Kuzminov2000}.
We developed a generator which allows to simulate the \C\ beta decay spectrum with a shape factor, i.\,e. a quantification of the deviation from the allowed shape,
either from Ref.~\cite{bib:Kuzminov2000} or from Ref.~\cite{montara}.

The \Bi\ decay is a first-forbidden beta decay and thus its spectral shape cannot be predicted in a straightforward way. In our simulation, the \Bi\ event generation 
is handled in such a way that the shape factor can be modified, in order to use the differences between the various models to evaluate
systematic uncertainties. 
The standard spectral shape which we use is that reported in Ref.~\cite{bib:DANIEL1962293}. An additional spectral shape is the one proposed in Ref.~\cite{Flothmann1969},
which differs by a few $\%$ from the standard one.

 $^{11}$C is simulated considering the probability of ortho-positronium formation as discussed in Sec.~\ref{sec:energy_loss}.
 
  \subsubsection{Solar neutrinos}
We developed a custom generator for simulating solar neutrino elastic scattering off electrons. 
The primary solar neutrino energy spectra are those computed in Ref.~\cite{bahcall, Winter:2004kf}.
The electron solar neutrino survival probability is computed according to Ref.~\cite{deHolanda:2004fd} and the mixing parameters are those reported in Ref.~\cite{Capozzi:2016rtj}. 
The $\nu_{e}-e$ and $\nu_{\mu/\tau}-e$ elastic scattering cross sections as a function of energy are computed following Ref.~\cite{Bahcall:1995mm}. 
Finally, the electron recoil energy is sampled including radiative corrections at next-to-leading order as calculated in Refs.~\cite{Bahcall:1995mm, pdg, Passera:2000ug}.

\subsubsection{$^{241}$Am-$^{9}$Be neutron source}
The \AmBe\ neutron source provided calibration events up to about 10\,MeV, due to neutron captures on different nuclei. 
Neutrons are mainly produced by the following reactions: $^9\mbox{Be}(\alpha,n)^{12}\mbox{C}_{\mbox{gs}}$ and $^9\mbox{Be}(\alpha,n)^{12}\mbox{C}^{*}$. 
The output neutron energy spectra are described in Refs.~\cite{bib:Vijaya, bib:Vilaithong}. They are used for simulating the primary neutrons emitted by the source. The $^{12}\mbox{C}^*$ de-excitation through the emission of $4.44$\,MeV \gr\ is simulated as well. 
The neutron scattering during the thermalization and its subsequent capture by protons or nuclei are directly managed by the Geant4 libraries as well as the de-excitation of the daughter nucleus. 

\subsubsection{Geo and reactor anti-neutrinos}
 Geo-neutrino energy spectra are computed from the $\beta^{-}$ spectra of the \U\ and \Th\ decay chains. Reactor anti-neutrino spectra are obtained from Ref.~\cite{PhysRevD.91.065002}. 
 Since anti-neutrinos are detected via inverse beta decay, the primary vertex is simulated as a positron-neutron pair emitted simultaneously from the same vertex.
    
\subsubsection{SOX anti-${\nu}$  source}
 A custom generator was developed to simulate the SOX anti-neutrino source~\cite{bib:thesisMikko}. 
 Anti-neutrinos emitted by \ce{^{144}Ce}-\ce{^{144}Pr} are generated according to Fermi's theory of $\beta$ decay, including corrections for the finite size and mass of the nucleus, 
 weak interaction finite size corrections, radiative corrections and screening corrections.  
 Further studies are currently under way to measure the shape factor of the forbidden beta decay.
    
\subsubsection{External background}
\label{sec:EB}
As described in Ref.~\cite{bib:BxBe2}, the SSS, the PMTs, and the light concentrators are
contaminated by non-negligible amounts of \Th, \U, and \ce{^{40}K}. The long-ranged \grs\ from the daughter nuclides \Tl\ (2.61\,MeV) and $^{214}$Bi (up to 3.27\,MeV) 
are the main contribution to the ``external background'' reaching the inner part of the detector from further-out components.
The knowledge of their spectral shapes is of paramount importance for several solar neutrino analyses, 
and particularly for the measurement of the \pep\ and \bor\ neutrino interaction rates~\cite{bib:BxPep, bib:BxB8}, as well as for constraining the CNO neutrino interaction rate. 
An appropriate understanding of the external background was achieved by combining the information obtained from the external calibration of the detector~\cite{bib:BxCalib} 
with the development of a special simulation procedure for the \grs\ originating in the PMTs and in the SSS. 

The rate of the external \grs\ reaching a fiducial sphere of 3\,m radius is reduced by a factor $\sim10^7$ with respect to that directly generated by the PMTs and the SSS.
This is estimated by considering the attenuation length of about 23\,cm measured with the calibration data.
The multiple interactions of \grs\ in the liquids produce an energy spectrum that cannot be modeled analytically.
However, the attenuation factor is so large that it is impossible in practice to perform a brute force simulation, i.\,e. it is not feasible to generate events in the periphery of the detector 
and then to trace the minute fraction of \grs\ reaching the fiducial volume. 
Therefore, aiming at a considerable reduction of the computing time, we applied the {\it importance sampling algorithm}~\cite{Geant4Manual} (implemented inside the Geant4 libraries) as variance reduction method 
to our MC simulation. These methods are typically used to simulate radiation transport through thick shields of radioactive sources. 

The algorithm considers the volume surrounded by the SSS as divided into spherical concentric shells. A number called ``importance'' is associated to each shell.
Every time a photon crosses a boundary between two shells, the particle is either split or killed with a given probability depending on whether 
the importance of the entering volume is higher or lower than the exiting one.

The simulation is performed in three steps. First, radioactive decay events are simulated on the PMT photocathode surface and on the SSS
and the importance sampling algorithm is applied. All \grs\ are propagated until they either are fully absorbed or reach the inner vessel. 
In case a $\gamma$-ray reaches the inner vessel, the kinematics of the first interaction inside this volume is saved. 
In the second step, the importance sampling is not applied and neither scintillation nor \v{C}erenkov light emissions are simulated. All \grs\ found to interact at least one time in the inner vessel
are generated according to the previously saved vertices (step one) and all the energy deposits are simulated until the \gr\ is fully absorbed. 
If some of the energy deposits are contained inside a sphere whose radius is typically 0.5\,m greater than the FV radius, 
the third and final step is performed: optical photons are generated according to the previously computed energy deposits and tracked until they reach the PMTs, like in standard simulations.
This algorithm simulates external background events interacting inside the 3\,m spherical FV at a rate of about $0.5$ events per second.
More details can be found in Ref.~\cite{bib:tesisimone}.

\subsubsection{GENIE and other event generators interface}
A custom interface was written to simulate events produced by GENIE Neutrino Monte Carlo Generator~\cite{GENIE}. 
Its output data format is a ROOT TTree~\cite{root} with
information on the particles produced and their interactions. The interface reads the characteristics of
the particles, such as type, energy, direction, position, and time, and generates
corresponding Geant4 events. In addition to GENIE, any external event generator
using TTree data format can be used.

\subsection{Generation of scintillation light}
\label{sec:lighttracking}
\begin{figure}[tb]
\centering
\includegraphics[width=\columnwidth]{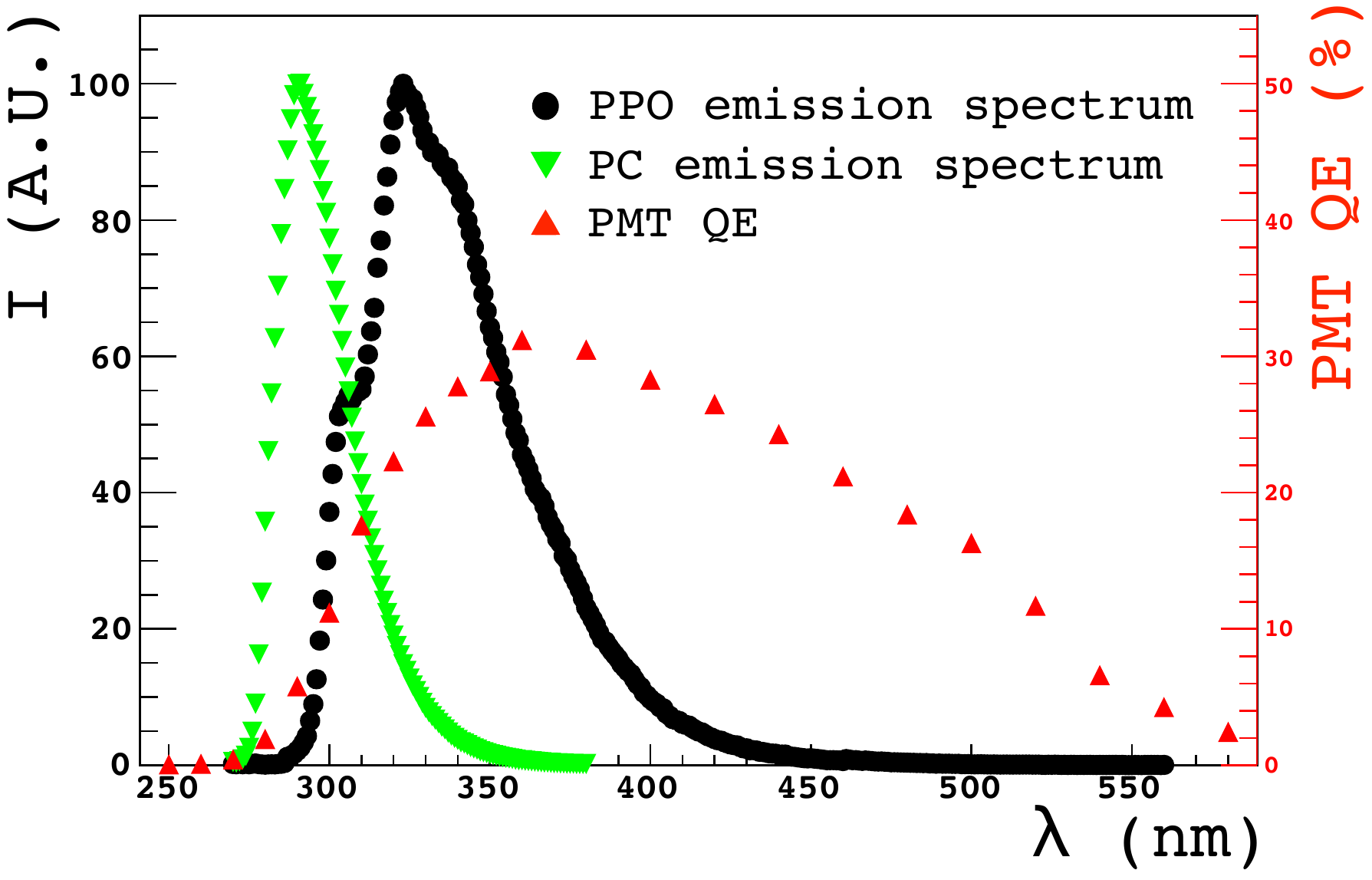} 
\caption{PPO (black) and PC (green) emission spectra. The two curves are normalized to 100 at the maximum.
The red points show the PMT Quantum Efficiency (QE) as a function of the wavelength.}
\label{fig:ppoemission}
\end{figure}

Scintillation photons are sampled according to the emission spectrum of PPO if they are generated in the inner vessel and according to that of PC if they are generated in the buffer region ~\cite{bib:BxEmission}. 
These spectra are shown in Fig.~\ref{fig:ppoemission}.

The number of emitted photons and their time distribution in the PC+PPO mixture depend on the details of the charged particle energy loss processes and on the molecular interactions 
between solvent (PC) and fluor (PPO) in the scintillator. We briefly remind that the fluorescence is a property of single molecules in organic scintillators and 
it depends on the features of the carbon bonds called $\pi$ bonds.
The $\pi$ electron energy levels are quantized in a series of singlet $S_{ij}$ and triplet $T_{ij}$ states, where $i=0,1,2..$ denotes the electron energy levels 
and $j=0,1,2,..$ vibrational sub-levels. $\pi$ electrons are those responsible for the fluorescence mechanism. Scintillation light is emitted when $\pi$ electrons decay from the
first excited state $S_1$ to one of the vibrational energy levels of the fundamental state. More information on the organic scintillation processes can be found in Ref.~\cite{knoll}.


Charged particles lose energy in the scintillator mainly exciting or ionizing the solvent. 
The energy is then transferred to the fluor either by radiative (emission of photons from the solvent and subsequent absorption from the fluor) or non radiative 
processes (dipole-dipole interactions between excited solvent and solute molecules). 
For fluor concentrations like the one of Borexino, non radiative transfer dominates. Once PPO is excited, it can emit fluorescence light only through the de-excitation of $\pi$ 
electrons. Excited electrons of the fluor promptly reach (through non radiative processes) the first excited state $S_{10}$, from where they decay to the fundamental state $S_{0j}$ emitting
scintillation light, following an exponential time distribution with a time constant of $\sim$1.6~ns. 
If some of the excitation energy is lost by collisions, and the electron is transferred to a triplet excitation state, $T_{10}$, it is impossible for the electron to  de-excite emitting light
but also return to the singlet state by non-radiative transition (being $E_{T_{10}}<E_{S_{10}}$). The only process available is the interaction with another solvent molecule in the same triplet state and this is the mechanism generating the delayed fluorescence. 

The time distribution and the light yield of the emitted fluorescence photons depend on the various molecular processes taking place as a consequence of the particle energy loss.
In addition, they do not only depend on the energy deposit in the scintillator but mainly on how the energy is released, i.\,e. on the value of $dE/dx$ and on the type of incident particle.
Heavy ionizing particles like $\alpha$'s feature a large $dE/dx$, and produce large ionization or excitation density, thereby increasing the probability to get the triplet excitation state $T_{10}$ and thus 
the presence of delayed fluorescence. In addition such large ionization or excitation densities  favor molecular processes in which the energy is dissipated in non radiative ways, resulting in the quenching of 
the scintillation light (ionization quenching)~\cite{bib:Birks}.

The light emission cannot be simulated at a molecular level, because too many molecules and processes should be considered, resulting in a huge simulation time. 
We modeled the emission of fluorescence light by considering the difference in the effective emission time profiles induced by different types of particles and 
by including the ionization quenching effect.

The light emission time $\tau$ in the PC+PPO scintillator is generated in the MC according to the following phenomenological distribution:
\begin{equation}
P(\tau) = \sum_{i=1} ^4 \frac{w_i}{\tau_i} \exp{-\tau/\tau_i},
\label{tau}
\end{equation} 
where the $\tau_i$ values and their weights $w_i$ were obtained fitting the data from a dedicated experimental setup and then optimized during the tuning of the MC (see Sec.~\ref{sec:tuning_timeresp}).

The light quenching due to ionization is modeled by the Birks formula~\cite{bib:Birks} that links 
the scintillation light yield $dY^{ph}$ produced when a particle loses energy over a distance $dx$ with a stopping power $dE/dx$:
\begin{equation}
\frac{dY^{ph}}{dx} = \frac {Y_0^{ph} dE} {1+k_B \cdot dE/dx},
\label{eq:Birks}
\end{equation} 
where the material-dependent Birks factor k$_B$ (of the order of $10^{-2}$ cm/MeV) and the primary scintillation yield $Y_0^{ph}$ have to be determined for every particular scintillator and incident particle. 
The total number of emitted scintillation photons is obtained by integrating Eq.~(\ref{eq:Birks}).
We define the quenching factor $Q_p(E)$ as
\begin{equation}
Q_p(E) = \frac {1} {E} \int_0^E   \frac {dE} {1+k_B dE/dx}.
\label{eq:birks}
\end{equation} 
$Q_p(E)$ is always lower than 1. The suffix $p$ recalls that $Q_{p}(E)$ depends on the particle type $p$ ($\alpha$, $\beta$, or $\gamma$) through k$_B$ and $dE/dx$.
Only if k$_B \cdot dE/dx \ll 1 $ (for instance for electrons of some MeV) the light yield $Y_p^{ph}(E)$ is approximately proportional to the initial particle energy, while in general the following relation holds:
\begin{equation}
Y_p^{ph}(E) = Y_0 Q_p(E) E.
\label{Nolinear}
\end{equation}
The above equation evidences the intrinsic non linearity between the deposited energy $E$ and the emitted scintillation $Y_p^{ph}(E)$.
Deviations of the light yield as a function of the energy deposit from a linear law are increasingly important for protons, $\alpha$-particles, and nuclear fragments,
because of their high ionization yield per unit length. The quenching effect is relevant also for \grs.
As discussed in Ref.~\cite{bib:BxLong}, the amount of scintillation light emitted when a \gr\ with energy $E$ is fully
absorbed in the scintillator is slightly lower than that emitted by an electron with the same energy $E$ (e.\,g. the light emitted by a 256\,keV $\gamma$-ray is similar to the light emitted by a 220\,keV electron \cite{bib:BxEDec}).
The scintillation yield $Y_{\gamma}$ and the quenching factor $Q_\gamma(E)$ for \grs\ are given by 
\begin{equation}
Y_{\gamma} = Y_0 \sum_i E_i Q_\beta(E_i) \equiv Y_0 \cdot Q_\gamma(E) \cdot E,
\label{eqn:Qgamma}
\end{equation}
where the sum is performed over all the electrons generated during the \gr\ interaction.
As $Q_\beta(E)$ decreases as a function of the energy, it results that $Q_\gamma(E)$ is smaller than $Q_\beta (E)$ for the same energy $E$.
As a result, the quenching factor is not negligible for \grs\ with $E$ in the MeV range.

The Birks parametrization is a macroscopic description of quenching and it can not be used directly in a stochastic approach, such as a MC simulation. 
In the Birks model, the quenching effect is obtained by comparing the initial energy of the primary particle with the energy deposited in the detector. 
In particular, all energy deposits due to secondary particles (like $\delta$-rays or \xrs) are assumed to belong to the primary. 
On the other side, in the MC approach each particle is treated independently. 
Therefore, a correct parametrization of the Birks formalism requires to make the model compatible with the Geant4 framework, by evaluating the quenching factor for the primary ionizing particle, 
and making each daughter inherit the same factor. The ingredients for this simulation approach are: an \emph{a priori} parametrization of the energy loss (as required by Eq.~(\ref{eq:birks})) 
and a table of quenching factors as function of the energy (built at the initialisation phase, to speed up the simulation).

The light generation in the detector buffer medium (PC+DMP) is modeled according to the available measured data~\cite{bib:DMP}.
The total scintillation yield in the buffer is about $\sim$4\% of that of Borexino scintillator. The scintillation time constant in the buffer is 2.8\,ns and the wavelength spectrum 
of buffer scintillation photons (which coincides with the PC emission spectrum) is presented in Fig.~\ref{fig:ppoemission}. 

\subsection{Generation of \v{C}erenkov light}
\label{Cerenkov}

The primary spectrum of \v{C}erenkov light is simulated by generating a number of photons $N_{\mbox{\small \v C}}$ per unit length and wavelength according to the Frank-Tamm equation~\cite{pdg}:
 \begin{equation}
 \frac{d^2N_{\mbox{\small \v C}}}{dx d\lambda} =
 \frac{2\pi\alpha}{\lambda^2} \left( 1- \frac{c^2}{\rm{v}^2 \cdot n^2(\lambda)}
 \right),
 \label{eq:Cherenkov}
 \end{equation}
 where $\alpha$ is the  fine structure constant, \noflambda\ is the wavelength dependent refraction index in the scintillator and $v$ is the particle velocity in the scintillator.  
 Note that the refractive index dependence on the wavelength $\lambda$ causes a dependence  of the \v Cerenkov  spectrum on the particle velocity  since the condition
\begin{equation}
 \left( 1- \frac{c^2}{\rm{v}^2 \cdot n^2(\lambda)}  \right) >0 
 \end{equation}
 must be satisfied.
 
 \begin{figure}[tb]
\centering
\includegraphics[width=1\columnwidth]{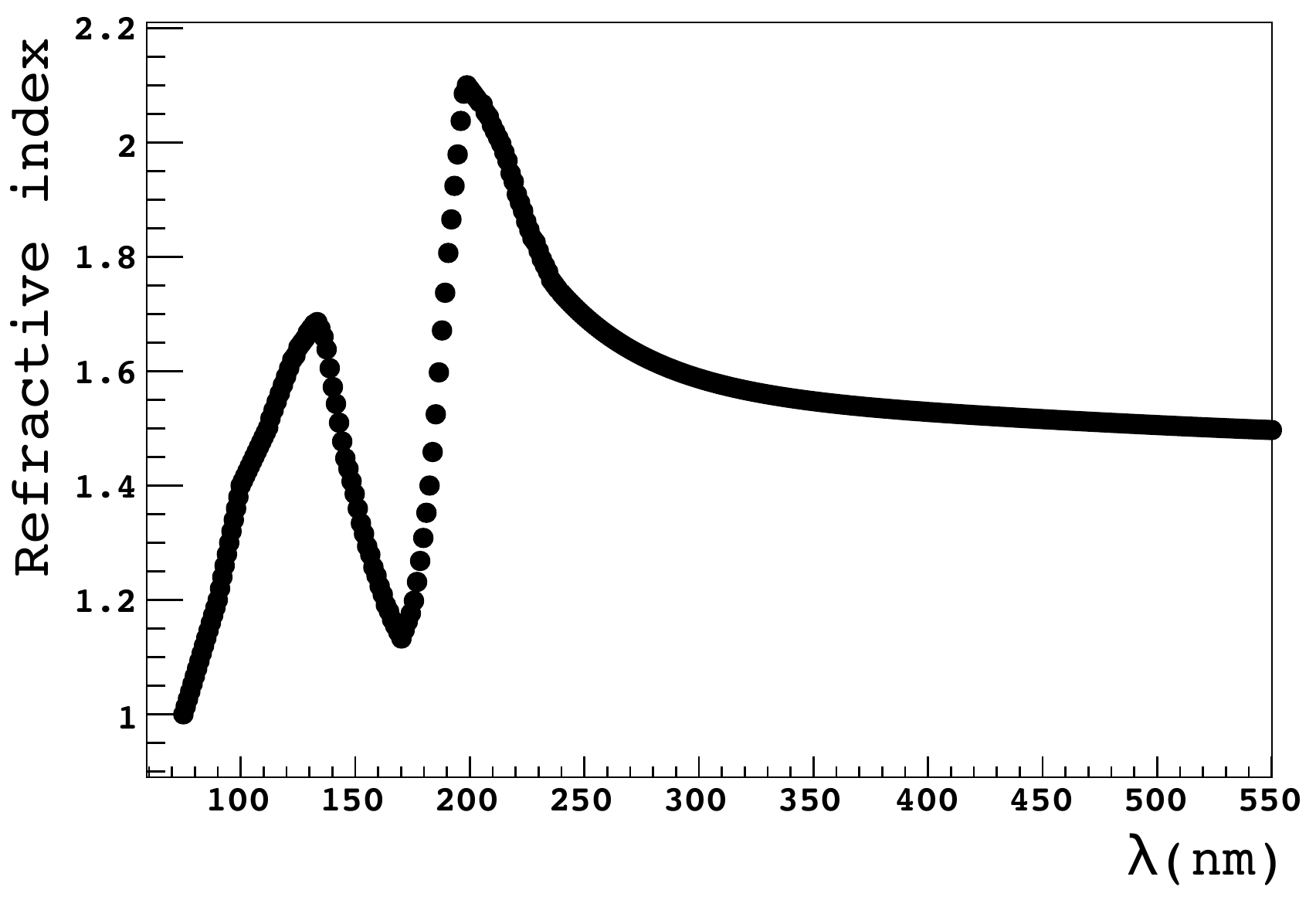}
\caption{Refractive index of the Borexino scintillator as a function of the optical photon wavelength.}
\label{fig:refraction}
\end{figure}

  The spectrum of \v{C}erenkov photons extends into the ultraviolet region, which is not directly detectable by the PMTs. The mean free path of this ultraviolet light in the scintillator is very short (sub mm),
  therefore, these photons are almost completely absorbed by the scintillator. However, the scintillator reemits a fraction of these photons at longer wavelengths, allowing for an indirect detection.

 As evident from Eq.\ (\ref{eq:Cherenkov}), the knowledge of the dispersion relation \noflambda\ is essential in order to properly simulate the \v{C}erenkov effect.
 Direct measurements allowed to obtain \noflambda\ for the Borexino liquid scintillator in the wavelength range $\lambda \in (245, 1688)$\,nm.
 The number of generated \v{C}erenkov photons is proportional to $\lambda^{-2}$ and therefore an accurate knowledge of \noflambda\ in the UV region is important. 
 Since direct measurements at so short wavelengths are not available for the Borexino scintillator, the measured \noflambda\ was extended in the UV region using the benzene data in 
 literature~\cite{benzene, nistbenzene}, considering that benzene and PC have similar refractive indices. The curve was furthermore extrapolated to the value $\mbox{n}=1$ in the 
 deep ultraviolet region. The knowledge of the scintillator properties in the deep ultraviolet region is not accurate and thus both the refractive index and the probability $P_{rem}$ of light reemission
 by PPO in this photon energy region are experimentally unbounded parameters, which are fixed at the tuning stage of the simulation optimization (see Sec.~\ref{sec:tracking} and Sec.~\ref{sec:tuning}).
 Figure~\ref{fig:refraction} shows the resulting \noflambda, which is used both for \v Cerenkov photon generation and optical photon tracking (see Sec.~\ref{sec:tracking}).

The standard Geant4 G4Cerenkov library assumes that \noflambda\ is an increasing function of the photon energy. 
As it is shown in Fig.~\ref{fig:refraction}, this is not our case and thus the standard library was extended to allow the usage of an arbitrary \noflambda.

\subsection{The light tracking}
\label{sec:tracking} 
\begin{figure}[bt]
\centering
\includegraphics[width =\columnwidth]{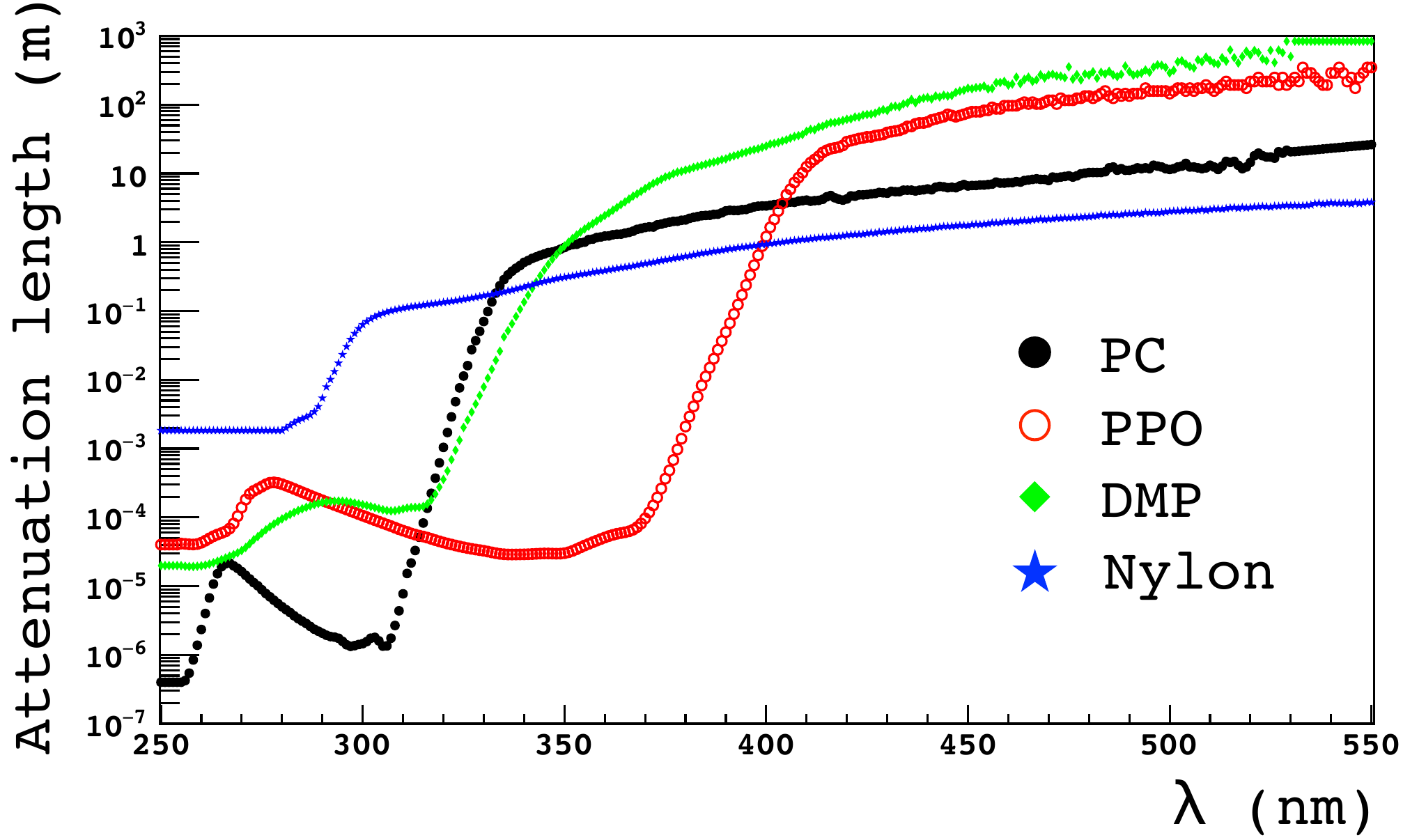}
\caption{Attenuation length of PC (black disks), PPO (red circles), DMP (green diamonds), and nylon (blue stars) as a function of the wavelength.
As a reference, the PPO emission spectrum peaks between 300 and 400 nm (see Fig.\ \ref{fig:ppoemission}).
}
\label{fig:attenuation}
\end{figure}

The Borexino MC tracks each optical photon individually, considering its interactions with the single components of the scintillator and the buffer. 
These processes include elastic Rayleigh scattering, absorption and reemission of photons by PPO molecules, absorption of photons by DMP, and also photon absorption in the thin nylon vessels.
The cross-sections (or equivalently the wavelength dependent attenuation length) for these interactions were obtained with dedicated spectrophotometric 
measurements as  shown in Fig.~\ref{fig:attenuation}.

Photons emitted by scintillation or \v Cerenkov processes can interact with PC, PPO, DMP, or nylon molecules on their paths to the PMTs. 
In the model implemented in our MC simulation, if photons interact with DMP or nylon molecules, they are always absorbed. 
In case of interaction with PC, two distinct cases are considered, and their definitions are related to the PC emission spectrum shown in Fig.~\ref{fig:ppoemission}. 
Optical photons with a wavelength  $\lambda>310\,\mbox{nm}$ undergo Rayleigh scattering with an angular distribution 
$P(\theta) = 1 + \cos^2\theta$, without time delay and without shift in energy. At shorter wavelengths, the interaction with 
PC is simulated as absorption by PC molecules, followed by the energy transfer to PPO (if the interaction happens in the inner vessel). 
With a probability of $82\%$, the PPO subsequently reemits the photon 
with an exponentially distributed emission time with a constant $\tau_{PCtoPPO} = 3.6\,\mbox{ns}$.
In the buffer medium, where PPO is not present, the interaction with PC for $\lambda<310\,\mbox{nm}$ is followed by subsequent PC scintillation according to the PC spectrum of Fig.\ \ref{fig:ppoemission}
with a probability of $0.04$ and with an exponential time distribution with $\tau_{PC} = 2.8\,\mbox{ns}$. 

\begin{table}
\centering
\begin{tabular}{c||c} 
 Optical photon                                                      &  Reemission   \\
 wavelenght $\lambda$  (nm)                                                     &  probability $P_{rem}$ \\[0.15cm]  \hline \hline
   $\lambda<320\,\mbox{nm}$ & 0.53* \\
   $320\,\mbox{nm}<\lambda<375\,\mbox{nm}$ & 0.839 \\
   $\lambda>375\,\mbox{nm}$ & 0.15 \\
\end{tabular}
\caption{Model for the three-plateau reemission probability function after absorption by PPO molecules as a function of the wavelength, as it is implemented in the MC.
The value highlighted with `*' is obtained through the tuning procedure (see Sec.\ \ref{sec:MCtuningenergy}).}
\label{tab:reemission}
\end{table}

Even if PC molecules are the most abundant, most of the optical photons interact with PPO.
This is caused by the substantially shorter attenuation length  of PPO in the wavelength range (300-400 nm) corresponding to the PPO emission spectrum (compare Fig.~\ref{fig:attenuation} with Fig.~\ref{fig:ppoemission}).
Therefore, the absorption and eventual reemission of optical photons by PPO is the most relevant effect on the light propagation in Borexino. 
While the wavelength-dependent attenuation length of PPO was measured with great accuracy (Fig.~\ref{fig:attenuation}),  an experimental determination of the 
reemission probability as a function of the optical photon energy $P_{rem}(\lambda)$ is challenging. The optical model describing $P_{rem}$
is summarized in Table~\ref{tab:reemission}. In practice, it is a three-plateau function in the wavelength range of interest.
No experimental data are available for $\lambda<320\,\mbox{nm}$,
and thus the value of $P_{rem}$ for this wavelength range was determined through the tuning with calibration data described in Sec.~\ref{sec:MCtuningenergy}.
Photons reemitted by PPO are produced with an isotropic distribution, according to the standard PPO energy spectrum 
and with an exponential time distribution with a time constant of $\tau_{PPO} = 1.6\,\mbox{ns}$. 

The MC simulation considers also the interactions of optical photons with interfaces (e.\,g. the refraction induced by the presence of the nylon vessels)
and surfaces (reflections on the light concentrator surfaces, on the SSS, on the PMT photocathodes).

This light propagation model was already validated in the CTF detector (Borexino's prototype)~\cite{bib:CTFLight} except for the treatment of UV photons due to \v Cerenkov production,
which is a novel development introduced in the Borexino MC.

Photons reaching the photocathodes are detected with a probability corresponding to the wavelength-dependent PMT quantum efficiency depicted in Fig.\ \ref{fig:ppoemission}.
The same spectral dependence of the quantum efficiency (QE), but different peak values, are assumed for all PMTs. 

\subsection{Effective quantum efficiencies}
\label{sec:qe}

\begin{figure}[tb]
\centering
\includegraphics[width =0.9\columnwidth]{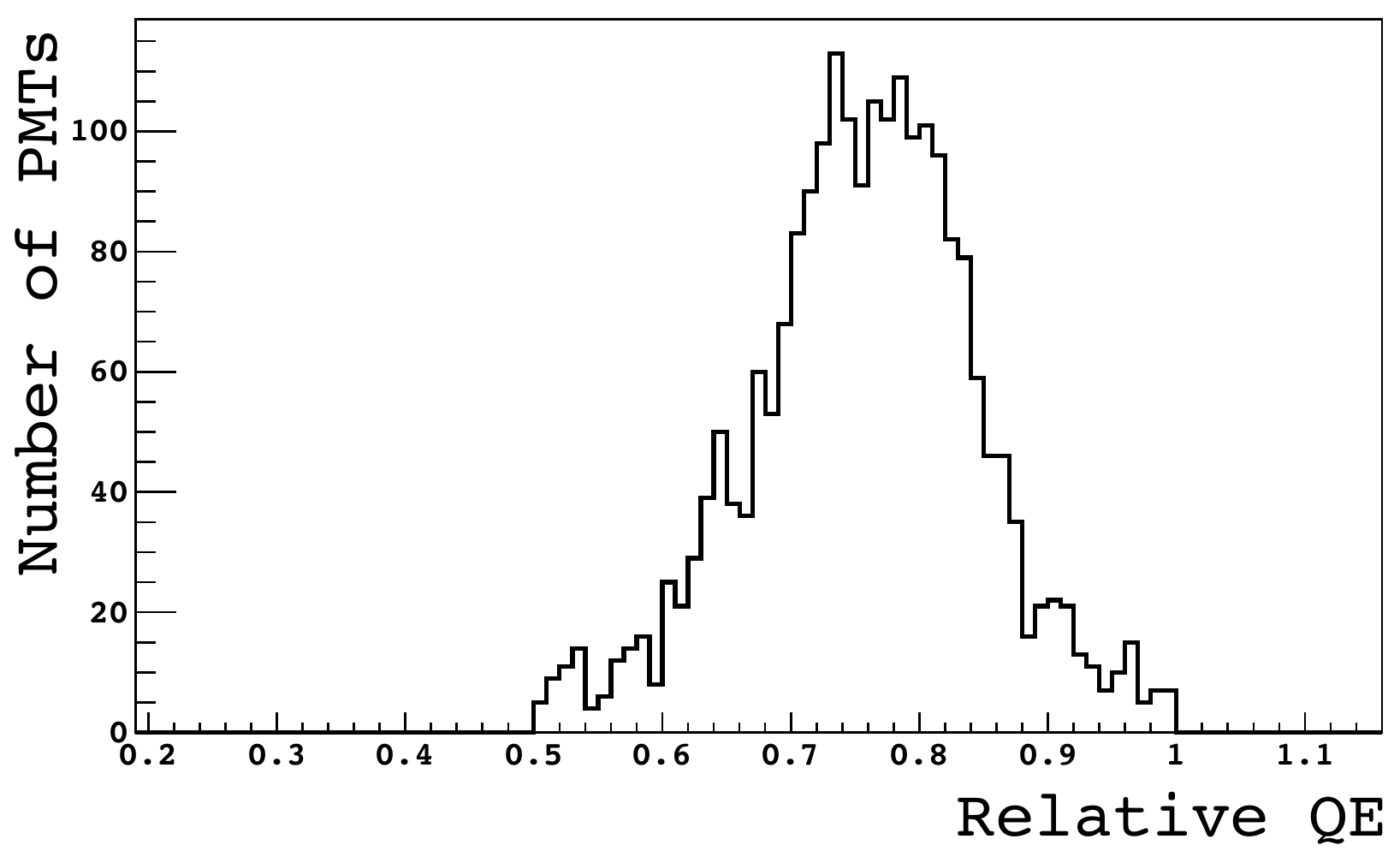}
\caption{Relative QE among the whole set of Borexino PMTs. The values are normalized so that $1$
corresponds to the highest measured QE.}
\label{fig:relqe}
\end{figure}

All Borexino PMTs have slightly different light detection efficiencies, caused by different intrinsic properties of the photocathodes. These relative quantum efficiencies
were measured prior to the PMT installation in Borexino. 
The peak values of the PMT quantum efficiency versus the wavelength (shown in Fig.\ \ref{fig:ppoemission}), normalized to the best PMT in the set, are shown in Fig.\ \ref{fig:relqe}.
The intrinsically different quantum efficiency values for the PMTs are not sufficient to properly characterize the time behavior of the detector, since the global 
detection efficiency of a single channel might vary with time. For this reason, we introduced an ``effective quantum efficiency''  that globally describes the efficiency of the PMT,
of the front end, and of the digital electronics.

The ``effective quantum efficiency'' is the frequency of a signal detection in case of a photoelectron, normalized to the solid 
angle of observation and to the total number of photons in that event. This definition, besides considering the intrinsic quantum efficiency differences among the PMTs,
also describes light propagation non uniformities, electronic channel properties, and similar effects.  

The effective quantum efficiencies are computed based on a selection of real data reconstructed close to the center (within  a spherical radius of $2\,\mbox{m}$). 
In this way, solid angle differences among the PMTs are negligible. PMTs with and without concentrators are treated separately, as the detection efficiency is much higher for PMTs 
equipped with a cone.
More specifically, we use \C\ events because of their relatively high statistics, uniform volume distribution,  steady rate and low probability of multiple hits on the same PMT. 
The effective quantum efficiency is assumed proportional to the physical count rate on the respective channel, caused by \C\ selected in the innermost $2\,\mbox{m}$ sphere. 
The dark noise of each PMT, which is evaluated weekly during the electronics calibration campaign, is statistically subtracted from the count rate. 
While some time dependent non-uniformities are observed for the resulting effective quantum efficiencies, they are negligible compared to the intrinsic differences among the PMTs, shown in Fig.\ \ref{fig:relqe}.

\section{Simulation of the readout electronics}
\label{sec:MCbxelec}

The Geant4 based tracking code simulates optical photons until they are detected by the PMTs.  
Our electronics simulation code (developed in \texttt{C++}) reproduces the electronics chain and the trigger system response, based on the information of the PMT pulse times. 
One of its most important features is the ability to simulate the detector electronics following its operating status over the entire data taking period. 
The detector is continuously monitored over time to assure the data quality and its stability. 
Because of the intrinsic loss of PMTs and since some acquisition boards might be disabled during part of the data taking, it is crucial that the electronics simulation follows
the real time evolution of the operating PMTs. This is evaluated on an event-by-event basis, thanks to the information stored in a database during the standard data taking.
The dark rate of individual PMTs, the effective quantum efficiency for each channel, and the PMT gains are also saved in the database and are injected in the simulation on a weekly basis. In addition,
the code includes the effect of bad channels and detector inefficiencies by default in the same time dependent way. 

Dark counts are generated according to the measured dark rate of each PMT and are superimposed to the signals due to scintillation. 
The typical dark noise rate is around a few hundreds of Hz for most of the PMTs.
During data taking, time alignment is achieved by a common laser pulse, distributed to all PMTs via fiber optics with a timing precision of better than 0.5\,ns. 
Intensities are chosen to detect an average of less than one photon per PMT per pulse, allowing the single photoelectron calibration in the sub-Poissonian regime.
The accuracy of the relative time alignment is better than the PMT time jitter and so no other effects must be included in the simulation. 
In the electronics simulation the effective quantum efficiency is applied to each optical photon reaching a PMT. In case of detection,
the code fully simulates the response of the analog and of the digital electronic chains connected to every single PMT.
 
The front end analogue board generates two signals: a linear amplification of the PMT pulse, and a gate-less integrated signal (as described in Ref.~\cite{Lagomarsino:1999pd}). 
The linear signal is discriminated by the digital board by means of a double threshold discriminator. 
The higher threshold allows to discriminate if the PMT was fired, while the lower one is used for timing purposes (with 0.5\,ns resolution),
thus reducing the time jitter dependence on the pulse height. 
We commonly define as ``hit'' a PMT pulse which crosses the higher discriminator threshold. 
The integrated signal is sampled by an Analog to Digital Converter (ADC) thus  measuring the charge of each pulse. 
The integrator circuit is designed to keep the signal almost constant for 80\,ns after the pulse, followed by an exponential decay with a time constant of 500\,ns due to 
the capacitive coupling between PMTs and front-ends. 
In order to minimize the noise in the charge measurement, for each hit the integrated signal is sampled twice with an 8-bit ADC, 
once before the signal rise and a second time after 80\,ns. The charge is proportional to the difference of these two ADC values. 

\begin{figure}[tb]
\vspace{0.2cm}
\includegraphics[width=0.756\columnwidth]{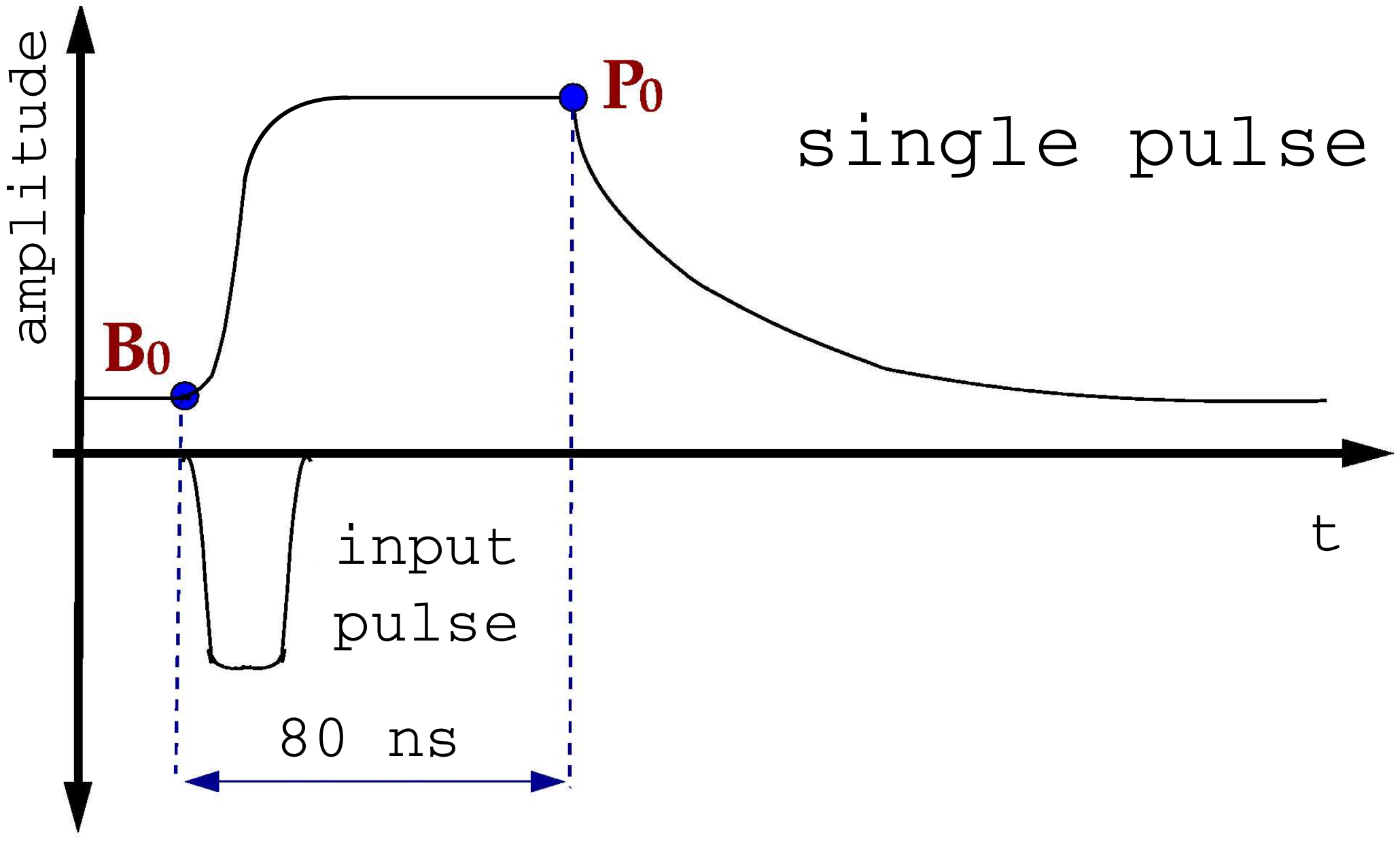} \vspace{0.2cm}
\includegraphics[width=0.7418\columnwidth]{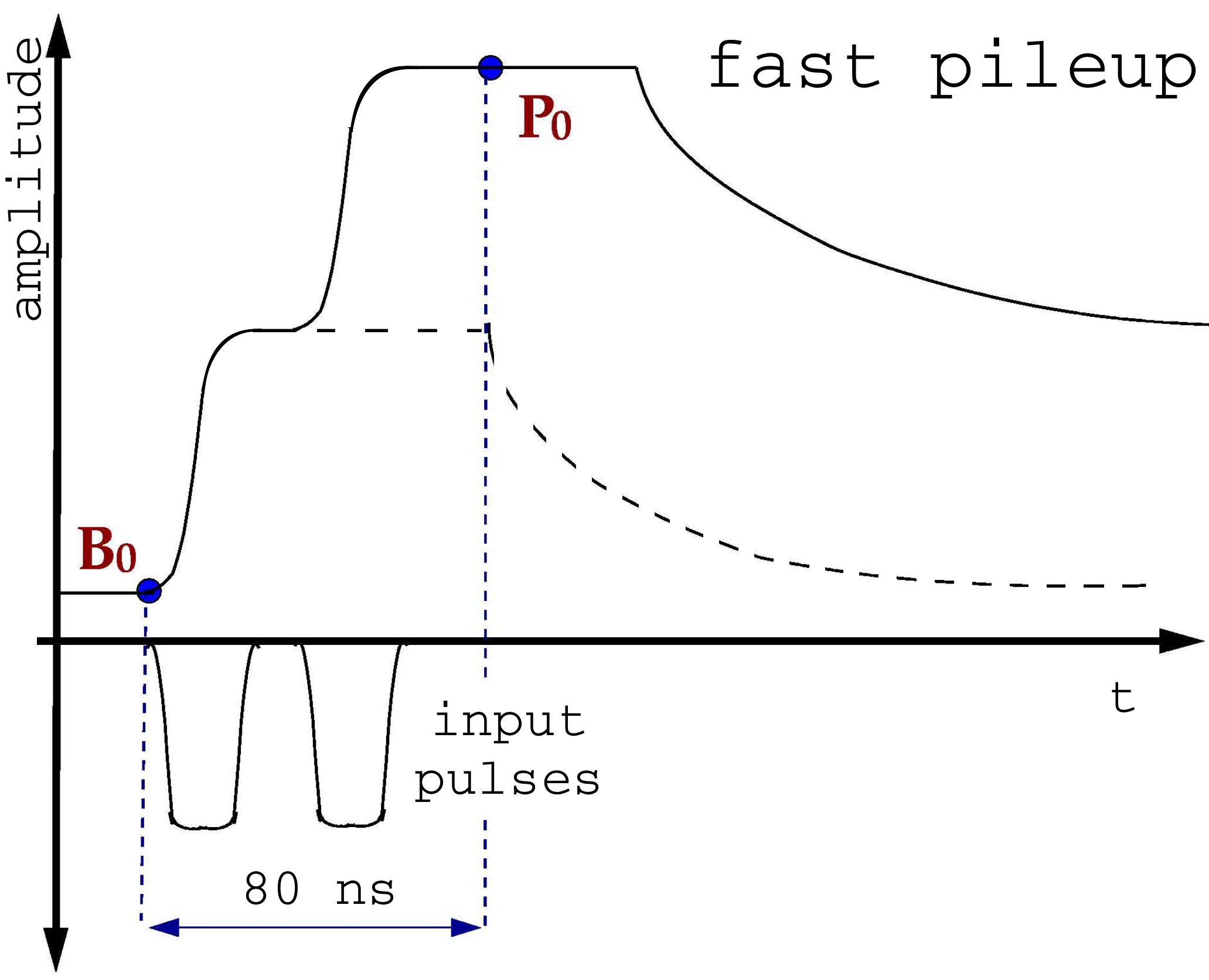}
\includegraphics[width=0.7476\columnwidth]{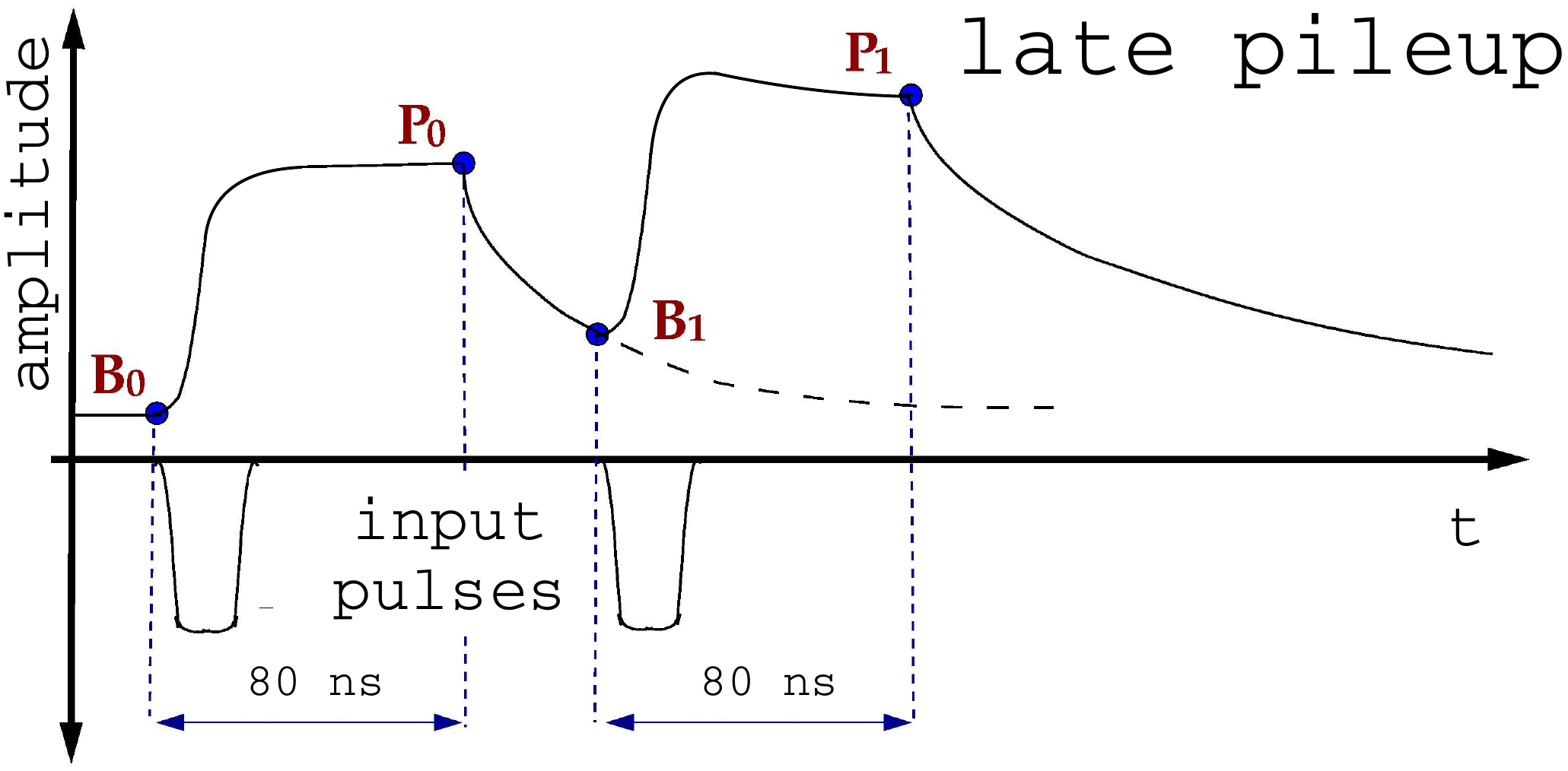}
\caption{Principle of operation of the gate-less charge integrator used in Borexino. The integrated output is shown as a function of time for various photoelectron arrival time combinations.
Note that in the last case, after the second photoelectron arrival, the integrated output is not exactly constant. This effect is both taken into account in data analysis and in MC simulations.}
\label{fig:gateless}
\end{figure}

After a channel detects a hit, the digital board has a dead time of 140\,ns. This means that if two hits are closer than 140\,ns in time for the same PMT, the time information of the second hit is 
lost. However, the system is able to measure its charge if it occurs less than 80\,ns after the first hit.
The occurrence of multiple hits on the same PMT is small but not null and its value must be accurately simulated to reproduce the energy response of the detector. 
The probability of double hits depends on the event position (if the scintillation event is not in the center of Borexino, the PMTs closer to the event detect more light) and on the energy, 
introducing a small non linearity and non uniformity which is included in the simulation. The response of the integrator to multiple hits is shown in Fig.~\ref{fig:gateless} for a few example
cases and it is implemented in the simulation.

\begin{figure*}[thb]
\centering
\includegraphics[width = 1\columnwidth]{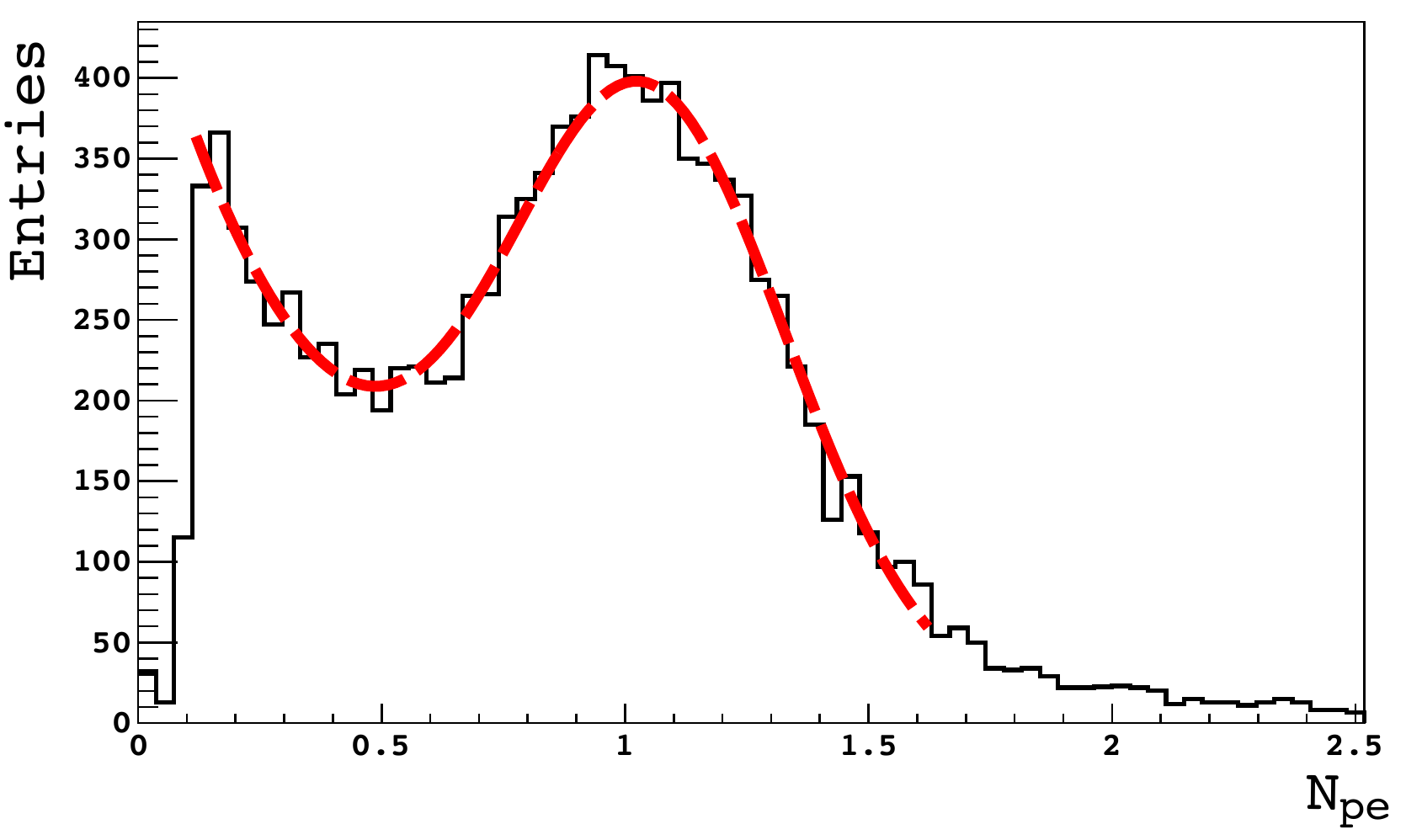}
\includegraphics[width = 1\columnwidth]{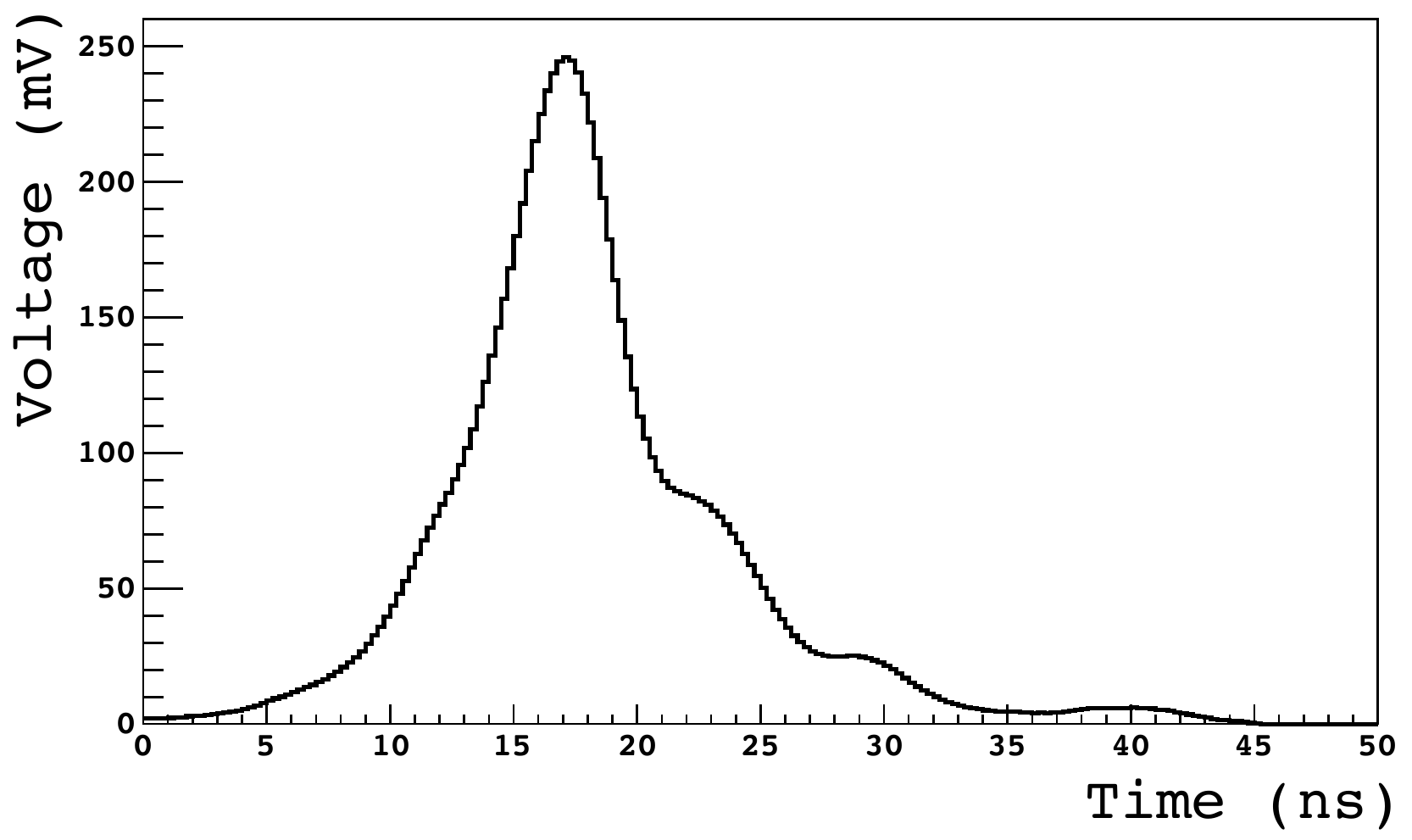}
\caption{\emph{Left Panel:} Example of single photoelectron charge response of a PMT  measured with a special laser calibration run with  a particularly low discriminator threshold.
\emph{Right Panel:} Mean single photoelectron
waveform, output of the front end, rescaled according to the specific hit charge and used for the discriminator threshold condition evaluation.
}
\label{fig:bxelecq}
\end{figure*}

In the electronics simulation code, the charge associated to each photoelectron is sampled from the single photoelectron response of each PMT modeled as an exponential plus Gaussian distribution. 
The specific parameters describing this curve for each PMT (different in the case of dark noise events and real scintillation events) were measured channel by channel.
One example is shown in the left panel of Fig.\ \ref{fig:bxelecq}. 
After the value of the charge is sampled, a corresponding PMT pulse is generated according to a reference shape for the 
signal which was acquired directly from the output of the front end. An averaged shape is applied to all the PMTs and it is shown in the right panel of Fig.\ \ref{fig:bxelecq}.
The generation of the analog output of the inverted and amplified PMT pulse, as it is produced by the front end, allows the simulation of the double threshold logic of the discriminators.

\begin{figure}[tb]
\centering
\includegraphics[width = 1\columnwidth]{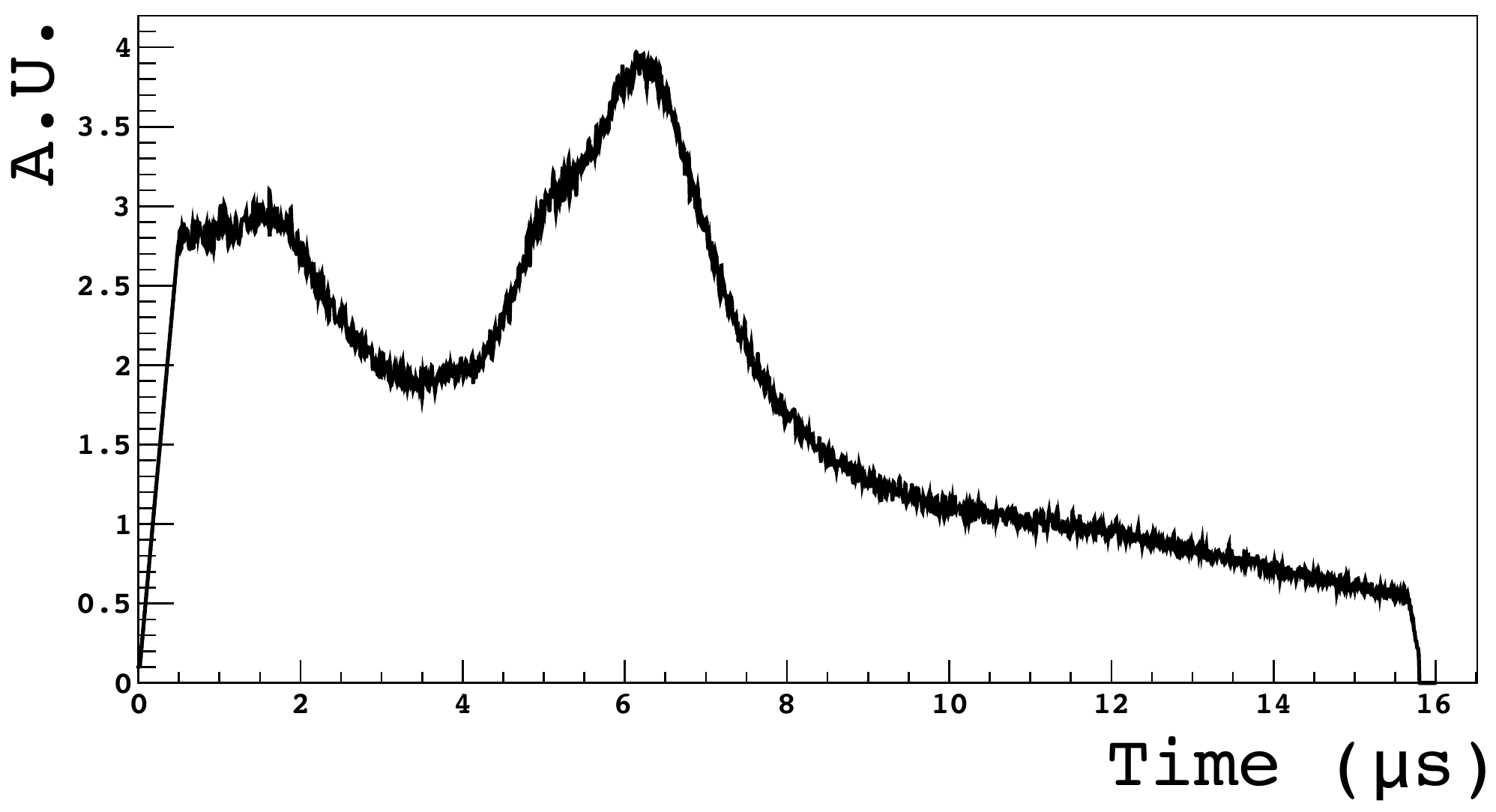}
\caption{Probability density function of after-pulse hit delays. This curve was measured for Borexino's PMTs in a time window of 16 $\mu$s, which corresponds to the trigger acquisition gate duration.
}
\label{fig:afterpulse}
\end{figure}

For each hit, the PMT transit time spread is simulated. The generation of an after-pulse is also considered with a fixed probability of $0.028$.
The probability density function used for extracting the after-pulse hit delay is shown in Fig.\ \ref{fig:afterpulse}.
The program also reproduces the gate-less integrator circuit previously described and the $140\,\mbox{ns}$ dead time of the digital board is included in the simulation. 
The correct reproduction of the response of the front end and of the digital modules is crucial to obtain an accurate simulation of the position-dependent energy response of the detector.

A physical trigger in Borexino and in the MC simulation happens when $N$ hits are detected within a time window $\Delta t$. 
Typically, trigger threshold and time window are set to $N=20\div30$ and $\Delta t=100\,\mbox{ns}$. The threshold $N$ changed within 
Borexino's history (mainly because of the loss of live PMTs) and is correctly taken into account in the MC.

The electronics simulation also reproduces the time structure of the acquired data which is due to the design details of the digital boards, providing an accurate reproduction of the output of the real system.
This allows the trigger evaluation conditions to be the same both in the detector and in the simulation. 
There are two different strategies for the search of the trigger condition in the simulation: a time flow based and a ``boosted'' one. The former consists in letting the time flow with
the steps of the time resolution of the trigger board. The condition for the occurrence of the trigger is evaluated as the time flows. 
This has the advantage of being able to generate a trigger event even if the ``real'' event generated with the tracking code is not present and the hits are due to the dark rate or the noise. 
The disadvantage consists in a very slow simulation. The boosted technique, instead, looks for the occurrence of the trigger condition in a time window around the 
physical Geant4 event. This is much faster, but the generation of the triggered event is related to the presence of the simulated vertex.
In normal simulations, the boosted technique is the working condition, while the first could be used for trigger efficiency studies. 

When the trigger is generated, all the hits within the acquisition gate of $\sim$16.5\,$\mu\mbox{s}$ ($\sim$ 6.9$\, \mu\mbox{s}$ before Dec. 2007) are saved. 
The information of each hit is stored using the real data format and is processed by the same reconstruction code.
For each event, in addition to the electronics response of each channel, the information of the ``MC truth'' (information regarding some true input parameters of the simulated event)
is stored in a dedicated structure, that can be accessed after the processing of the simulated events through the reconstruction code.

The primary events generated by the Geant4 code are not correlated in time. The only exceptions are radioactive decay chains and neutron captures.
The time information is added by the electronics simulation, which assigns a time-stamp to each event, 
by sampling from an exponential distribution with a custom time constant, usually chosen to reproduce the observed event rate in real data. 
Coincident scintillation events caused by fast radioactive decays are sometimes detected within the same acquisition gate.
The MC simulation implements this effect by copying the timing information from the Geant4 tracking code to the electronics simulation, which correlates the events in time.

The effects and asymmetries induced by dead PMTs and effective quantum efficiencies can be studied best by comparing different configurations with an ideal detector featuring a uniform distribution of equal PMTs, an option implemented in the MC.

\section{The reconstruction code}
\label{sec:reco}

The reconstruction in Borexino is based on the amount of light collected by each PMT (related to the event energy) 
and on the relative detection times of the hits on the PMTs (used for event position and particle identification).
A dedicated algorithm (called ``clustering'') identifies the number of hits within the gate that correspond to a single scintillation event. Occasionally more than one cluster may be present during the gate duration.
We define (both in data and in MC) different energy estimators called $N_p$, $N_h$, and $N_{pe}$~\cite{bib:BxLong}.
They correspond to measured quantities such as the number of hit PMTs ($N_p$), the total number of hits ($N_h$), or the number of photoelectrons ($N_{pe}$)
detected during the duration of a cluster or within proper fixed time interval. 
$N_p$ is basically the number of PMTs that detected at least one hit, and in general it differs from $N_h$.
Both $N_p$ and $N_h$ are computed starting from the measured values $N_p^{m}$ and $N_h^{m}$~\cite{bib:BxLong}:
\begin{align}
N_p^{m}&=  \sum_{j=1}^{N^{\prime}} p^j  \\
N_h^{m}&=  \sum_{j=1}^{N^{\prime}} h^j, 
\end{align}
where $p^j$ = 1 when at least one photon is detected by the PMT $j$ and $p^j$ = 0 otherwise, $h^j=0,1,2...n$ corresponds to the number of detected hits and
$N^{\prime}$ is the number of operating channels and it is evaluated on a nearly event--by--event basis using calibration events acquired during the run. 
$N_p$ and $N_h$  are then obtained after normalizing the measured values to $N_{tot}= 2000$ working channels through the relations~\cite{bib:BxLong}:
\begin{equation}
N_{p,h}= \frac{N_{tot}}{N^{\prime}(t)} N_{p,h}^{m}.
\label{eq:equa}
\end{equation}
$N_{p}$ and $N_{h}$ are obtained by the clustering algorithm without choosing a priori the time duration of the cluster (the maximum allowed value is $1.5\,\mu\mbox{s}$).
For some particular analyses (e.\,g. those extending to ``low energy'' such as the \pp\ neutrino flux measurement~\cite{bib:BxPp}), additional energy estimators were introduced to perform the sum
of Eq.\ (\ref{eq:equa}) within a fixed time window measured from the first hit belonging to the cluster. We refer to them as $N_{p,h}^{dti}$ ,where $dti=\,230$ or  400\,ns and it corresponds to the fixed time interval over which the sum is computed. The time interval begins with the first hit of the first cluster and if needed it includes hits belonging to any additional clusters if present.

The third energy variable, $N_{pe}$, is the total number of collected photoelectrons normalized to $N_{tot}$ channels. 
Following the same procedure explained above for $N_{p,h}$, the measured charge $N_{pe}^{m}$ is calculated by summing the hit charges $q_i^j$ expressed in photoelectrons:
\begin{equation}
N_{pe}^{m}= \sum_{i=1}^{N_h^{m}} q_i^j.
\end{equation} 
The number of channels with working ADC's and charge readout, $N^{\prime\prime}$, is used for normalizing $N_{pe}^{m}$ to $N_{tot}$ working channels:
\begin{equation}
N_{pe} = \frac{N_{tot}} {N^{\prime\prime}(t)} N_{pe}^{m} .
\label{eq:Npe}
\end{equation} 
$N^{\prime\prime}$ is normally lower than $N^{\prime}$ by a few tens of channels.

The typical light yield in Borexino is of the order of 500 photoelectrons per MeV for events uniformly distributed in the fiducial volume. 
Therefore, the number of hits recorded by each PMT approximately follows a Poisson distribution 
whose mean value is much lower than 1 up to an energy of a few MeV. Only for relatively high energy events ($E\gtrsim3$\, MeV) the mean value is close to 1 or greater. 

The position reconstruction algorithm determines the most likely vertex position $\vec{r}_0$ for the scintillation event based on the knowledge of the arrival times $t_i^j$ of the detected hits on the PMTs.
For each measured $t_i^j$, the algorithm subtracts a position dependent time-of-flight $T_{flight}^j$ from the interaction point to the PMT $j$ placed at $\vec{r}^j$, i.\,e.:
\begin{equation}
T_{flight}^j(\vec r_0, \vec r^j)  = \mid \vec r_0 -\vec r^j \mid \frac{n_{\rm eff}}{c}.
\label{eq:tof}
\end{equation}
Then, the likelihood $L_E(\vec{r}_0, t_0 \mid (\vec{r}^j, t_i^j))$ that the event occurs at the time $t_0$ in the position
$\vec{r}_0$ given the measured hit space-time pattern $(\vec{r}^j,t_i^j)$ is maximized.
The probability density function for each hit used in the likelihood computation depends on the total number of collected photoelectrons in that same hit.
More details can be found in Ref.~\cite{bib:BxCalib}.

The quantity $n_{\rm eff}$ appearing in Eq.~(\ref{eq:tof}) is called ``effective refractive index" and it defines an effective velocity for the optical photons.
This single parameter globally considers  that photons with different wavelengths travel with different group velocities and that photons can be scattered or reflected from the emission to the detection points. 

The position reconstruction code is used in the same way both for real data and MC ones. 
However, a small difference in the optimal value of $n_{\rm eff}$ for data and MC is observed. Using calibration data (see Sec.~\ref{sec:calibration})
we determined $n_{\rm eff}^{\rm data}$~=~1.68 for real data, while for MC events the best value is $n_{\rm eff}^{\rm MC}$~=~1.66. 
This difference (about 1.2\,\%) is consistent with the instrumental error on the measured values of \noflambda, obtained with ellipsometric techniques.

\section{Tuning of the Monte Carlo}
\label{sec:tuning}

The MC simulation forms the basis of all data analyses in Borexino. The best agreement between MC and real data is required in case of the solar neutrino analyses, 
where all the rates different from that of $^8B$ neutrinos, are obtained through a fit of the energy spectrum of the detected events passing criteria of data selection and quality cuts~\cite{bib:BxLong}. The MC simulation is used to produce the probability density functions (PDFs) to be used in the fit. 
The background spectra or the neutrino induced electron recoil spectra are converted into PDFs 
dependent on one of the energy estimators discussed in Sec.~\ref{sec:reco},
using the MC to simulate the detector response. The accuracy of the conversion directly relates to the systematic uncertainty on the solar neutrino rate measurements.

\begin{figure}[thb]
\centering
\includegraphics[width = 1\columnwidth]{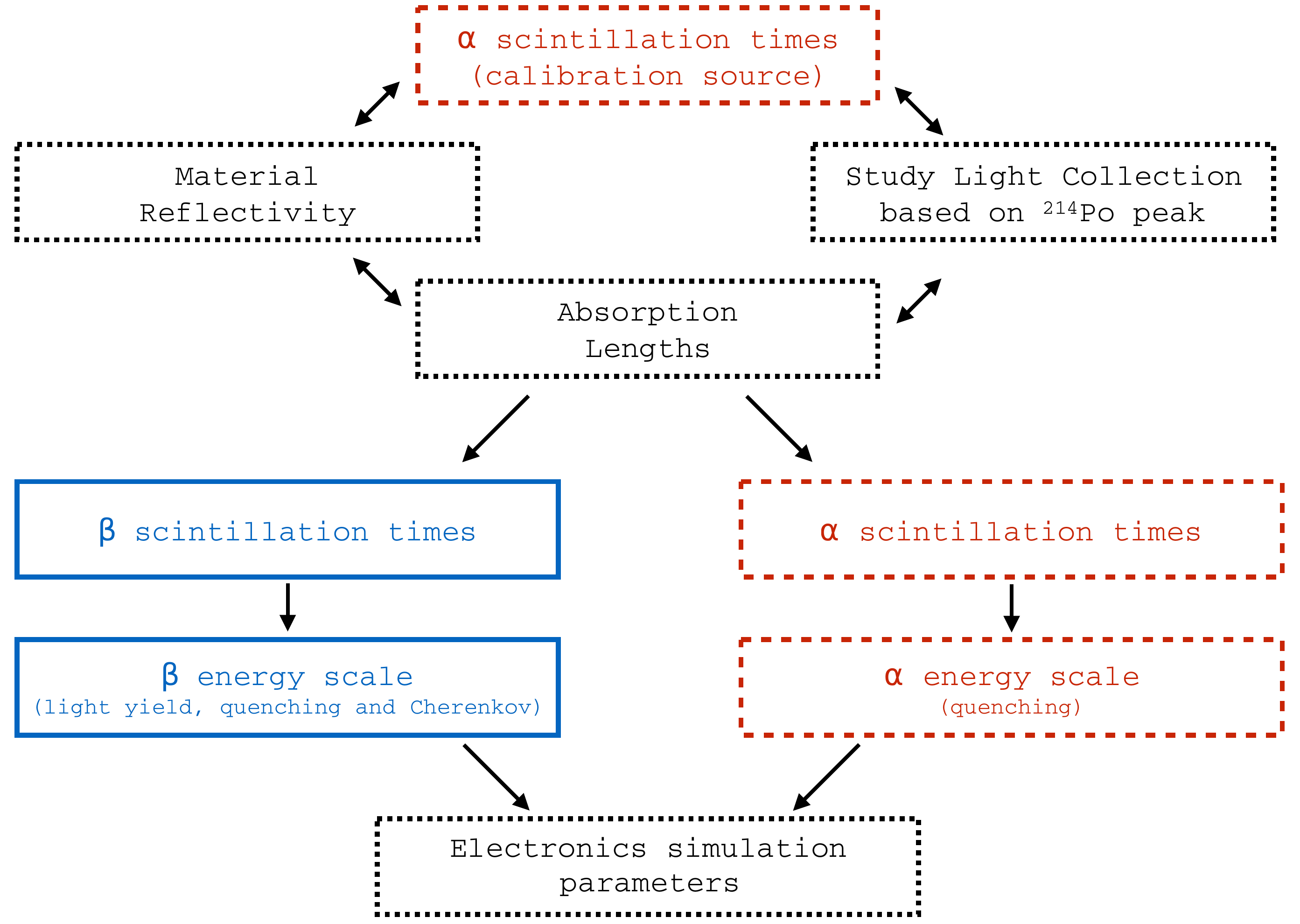}
\caption{Sketch of the ``MC tuning strategy''. Continuous contours (blue) are relative to $\beta$ particles, while dashed ones (red) refer to $\alpha$'s. 
Finer dashed contours (black) indicate sets of response parameters common to both $\alpha$'s and $\beta$'s. }
\label{fig:tuning}
\end{figure}

The MC was tuned and validated using data sets independent from the one on which the actual fit was performed, such as calibration data and service triggers. The most important features of the events, i.\,e. the time-related distributions and the position-dependent number of hits and photoelectrons detected by each PMT, are reproduced with an accuracy of better than $1\%$ within a 3\,m radius fiducial volume
in the energy range of solar neutrino analysis. 

The full simulation of the scintillation and light propagation requires the input of a large number of parameters describing various features of the scintillator and of the materials. 
Examples are the attenuation lengths, the reflectivity, the reemission probabilities and the refractive indexes.
The calibration campaigns~\cite{bib:BxCalib} provided clean data samples, which were used to check the accuracy of the MC simulation.
A portion of the calibration data was used as a tuning sample (to optimize some input parameters), while 
another independent set of calibration data (or data coming from the standard data taking) was used for testing the performance of the simulation.
Great care was devoted to simulating the source runs and the vial geometry, accounting for all the known non idealities of the detector and its operating status as a function of time, as discussed
in the previous sections. 

Most of the input parameters of the simulation were measured in dedicated laboratory setups. However, laboratory conditions might sometimes have deviated from the operating conditions of the Borexino detector. For instance, all the laboratory characterization of the Borexino liquid scintillator might result inaccurate,
simply because of the optical purity of the sample could be not the same as the one achieved inside Borexino.

The main difficulty in the tuning procedure is posed by the correlations among the parameters. Most of the physical effects (which are connected to the material properties)
are mutually dependent. 
As an example, the energy response is strictly related to the time response\footnote{This is mostly related to the electronics response, see Sec.~\ref{sec:MCbxelec}.}, 
which in turn depends on light propagation effects and transparencies.  
For these reasons, the tuning was performed iteratively and following the strategy outlined in Fig.\ \ref{fig:tuning}.

At the beginning, all the parameters are initialized to measured values and a global check of the simulation is performed by direct comparison with results from calibrations. 
In this condition, the simulated detector response differs from the measured one by at most a few tens of percent. As shown in Fig.\ \ref{fig:tuning}, the first step consists in varying 
attenuation lengths and reflectivity, in order to roughly reproduce the amount of light collected by the PMTs when the scintillation happens at various points inside the detector volume
(see Sec.~\ref{sec:MCtuninglightcollection}). Then, the time response for $\alpha$-particles (see Eq.~(\ref{tau})) 
contained in the vial used for calibrations (see Sec.\ \ref{sec:calibration}) is optimized, so that these calibration
points can be used for a precise tuning of the reflectivity and attenuation lengths. This part of the tuning of the parameters is referred to as ``$\alpha$ scintillation times (calibration source)'' and ``study light
collection based on \Pofour\ peak'' in Fig.\ \ref{fig:tuning}. 

Then, the procedure restarts from the beginning, using $\beta$-like events to cross check the time response of the simulation with the tuned attenuation
lengths (because of absorption and reemission, attenuation lengths influence the time response) and reflectivity. Then, the specific parameters of the scintillation times for
$\beta$'s and $\alpha$'s are tuned, as well as the energy scale. The energy scale for $\beta$ events is tuned together with some parameters of the electronics simulation, as the amount
of reconstructed photons depends both on the intrinsic scintillator response and on the electronics. In the next sections, more details on all the steps of the tuning
are described.

 \subsection{Non uniformity of the energy response within the inner vessel}
 \label{sec:MCtuninglightcollection}
The first step consists in the tuning of the optical attenuation lengths and of material reflectivity. 
These parameters impact on the uniformity of the energy response of the detector.
The mixed  \C-\Rn\ source, deployed in $\sim$200 different positions, was used to perform such a study.
Particularly, the chosen reference is the position of the \Pofour\ $\alpha$ energy peak. Because of the quenching of the scintillator contained in the vial, the precision tuning and study of the light 
collection was performed after a dedicated tuning of the time response of the scintillator contained in the source vial.
\Pofour\ events are point like with respect to the Borexino reconstruction resolution and produce about $\sim300$ $N_p$. Since the mean channel occupancy is $\sim0.15$, the \Pofour\
 allows to operate in the single photoelectron regime, where the $N_p$ and $N_h$ energy estimators have a negligible dependence on electronic effects,
 allowing to understand the light propagation almost independently of the electronics simulation.
 
\begin{figure*}[tb]
\centering
\includegraphics[width = 1\columnwidth]{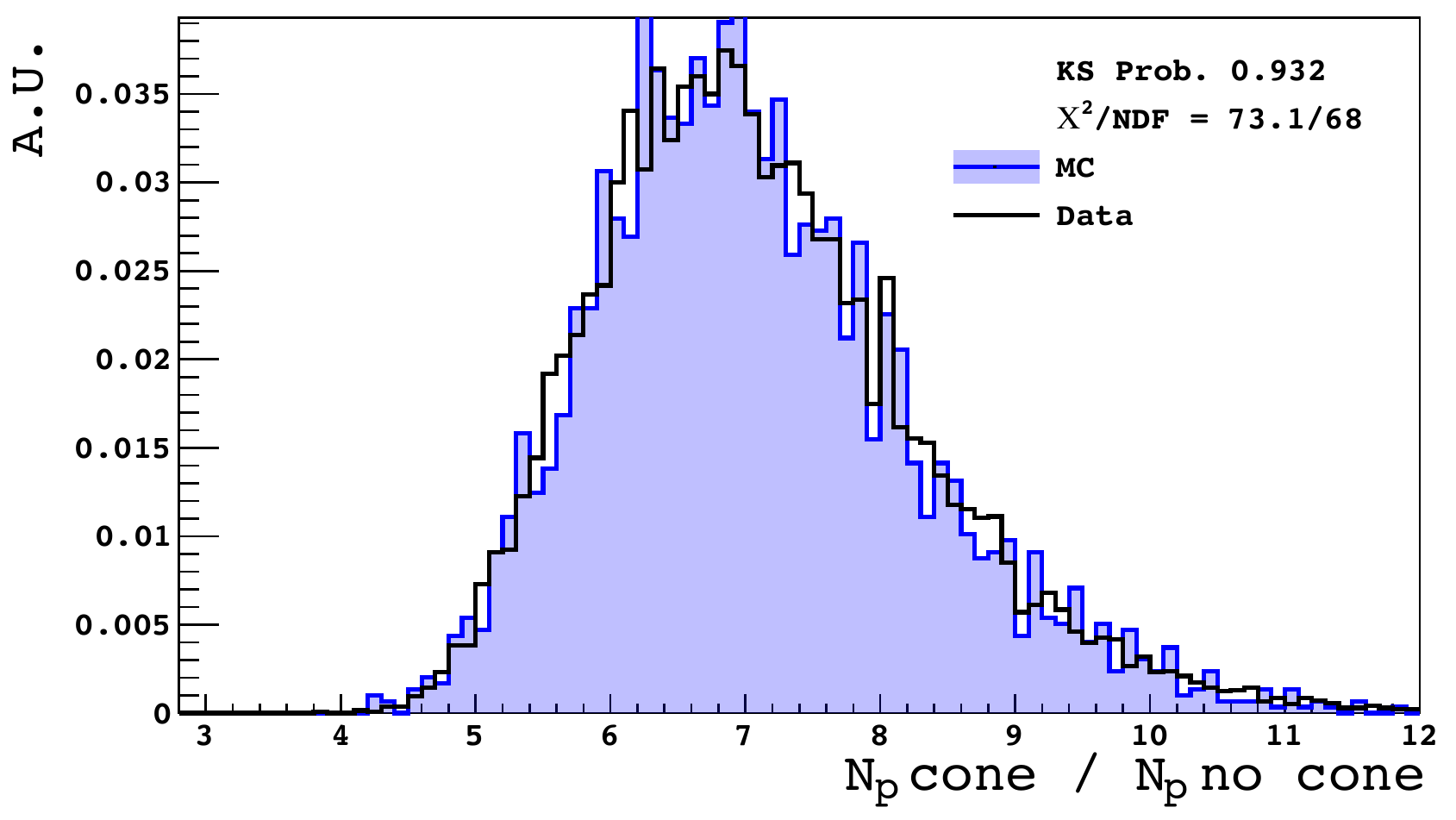}
\includegraphics[width = 0.98\columnwidth]{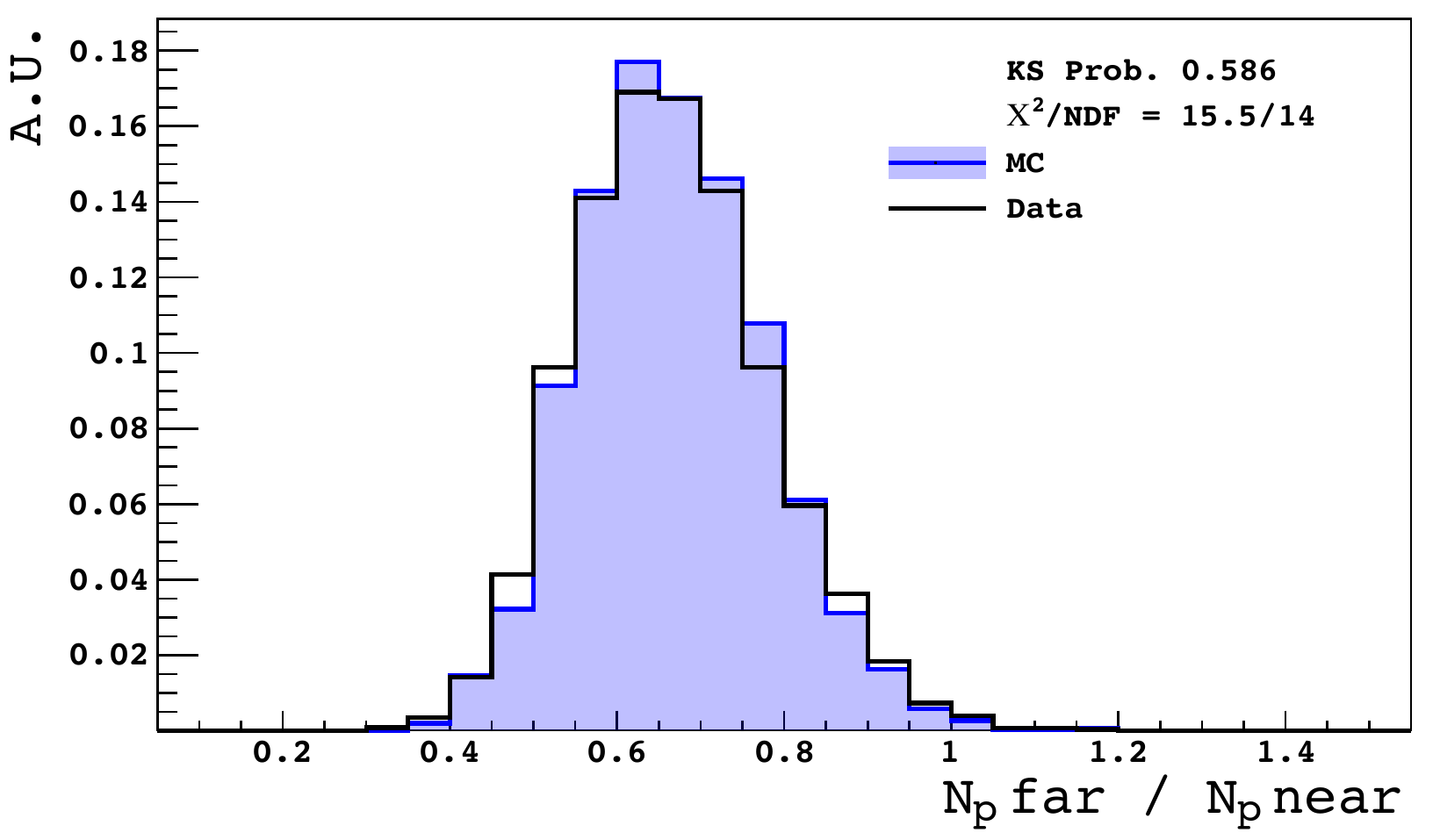}
\caption{\emph{Left Panel:} MC-data comparison of the cone-no cone ratio in a selected point in the inner vessel. The first entry in the legend indicates the outcome of a Kolmogorov-Smirnov test. \emph{Right Panel:} MC-data comparison of the far-near ratio in a selected point in the inner vessel. }
\label{fig:cnc}
\end{figure*}

\begin{figure*}[bt]
\centering
\includegraphics[width = 0.95\columnwidth]{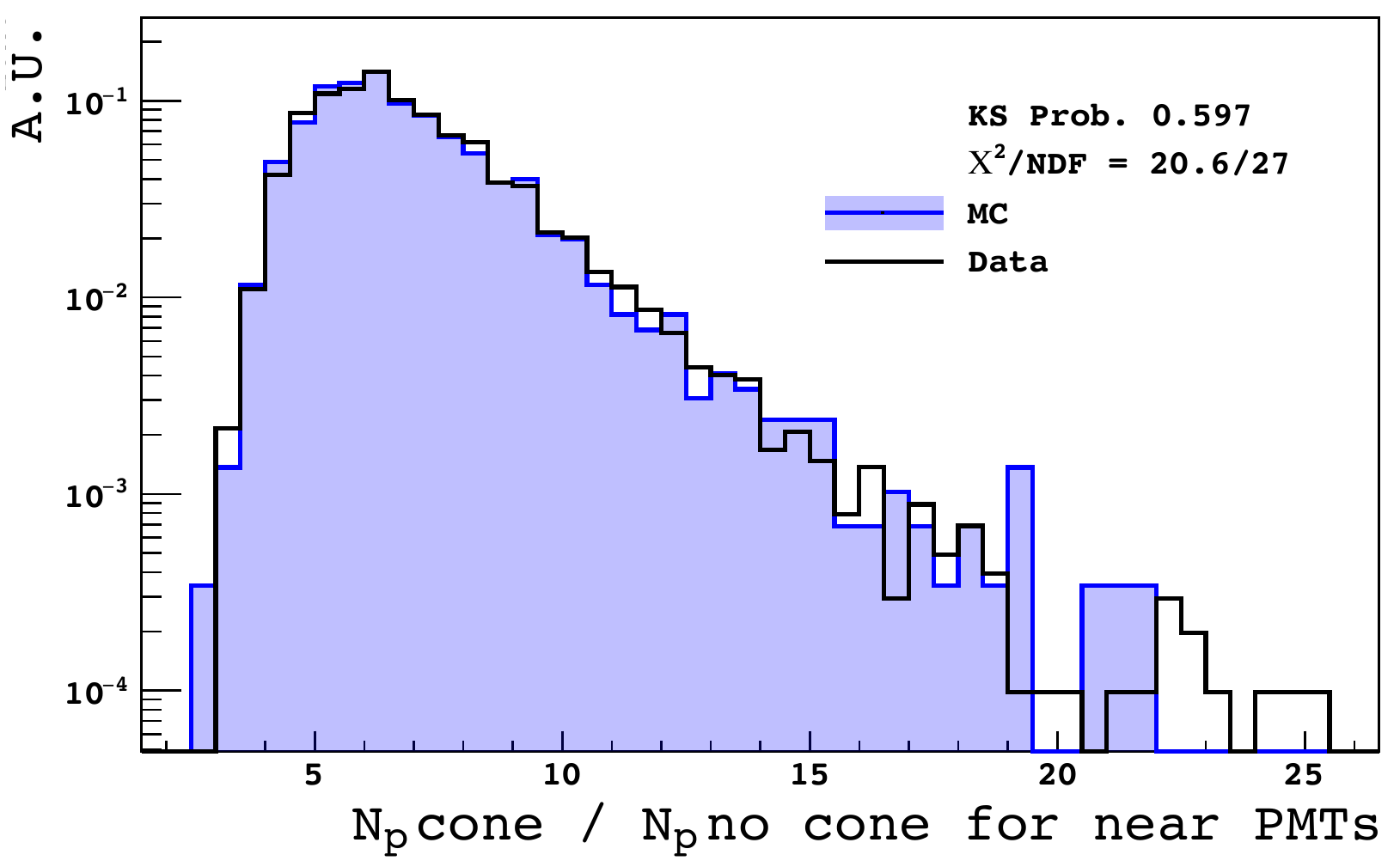}
\includegraphics[width = 1.02\columnwidth]{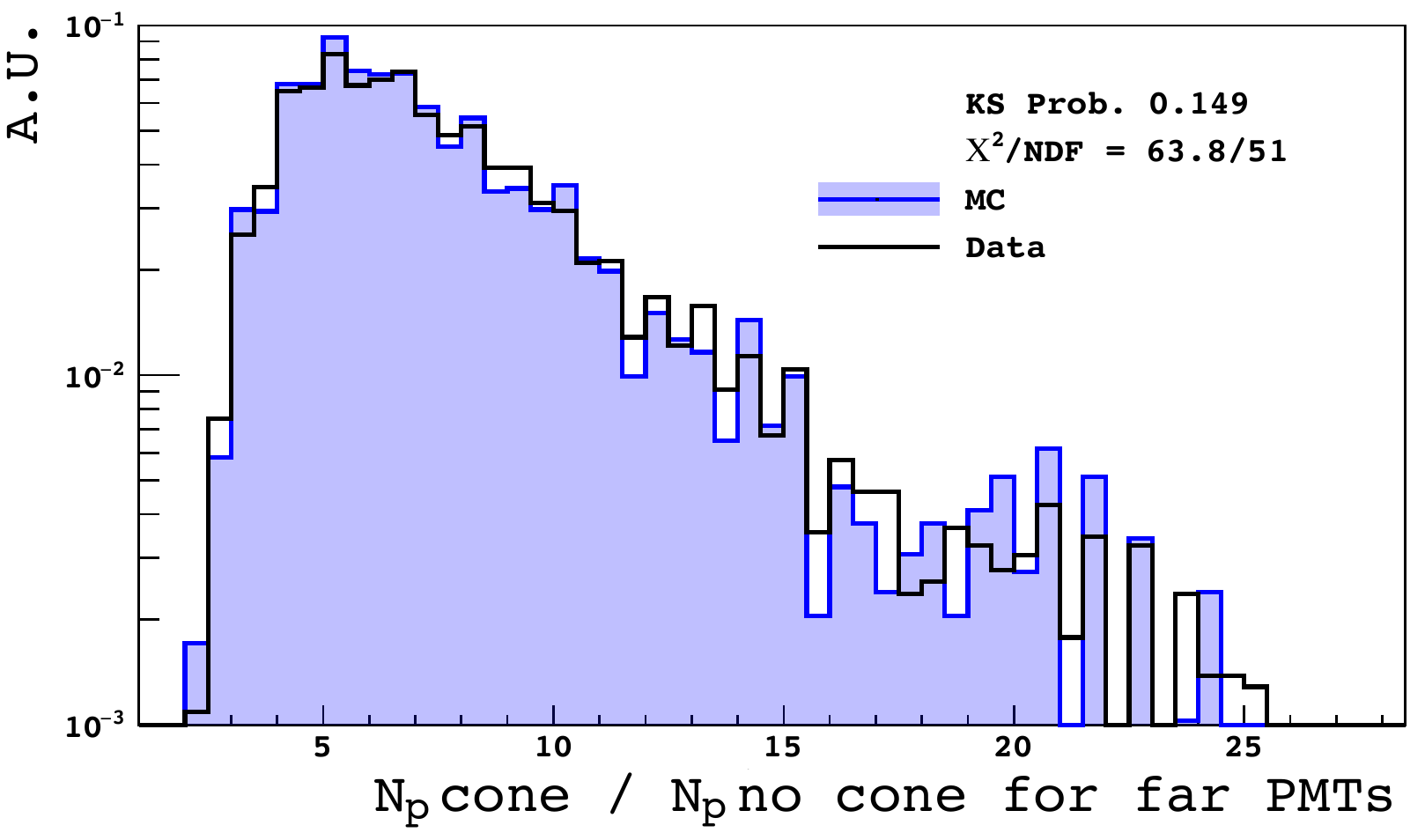}
\caption{Cone-no cone ratios for near (left) and far (right) PMTs. The first entry in the legend indicates the outcome of a Kolmogorov-Smirnov test.}
\label{fig:fncnc}
\end{figure*}

 The attenuation lengths for PC, PPO, DMP, and nylon were measured as a function of the photon wavelength with spectrophotometric techniques using samples with small volumes of liquid.
 During the tuning procedure, the absolute values of the attenuation lengths were slightly adjusted by scaling them with multiplicative factors ($\Lambda$). 
 Other important parameters responsible of the asymmetry in the light collection inside the detector are the values of reflectivity of the light concentrators and of the stainless steel sphere.
 Besides a good reproduction of the shape of the concentrators, 
 the aluminum cone reflectivity and the ratio  between specular and diffusive reflection\footnote{Diffusive 
 reflection is modeled as Lambertian reflection.} components play a crucial role and have to be determined with high accuracy. 
 The key observables allowing the tuning of these parameters are basically two:
 \begin{itemize}
 \item the ratio of the fired PMTs with and without light concentrators (``cone-no cone ratio'').
 \item the ratio of the fired PMTs far ($>4\,\mbox{m}$) and close ($<4\,\mbox{m}$) from the source (``far-near ratio'').
 \end{itemize}
 These observables have the advantage to be energy independent, allowing us 
 to avoid degeneracies due to the non linear behavior of the energy scale and also tolerate differences on the absolute energy scale at this stage of the tuning. 
 The cone-no cone ratio allows to determine the reflectivity of the inner and outer surfaces
 of the light concentrators and the specular/diffusive ratio. The same holds for the reflectivity of the steel ring mounted on the PMTs without concentrator.
 The far-near ratio allows to study the attenuation lengths. Actually, both the attenuation lengths and the reflectivity affect the two ratios at the same time, since the concentrator
 efficiency depends on the distance of the event from the PMT. 
 The two mentioned distributions and their correlation were compared with simulation data for $\sim$20 different source position inside the inner vessel.
 The best parameters  have been selected using the  $\chi^2$ and the Kolmogorov-Smirnov test statistics.
  Figure \ref{fig:cnc} shows examples of the cone-no cone ratio and far-near ratio distributions for selected points inside the inner vessel. In the plot, the results of the Kolmogorov-Smirnov
 and $\chi^2$ tests are also reported. Figure\ \ref{fig:fncnc} shows the combination of the two ratios, and particularly the different cone-no cone ratio distributions for near and far PMTs.
 The two distributions show different features, ideal candles to fully characterize the detector response as function of the vertex position.
 
\begin{figure}[htb]
\begin{flushright}
\includegraphics[width = 0.94\columnwidth]{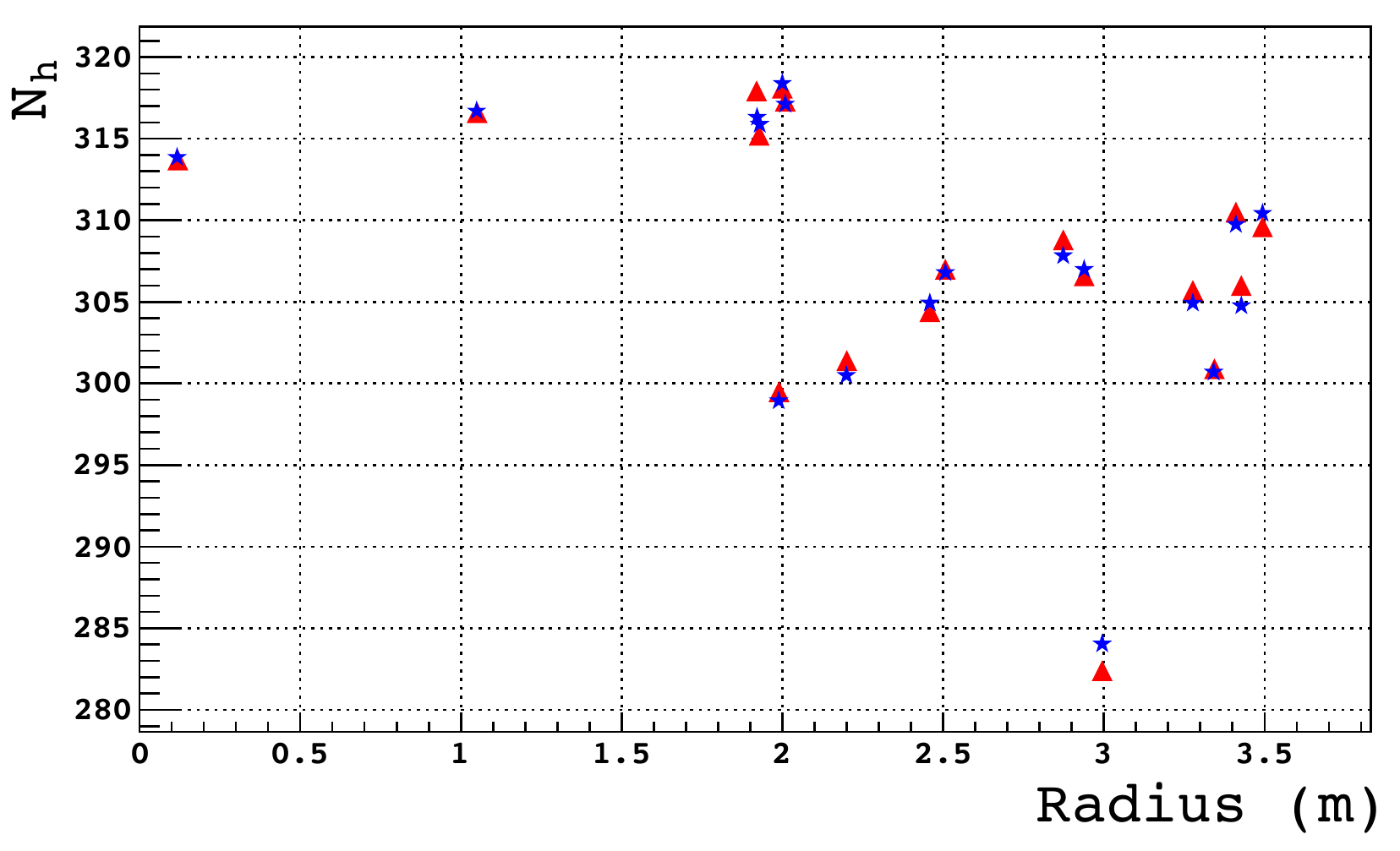}
\includegraphics[width = 1\columnwidth]{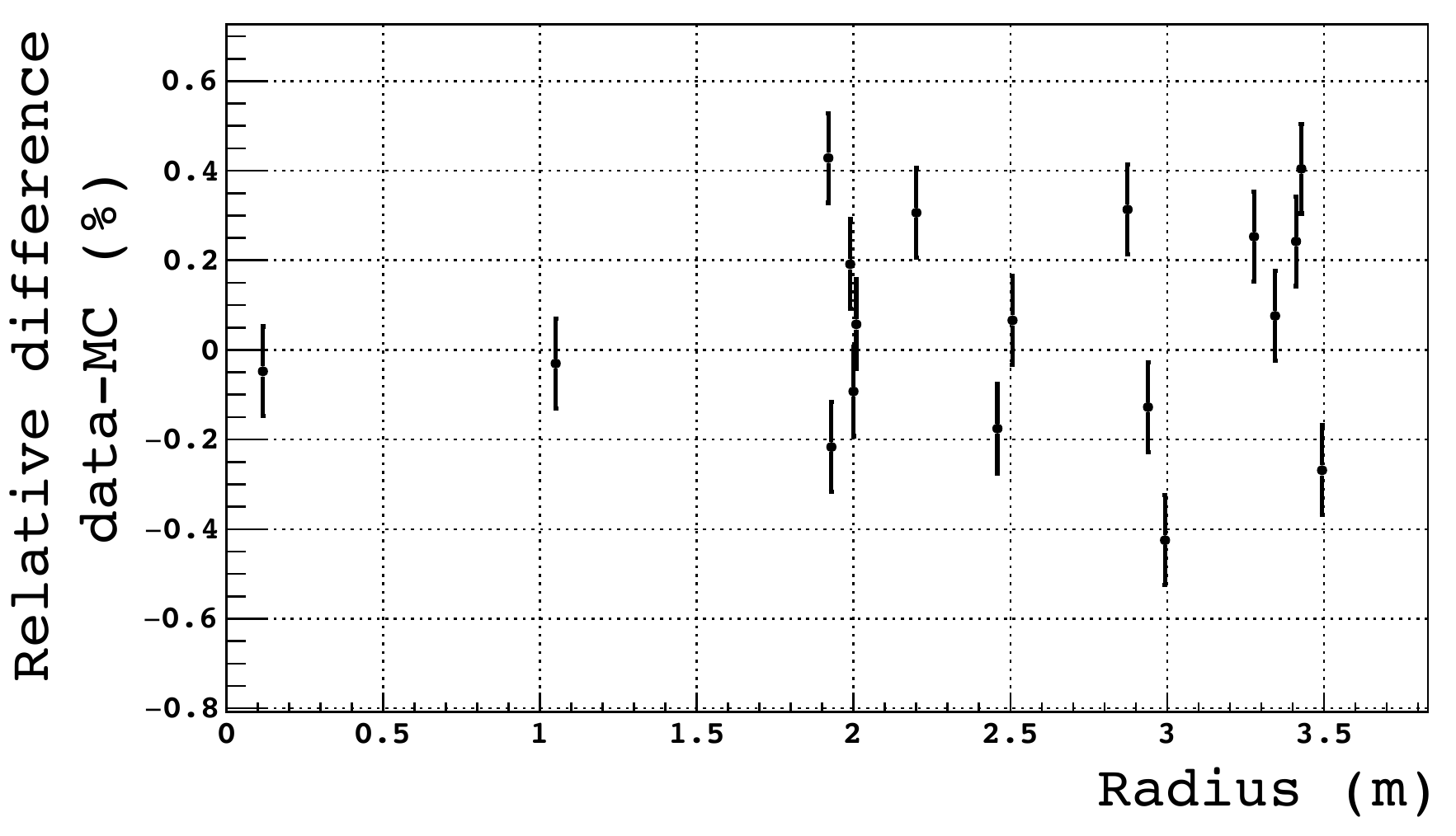}
\end{flushright}
\caption{
\emph{Top Panel:} \Pofour\ energy peak in MC (blue stars) and data (red triangles) as a function of the radial distance of the source from the center. 
The statistical uncertainties on the points are smaller than the markers.
\emph{Bottom Panel:} Relative difference of the \Pofour\ energy peak in MC and data ($N_{h}$ energy estimator) as a function of the radial distance of the source from the center. 
}
\label{fig:LC}
\end{figure}

The results of the attenuation length and reflectivity tuning are shown in Table\ \ref{tab:tuning_att}, where the nominal and tuned values are reported for various parameters. 
As already mentioned, the values of the reflectivity are also influenced by the time response of the detector (see Sec.\ \ref{sec:tuning_timeresp}) and thus 
the final result was reached through the iterative approach described in this section.
The goal set on the accuracy of the light collection is $\lesssim1\%$ and the comparison between data and simulation is shown in the bottom panel of Fig.\ \ref{fig:LC}, where the relative
discrepancy in the case of $N_h$ is plotted as a function of the source radial position. 
Inside the $3.5\,\mbox{m}$ sphere, all the points are contained in the $\pm0.5\%$ band, thus showing that the precision goal was reached. 
The top panel of Fig.\ \ref{fig:LC} shows the comparison between data and MC of the \Pofour\ peak position in the $N_h$ energy estimator. 
Similar results were achieved both with $N_p$~and $N_{pe}$.
These results demonstrate a strong improvement in the comprehension of the detector response with respect to the Borexino calibration results already published in  Ref.~\cite{bib:BxLong}.

\subsection{Time distribution of the collected light}
\label{sec:tuning_timeresp}
The main parameters describing the time response of the detector are the scintillation time constants, the absorption-reemission delays, and the electronics response of the detector.
In order to decouple scintillation and light propagation effects, ``low energy'' sources in the center of the detector were chosen both for $\alpha$ and $\beta$-particles .
This choice  allows to study the response of the system with all the PMTs in the single photoelectron condition. Particularly, \Po\ $\alpha$ decays from data
were selected using a pulse shape discriminator parameter~\cite{bib:BxLong} in the innermost $1\,\mbox{m}$ sphere. 
Data acquired with the \Sr\ source in the center of the detector were used as benchmark for the simulations of $\beta$-particles. 
Both \Sr\ and \Po\ have a visible energy of around $\sim$200 $N_h$, which is relatively
low and thus the electronics effects are minimized. 
Unfortunately, it is not possible to precisely tune the response to $\alpha$ particles using any of the data from the calibration campaign, since
in that case the scintillation happened in the vial and not in the actual Borexino scintillator.

\begin{figure}[htb]
\centering
\includegraphics[width = 1\columnwidth]{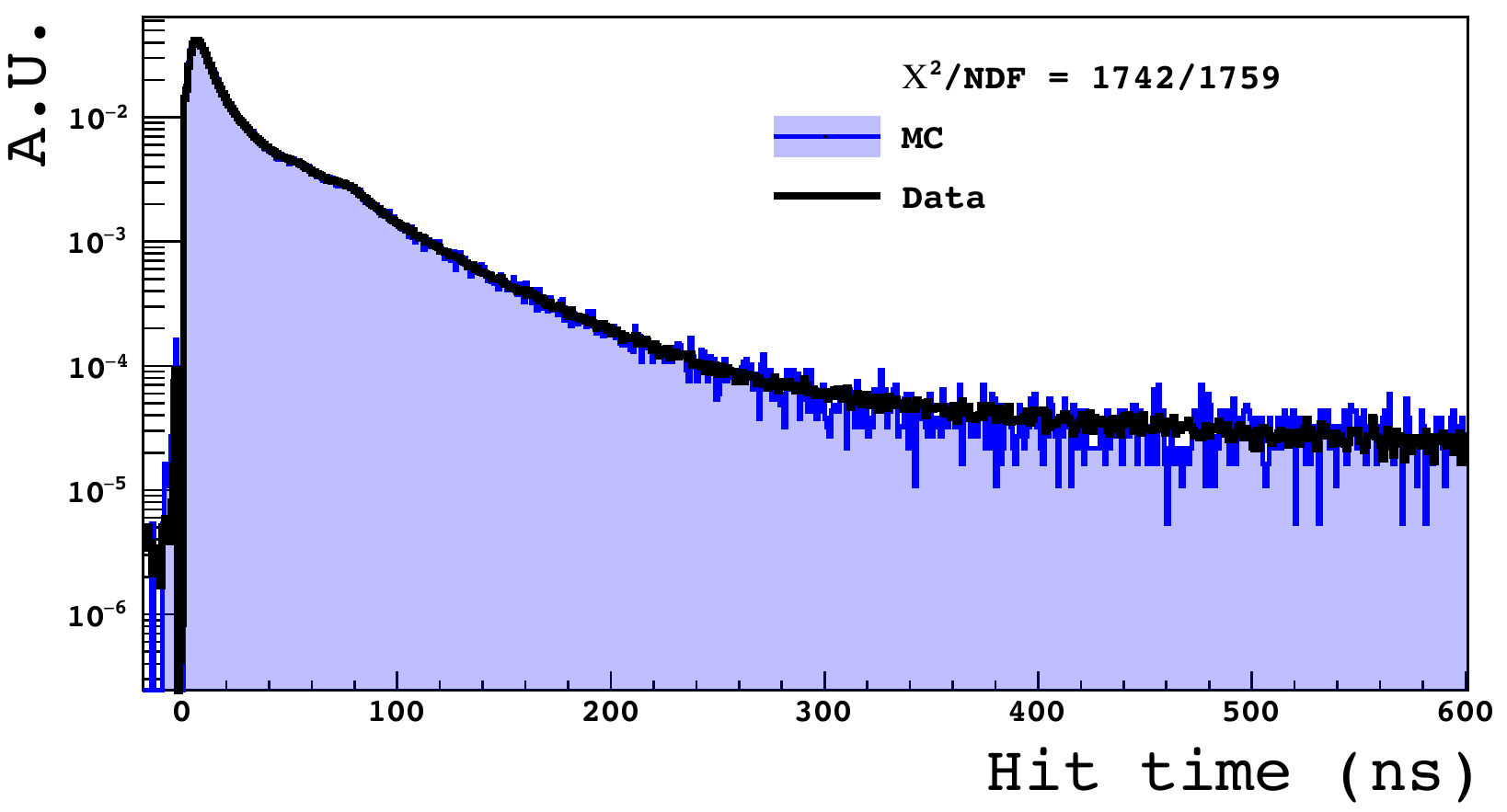}
\caption{Light curve (i.\,e. hit times with respect to the first hit in the cluster) for \Sr\ events in the center of the detector. 
}
\label{fig:tuninglightcurve}
\end{figure}

\begin{figure*}[htb]
\centering
\includegraphics[width = 1.8\columnwidth]{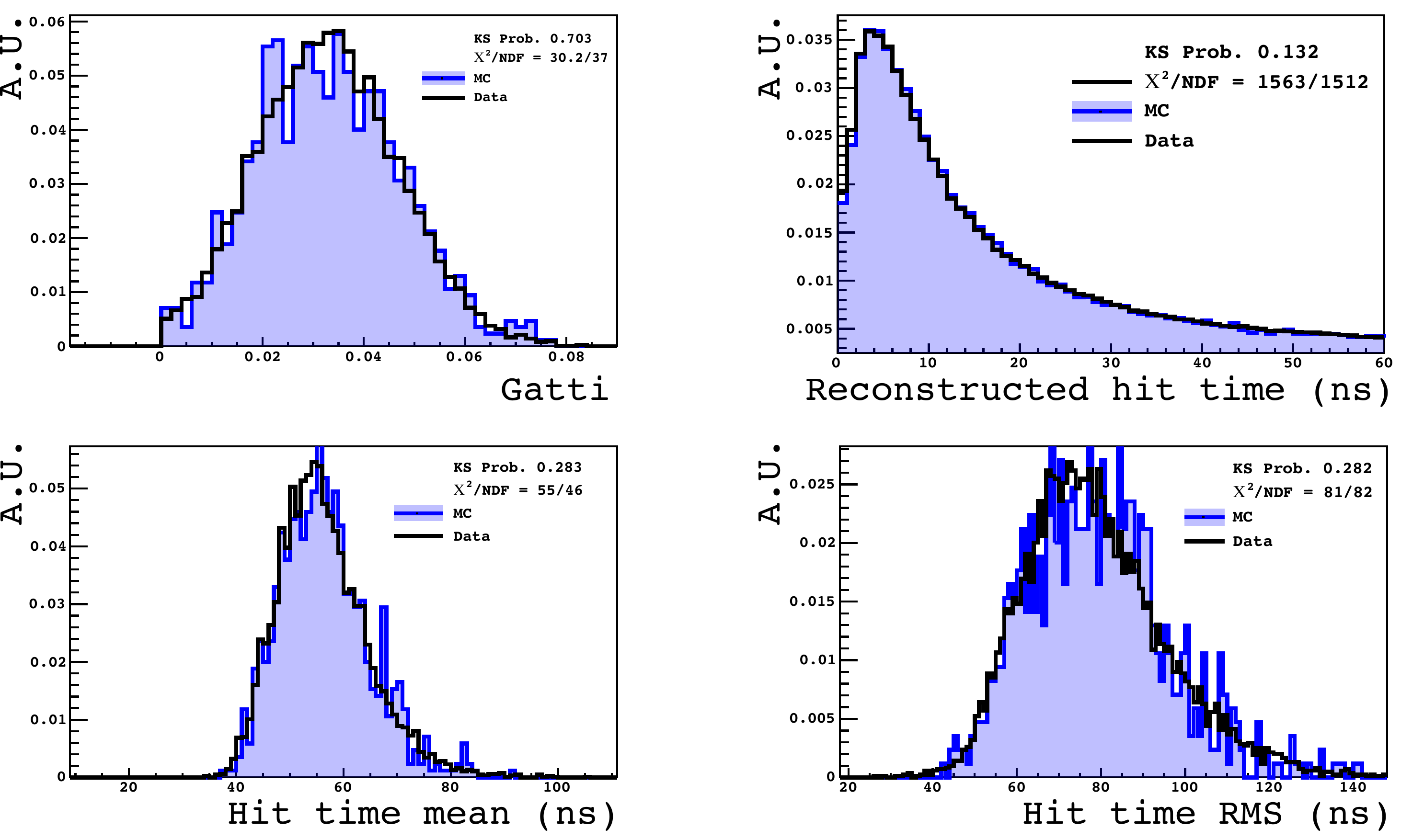}
\caption{Data-MC agreement for some timing variables for \Po\ events in the center. In the top left panel the $\alpha/\beta$ Gatti~\cite{bib:BxLong} distribution is shown, the top right panel
contains the reconstructed hit time distribution (hit time distribution corrected for the time of flight), the bottom left (right) panel shows the mean (RMS) time of the hits in the cluster.
The first entry in the legend indicates the outcome of a Kolmogorov-Smirnov test.}
\label{fig:tuningtempi}
\end{figure*}

The tuning is performed for these classes of events in the center, and then the agreement is checked for 
different source positions. This also allows to test whether the attenuation lengths were properly selected at the previous stage, since the time response is affected by the amount of absorptions
and reemissions. Figure\ \ref{fig:tuninglightcurve} compares the MC and data time distributions of the hits detected by the PMTs for events from the \Sr\ calibration source in the center.
This curve is the convolution of the four exponential response function used to describe the scintillation process (see Ref.~\cite{bib:BxLong})
and the simulated absorptions and reemissions. It is possible to distinguish the shape due to the sum of the four exponentials, but some additional features are evident, such as
the structure at approximately $\sim$100~ns, which is due to the reflection of photons on the stainless steel sphere, and the tail at larger times, which is due to the contribution of after-pulse hits
and dark noise. The correct reproduction of the hit time distribution ensures that other more complex variables reconstructed from the hit times 
(such as the reconstructed event position and the Gatti $\alpha$/$\beta$ pulse shape discrimination variable~\cite{bib:BxLong}) are also correctly reproduced by the MC, as shown for example in Fig.\ \ref{fig:tuningtempi}.

The statistical tests adopted  to select the best parameters were again the Kolmogorov-Smirnov and $\chi^2$ tests. The results of the tuning of the parameters 
are reported in Table\ \ref{tab:tuning_t}. The exponential parameters were only slightly adjusted with respect to the experimentally determined ones 
(coming from Ref.~\cite{Lombardi:2013nla}). This is acceptable, because of the optical purity of the laboratory scintillator sample (see beginning of Sec.~\ref{sec:tuning}). 
Furthermore, while the full Borexino MC is essentially an ab initio
simulation, the model adopted in Ref.~\cite{Lombardi:2013nla} for estimating the scintillation time constants is based on phenomenology. Therefore, a little discrepancy might also be due
to some small propagation effects which are not accounted for in the laboratory setup measurement.
However, it was not necessary to modify the characteristic times of absorption and reemission. In this respect, the most important parameter is $\tau_{PPO}$, which 
affects significantly the time distribution of photons (most of the primary light is absorbed and reemitted within a few $\mbox{cm}$ from the interaction point). $\tau_{PPO}$ is quite
well established in the literature, and thus the absence of the need of a modification is an argument for the solidity of the optical model of the detector in the tracking code.

\subsection{The energy scale}
\label{sec:MCtuningenergy}
The last step of the tuning of the simulation is the reproduction of the absolute energy scale. 
All the input parameters, apart from the ones describing the quenching and the scintillation times, are common to the $\beta$ and $\alpha$ energy response. 
Crucial parameters affecting the energy scale are the absolute scintillation yield $Y_0^{ph}$, the Birks quenching parameter \birks, and the yield of the \v Cerenkov light.
As discussed in Sec.~\ref{Cerenkov}, there are no experimental data available for $P_{rem}$ for $\lambda<320\,\mbox{nm}$.
In addition, the index of refraction in the near UV is not accurately constrained. As a consequence, $P_{rem}$ is considered as an effective parameter.
Furthermore, some electronics fine tuning was required for a high precision simulation of $N_{pe}$.

\begin{figure}[t!]
\centering
\includegraphics[width = \columnwidth]{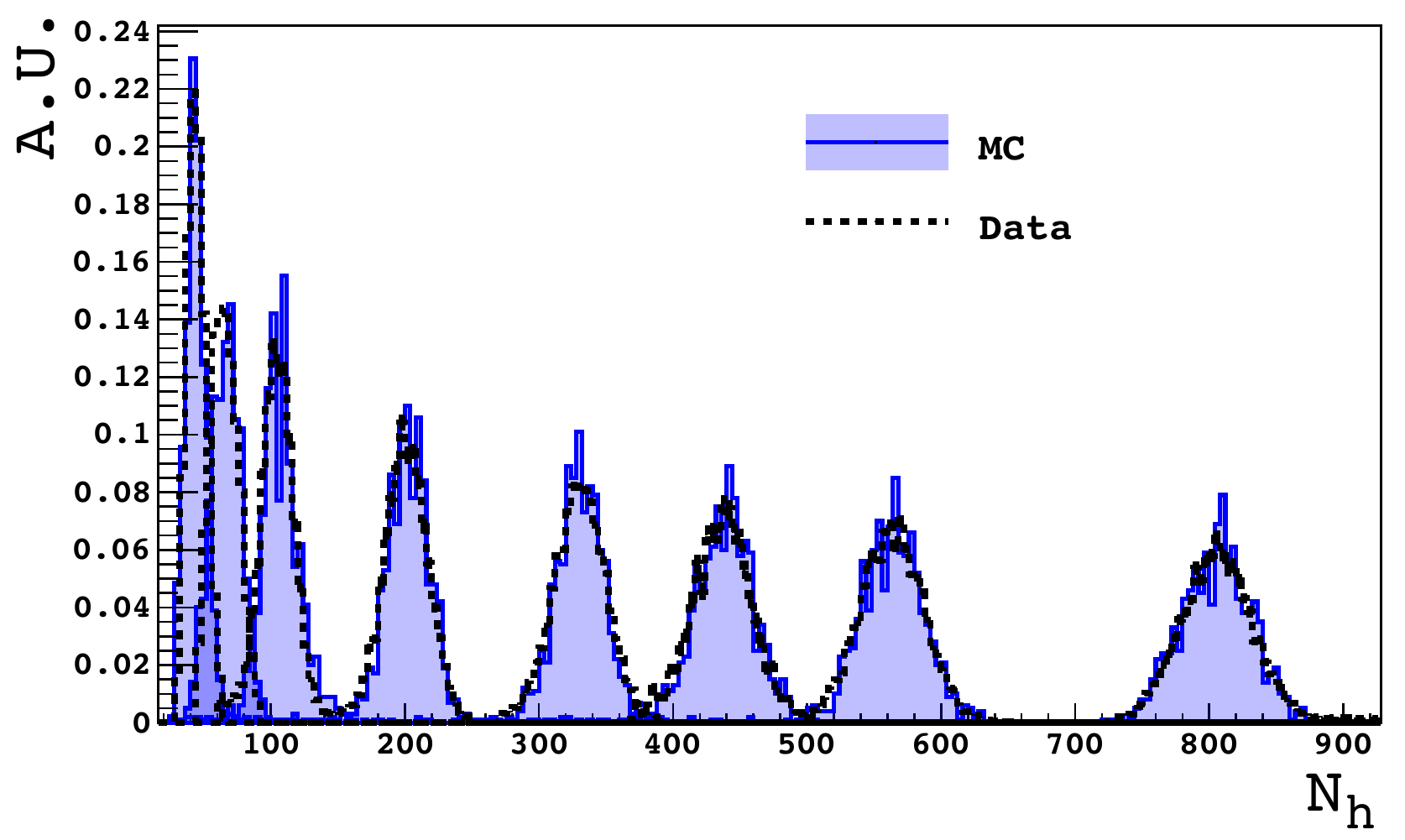} \\
\includegraphics[width = \columnwidth]{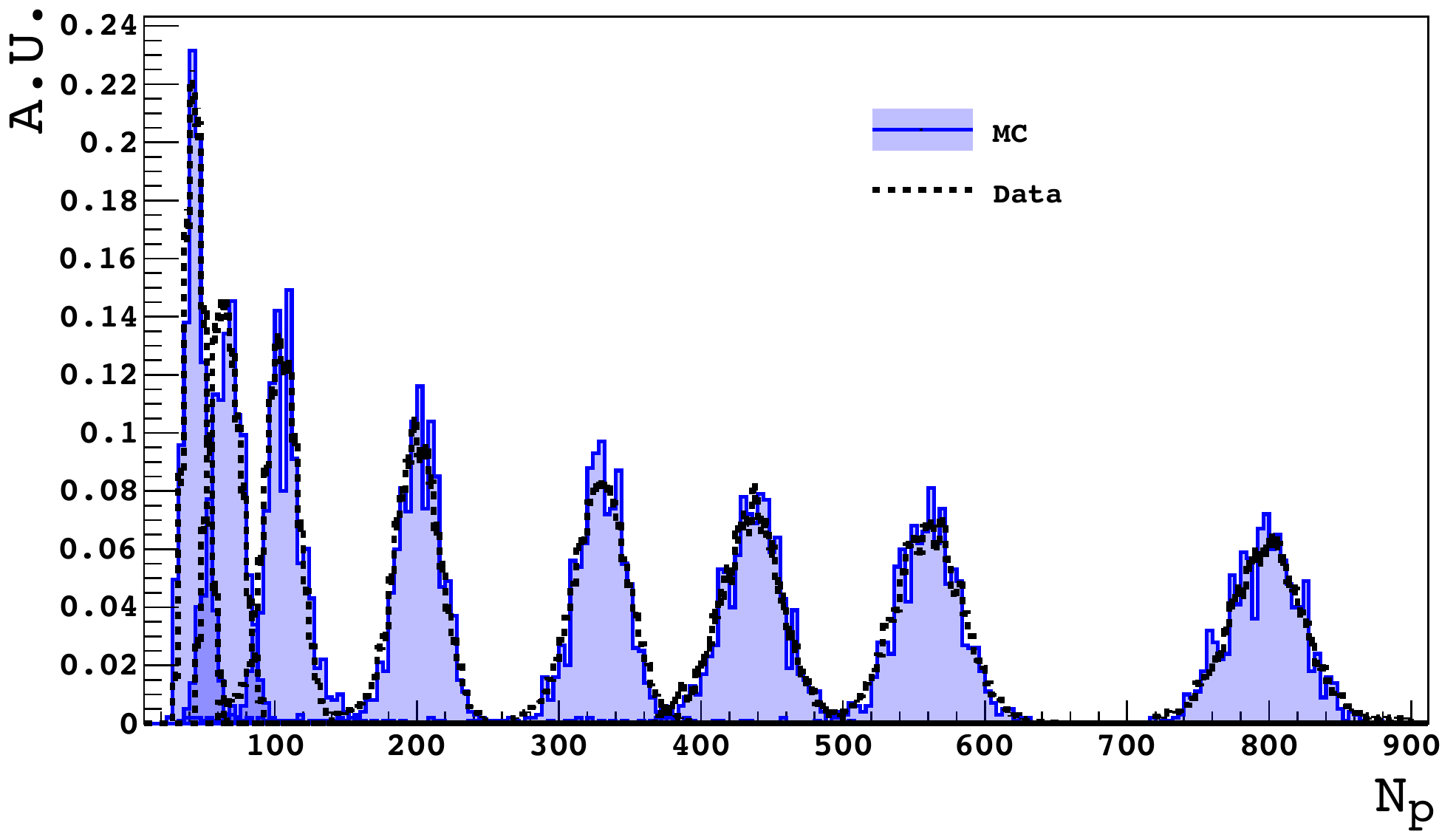} \\
\includegraphics[width = \columnwidth]{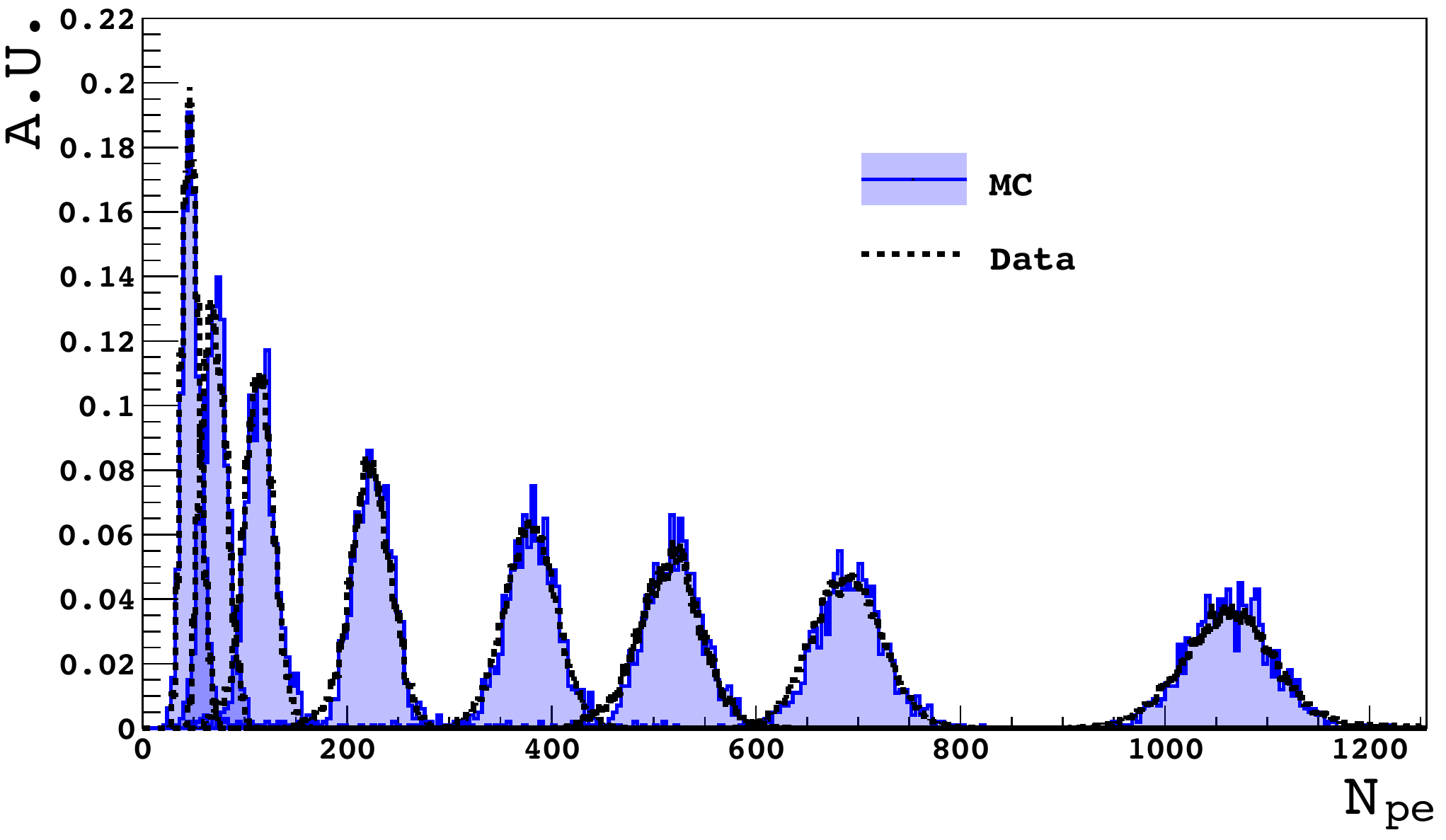}
\caption{\emph{Top panel:} Data-MC agreement for the $N_h$~energy estimator for the \gr\ sources listed in Sec.~\ref{sec:calibration} with the addition of the 2.22\,MeV \gr\ events from the neutron capture of the \ce{^{241}Am}-\ce{^9Be} source in the center of the detector. 
\emph{Middle Panel:} Data-MC agreement for the $N_p$~energy estimator for the \gr\ sources listed in Sec.~\ref{sec:calibration} with the addition of the 2.22\,MeV \gr\ events from the neutron capture of the \ce{^{241}Am}-\ce{^9Be} source in the center of the detector.
\emph{Bottom Panel:} Data-MC agreement for the $N_{pe}$~energy estimator for the \gr\ sources listed in Sec.~\ref{sec:calibration} with the addition of the 2.22\,MeV \gr\ events from the neutron capture of the \ce{^{241}Am}-\ce{^9Be} source in the center of the detector. }
\label{fig:tuningenergia}
\end{figure}

The reproduction of the energy response of $\beta$ particles was optimized scanning the light yield $Y_0^{ph}$, the Birks quenching factor \birks, the reemission 
probability at low wavelengths, and an overall multiplicative parameter ($G_{s}^{PMT}$) rescaling the simulated single photoelectron gains of the PMTs.
$G_{s}^{PMT}$ is needed since the lack of some details in the electronics simulation requires a small adjustment. In particular, it is
necessary to overcome the approximation used in the electronics simulation where the reference shape for the pulse of the PMTs is unique for all the channels.
All the other parameters in the simulation were kept fixed and the measured energy estimators were compared with the simulated ones for events in the center of the detector.
A comparison of the \gr\ peak positions for the $N_h$, $N_p$, and $N_{pe}$ variables is shown in Fig.\ \ref{fig:tuningenergia}. 

The energy spectra of \Coseven, \Cenine, and \Hg\ overlap with the tail of \C\ spectrum.  
A volume cut is not sufficient to obtain clean peaks and thus a proper subtraction of background events (acquired before and after the source insertion) 
was performed.
The agreement between data and simulation is evaluated comparing the reconstructed energy peaks of each \gr\ source.
They are fitted with a Gaussian function plus a line (the irreducible background due to radioactive decays or neutrinos) to obtain the mean value and the RMS of the energy estimator. 
The goodness of the energy simulation was defined with a custom estimator:
\begin{equation}
\label{eq:good_energy}
\epsilon= \sum_{sources} \frac{(\mu^{data} - \mu^{MC})^2}{e^2_{\mu^{data}} + e^2_{\mu^{MC} } + 1},
\end{equation}
where $\mu$ is the mean value of the Gaussian function from the fit (i.\,e. the peak position in the energy spectrum) and $e_\mu$ its error.
Such definition of the denominator avoids biases due to very small errors on $\mu^{data}$ and $\mu^{MC}$.
The optimized set of $Y_0$, \birks, reemission probability, and $G_{s}^{PMT}$ was found looking for the minimum value of $\epsilon$.
This procedure was followed for each energy estimator, showing that the results are independent on it.

The energy response (peak and resolution) of all the  sources located in the center of the detector is reproduced 
with a precision better than $0.8\%$. The resulting tuned parameters are given in Table~\ref{tab:tuning_ene}.
The procedure is then repeated once again for $\alpha$ particles. In this case, only the Birks quenching parameter (\birks\ for $\alpha$) is free, as all the others are fixed by the previous iteration.
\begin{table}
\centering
\begin{tabular}{l||r||r} 
 Parameter                                                       &  Expected     &            Tuned \\
 & value                                                           & value \\[0.15cm]  \hline \hline
$Y_0^{ph}$ (photons/MeV)                                      & $O(10^4)$             &               13600 \\[0.15cm] 
\birks ($\beta$, $\mbox{cm}\,\mbox{MeV}^{-1}$)     & $O(10^{-2})$           & 0.01098 \\[0.15cm] 
\birks ($\alpha$, $\mbox{cm}\,\mbox{MeV}^{-1}$)   & $O(10^{-2})$             & 0.01055 \\[0.15cm] 
$G_{s}^{PMT}$                                                   & 1                   & 1.015 \\[0.15cm] 
$P_{rem}(\lambda<320\,\mbox{nm})$           & -               & 0.53
\end{tabular}
\caption{Results from the tuning of the parameters describing the energy response. $G_{s}^{PMT}$ allows to slightly vary the gains of the PMTs in the simulation. $P_{rem}(\lambda<320\,\mbox{nm})$
is the parameter discussed in Sec.\ \ref{sec:lighttracking}. The orders of magnitude of the starting points of the light yield $Y_0^{ph}$ and of the quenching parameters \birks were inferred from standard properties
of organic liquid scintillators in the literature when available.}
\label{tab:tuning_ene}
\end{table}

\begin{table}[htb]
\centering

\begin{tabular}{l||r||r} 
 Parameter                              & Expected           & Tuned \\
 & value                                   & value \\[0.15cm] \hline \hline 
$\Lambda_{\mbox{PC}}$       & 1                         & 1.35 \\[0.15cm]
$\Lambda_{\mbox{nylon}}$  & 1                         & 0.75 \\[0.15cm]
$\Lambda_{\mbox{DMP}}$    & 1                         & 1.1 \\[0.15cm]
$\Lambda_{\mbox{PPO}}$ & 1 & 1.15 \\[0.15cm]
$R_{\mbox{SSS}}$ & 0.49* & 0.55 \\[0.15cm]
$R^{spike}_{\mbox{SSS}}$ & 0.4* & 0.12 \\[0.15cm]
$R_{\mbox{cathode}}$ & 0.1* & 0 \\[0.15cm]
$R_{\mu-\mbox{metal}}$ & 0.4* & 0.4 \\[0.15cm]
$R_{\mbox{PMT ring}}$ & 0.6* & 0.75 \\[0.15cm]
$R^{spike}_{\mbox{PMT ring}}$& 0.8* & 0.88 \\[0.15cm]
$R_{\mbox{conc. int.}}$ & 0.88~\cite{bib:BXLC} & 0.95\\[0.15cm]
$R^{spike}_{\mbox{conc. int.}}$ & 0.8* & 0.985\\[0.15cm]
$R_{\mbox{conc. ext.}}$ & 0.88~\cite{bib:BXLC} & 0.95\\[0.15cm]
$R^{spike}_{\mbox{conc. ext.}}$ & 0.8* & 0.975\\[0.15cm]
$R_{\mbox{nylon ring}}$ & 0.4* & 0.3
\end{tabular}
\caption{Results on the tuning of the parameters related to attenuation lengths and reflectivity. $\Lambda$ is the the multiplicative coefficient
which multiplies the measured curves of the attenuation lengths, $R$ is the reflectivity and $R^{spike}$ is the fraction of specular reflection with respect to
diffusive reflection. ``PMT ring'' denotes the stainless steel small ring mounted on the PMTs without concentrators.
$R_{\mbox{nylon ring}}$ is the value of the reflectivity of the nylon endcaps on top and bottom of the inner vessel.
The expected values for the parameters highlighted with ``*'' are inferred from typical values for those materials as found in the literature.
}
\label{tab:tuning_att}
\end{table}
 
 \begin{table}
\centering
\begin{tabular}{l||r||r} 
 Parameter & Measured & Tuned \\
  & value & value \\[0.15cm]  \hline \hline
$\tau_{1,\beta}\,\mbox{(ns)}$ & 3.95 & 3.7 \\[0.15cm]
$\tau_{2,\beta}\,\mbox{(ns)}$ & 23.56 & 24\\[0.15cm]
$\tau_{3,\beta}\,\mbox{(ns)}$ & 78.86 & 60\\[0.15cm]
$\tau_{4,\beta}\,\mbox{(ns)}$ & 546.39 & 600\\[0.15cm]
$q_{1,\beta}$ & 0.933 & 0.889\\[0.15cm]
$q_{2,\beta}$ & 0.024 & 0.055\\[0.15cm]
$q_{3,\beta}$ & 0.022 & 0.027\\[0.15cm]
$q_{4,\beta}$ & 0.021 & 0.029\\[0.15cm]
$\tau_{1,\alpha}\,\mbox{(ns)}$ & 4.15 & 3.9\\[0.15cm]
$\tau_{2,\alpha}\,\mbox{(ns)}$ &  19.90 & 26\\[0.15cm]
$\tau_{3,\alpha}\,\mbox{(ns)}$ & 99.91  & 110\\[0.15cm]
$\tau_{4,\alpha}\,\mbox{(ns)}$ & 617.96 & 630\\[0.15cm]
$q_{1,\alpha}$ & 0.679 & 0.674\\[0.15cm] 
$q_{2,\alpha}$ & 0.144 & 0.146\\[0.15cm] 
$q_{3,\alpha}$ & 0.102 & 0.103\\[0.15cm] 
$q_{4,\alpha}$ & 0.075  & 0.077\\[0.15cm] 
$\tau_{PC}$ (ns) & 28 & 28\\[0.15cm]
$\tau_{PPO}$ (ns)& 1.6 & 1.6\\[0.15cm]
$\tau_{PCtoPPO}$ (ns) & 3.6 & 3.6
\end{tabular}
\caption{Results on the tuning of the scintillation time parameters for $\alpha$ and $\beta$ particles. It has to be noted that all these numbers are correlated to some of the
reflectivity values (especially stainless steel sphere's and concentrator's). The $\tau_i$ and $q_i$ values are the time constants and the weights describing the four exponentials modeling the primary scintillation 
light generation (see Ref.~\cite{bib:BxLong}). The starting values for the exponential parameters come from Ref.~\cite{Lombardi:2013nla} (see the discussion in text for the explanation of the 
differences with the tuned values). 
The three last rows describe absorption and reemission time constants, as discussed in Sec.\ \ref{sec:lighttracking}. .}
\label{tab:tuning_t}
\end{table}

\section{Validation of the tuned Monte Carlo}
\label{sec:validation}
\subsection{$^{210}$Po events and energy resolution}
$\alpha$ decays of \Po\ in the scintillator represent a good benchmark for the validation of the tuned MC.
\Po\ is not uniformly distributed in the detector and its activity and space distribution change as a function of time because of its decay and of convective movements of the scintillator. 
Furthermore, Borexino experienced a quite significant loss of live PMTs along the almost 10 years of data taking, which resulted in a time varying energy response and in a time dependence of 
the non uniformity of the energy response~\cite{bib:BxLong}. The distribution of dead PMTs changed over time (at the beginning, most of the dead PMTs were at the bottom of the detector, while in
the last years the situation evolved towards a more uniform distribution). 
The accurate MC tuning allows to obtain a good agreement in the reproduction of the detector energy response to \Po\ $\alpha$ decays without any change of the input parameters found at the tuning stage.

\begin{figure}[htb]
\centering
\includegraphics[width = 1\columnwidth]{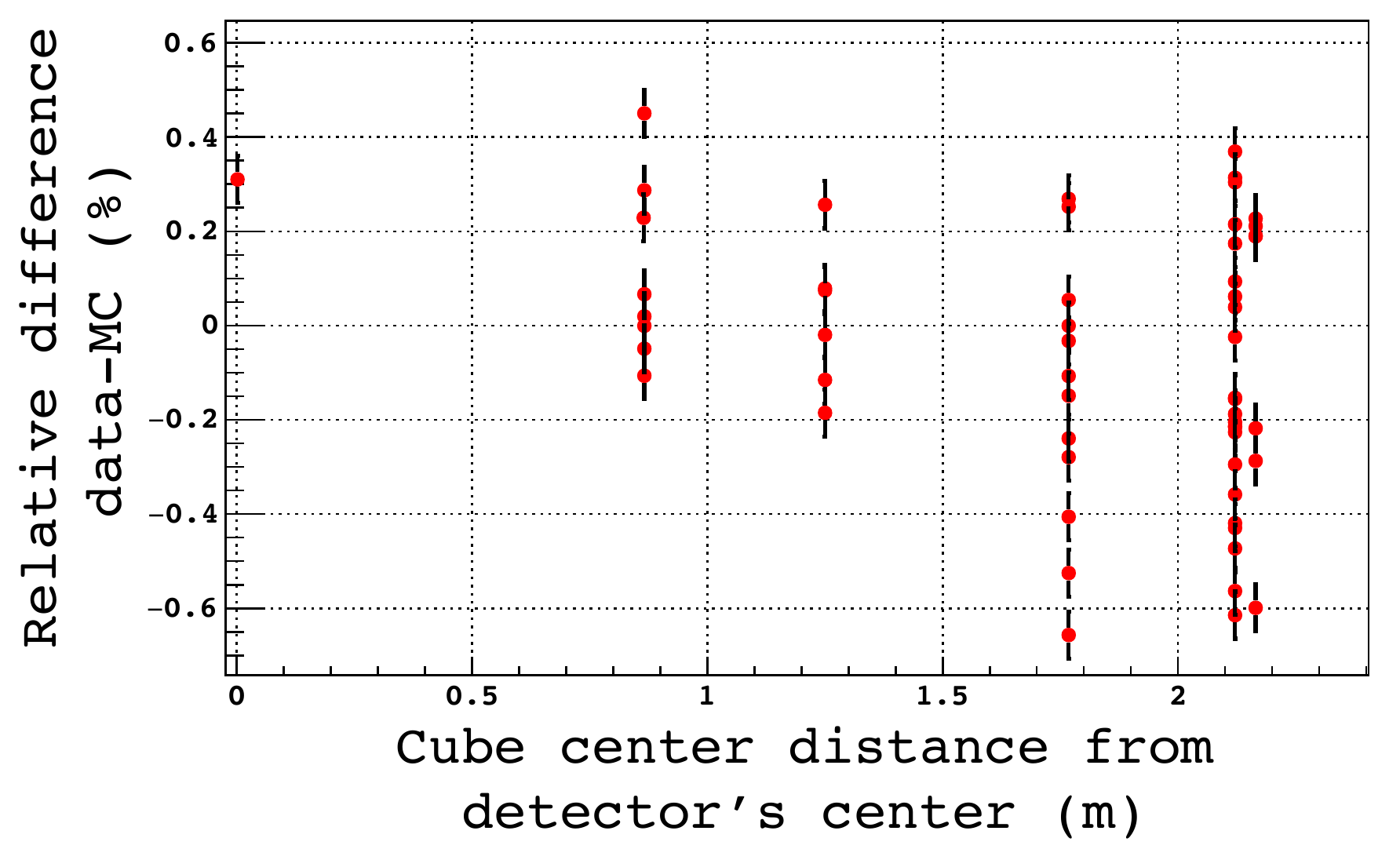}
\caption{Relative difference between the \Po\ energy peak (evaluated with the $N_{h}$ energy estimator) in MC and in data as a function of the cube center position (radius from the center of the detector). The cubes
cover completely the innermost $3\,\mbox{m}$ of Borexino's inner vessel.}
\label{fig:po210sim}
\end{figure}

We investigated the agreement between data and MC for the period 2012--13.
On a monthly basis, \Po\ events were selected in the data in cubic voxels of $\sim$1.5~m edge using a suitable pulse shape parameter and appropriate energy cuts~\cite{bib:BxLong}.
We separately regarded 59 cubic voxels within the fiducial volume. 
As a second step, we simulated \Po\ data sets reproducing the time and space distributions observed, including the number of operating channels in this data taking period.
We then compared the spectral position of the \Po\ peak in  data and MC and found an agreement (see Fig.\ \ref{fig:po210sim}) of better than $1\%$. 
This result demonstrates that the simulation that is tuned on discrete source positions is able to reproduce the non uniformity of the energy response observed in data. 
In addition, it shows that the scintillator response is stable over time as calibration data taken in the year 2009-2011 furnish MC parameters that accurately reproduce data measured several years later\footnote{Note that the \Po\ events selected for the $\alpha$ particle tuning are different from those used for this test and shown in Fig.\ \ref{fig:po210sim}.}. 

An important crosscheck of the MC modeling is based on the comparison of the predicted energy resolution with the one measured in data, as no additional parameter in the simulation is required for tuning the energy resolution.
The reproduction of the correct value is a by-product of the simulation of all relevant scintillation processes (charged particle 
energy loss, scintillation and \v Cerenkov light generation, absorption, reemission, and light propagation) and of the proper schematization of the electronics effects (PMT jitter, electronics features and so on). Testing the ability of the MC simulation to reproduce the \Po\ peak position and width allows to test the combination of all the above mentioned effects.
Both the peak position and its width are strictly correlated with the non uniformity simulation both in time and in space of the detector response. In addition, the peak width accounts for the energy resolution
effects providing a benchmark for testing the whole MC simulation.

\subsection{High energy light collection}

\begin{figure}[tb]
\centering
\includegraphics[width = 1\columnwidth]{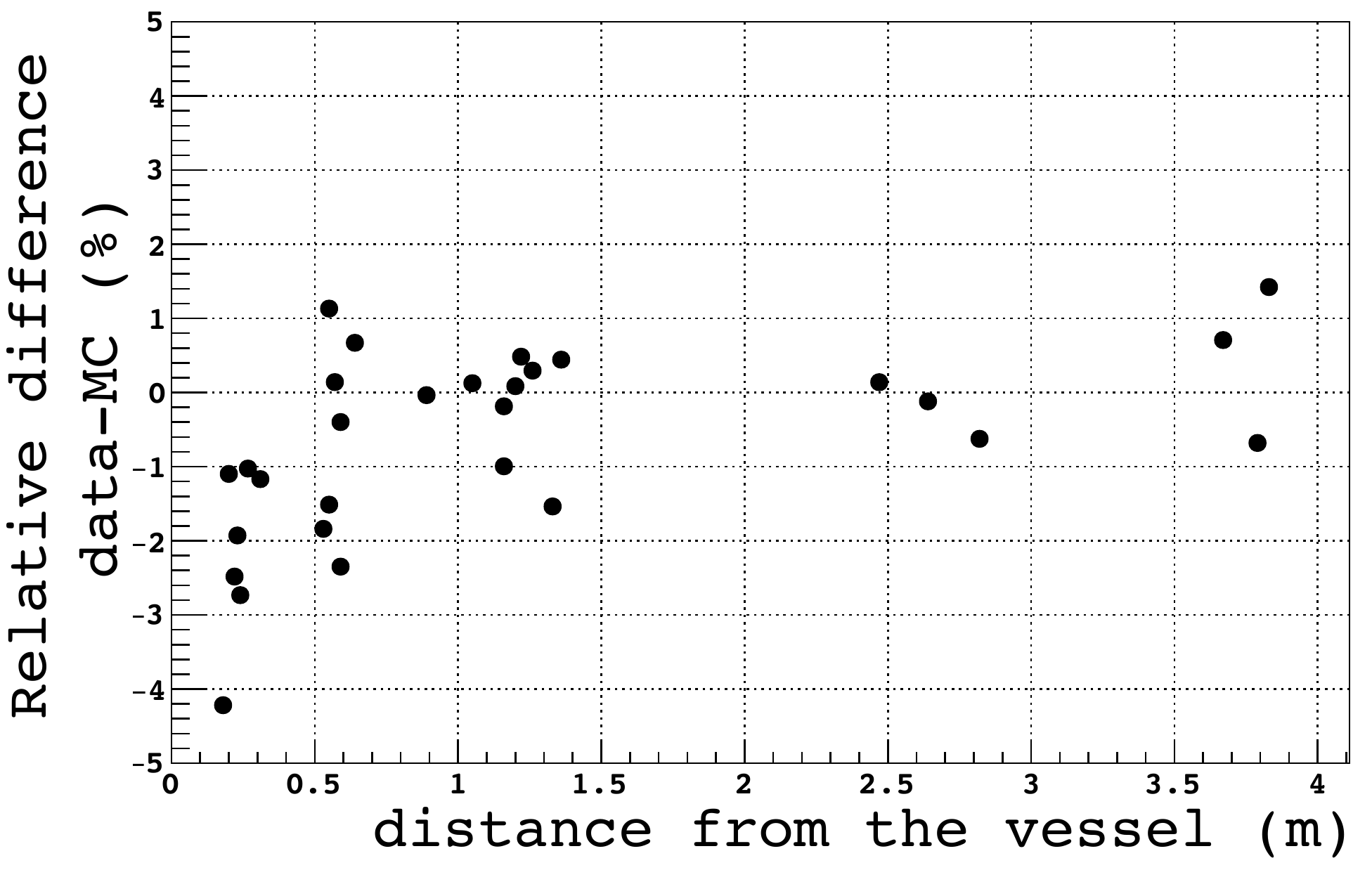}
\caption{Relative difference of the $2.2$~MeV neutron capture on hydrogen $\gamma$-ray energy peak ($N_{pe}$ variable) in MC and data as a function of the distance from the inner vessel. 
The agreement worsens at short distance from the vessel (distance similar to the $\gamma$-ray attenuation length on the scintillator), probably due to uncertainties in the reconstruction of the vessel shape~\cite{bib:BxLong}.}
\label{fig:LC_AmBe}
\end{figure}

Light collection tests were also performed studying the $2.2$ MeV $\gamma$-ray emission after neutron capture events on hydrogen from an \AmBe\ neutron source.
These $\gamma$-ray events have higher energy with respect to the typical energies of analysis for solar neutrino measurements\footnote{With the exclusion of the \bor~neutrino analysis.}.
However, reproducing the detector behavior at energies of a few MeV are of interest for geo-neutrinos and SOX analyses. 
Since for the former analysis the fiducial volume radius is increased~\cite{bib:BxGeo}, several positions close to the inner vessel were mapped with the \AmBe\ source. 
Figure~\ref{fig:LC_AmBe} shows the relative difference between data and MC as a function of the distance from the vessel using the $N_{pe}$ energy estimator. 
Although the tracking code was tuned at lower energy, the discrepancy is $\lesssim$2\% in the volume of interest. 
The worsening of the agreement at short distance from the vessel is probably due to uncertainties in the reconstruction of the vessel shape~\cite{bib:BxLong}.

\subsection{External background}
\label{sec:extcomparison}

The external \gr\ simulation procedure was validated by direct comparison with data from calibrations. No parameters were tuned or modified  to improve
the goodness of the simulations, thus the complete data set from external calibrations was used for testing the already tuned MC code. 
\begin{table*}
	\centering
	\begin{tabular}{l||llllllllll}
		& N1        & N2        &  N3        & N4        & N5        & N6        & N7        & S3        & S5        & S7        \\ \hline \hline
		X {[}m{]} &2.63      &4.16      &5.93      &6.53      &6.07      &4.37      &2.55      &-6.00     &-6.06     &-2.21     \\
		Y {[}m{]} &-0.62     &-0.81     &-1.44     &-1.65     &-1.08     &-0.80     &-0.43     &0.98      &1.02      &-0.27     \\
		Z {[}m{]} &6.29      &5.39      &3.11      &-1.24      &-2.99     &-5.22     &6.34     &3.15      &-3.03     &-6.48    
	\end{tabular}
	\caption{Positions of the external calibration \Theight\ source deployments. The coordinates are relative to the center of the detector. ``N'' (``S'') letter indicates the north (south) side of the detector while the numbers refer to the position's height. }
	\label{tab:extpos}
\end{table*}

\begin{figure*}
	\centering
	\includegraphics[width = 2\columnwidth]{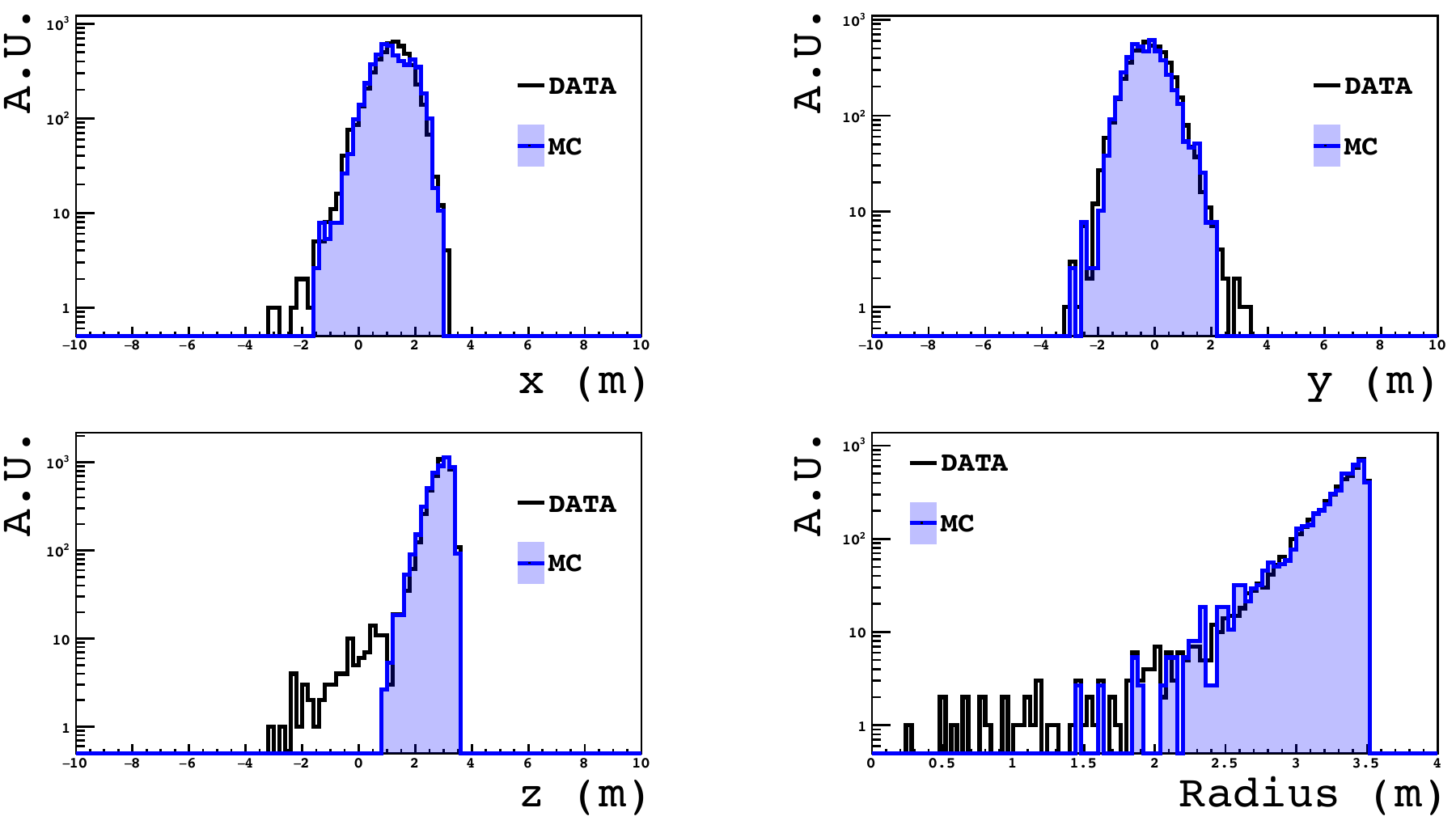}
	\caption{Comparisons of the reconstructed positions ($x$, $y$, and $z$ coordinates and radius) in MC and calibration data for the source deployed in the N3 position from Table~\ref{tab:extpos}. The discrepancy between data and MC, mainly visible at negative z values, is due to bulk events (e.\,g. radioactive decays in the scintillator) which are not external-source related and are not simulated in this specific example.}
	\label{fig:posrecoext}
\end{figure*}

\begin{figure*}
	\centering
	\includegraphics[width = 1\columnwidth]{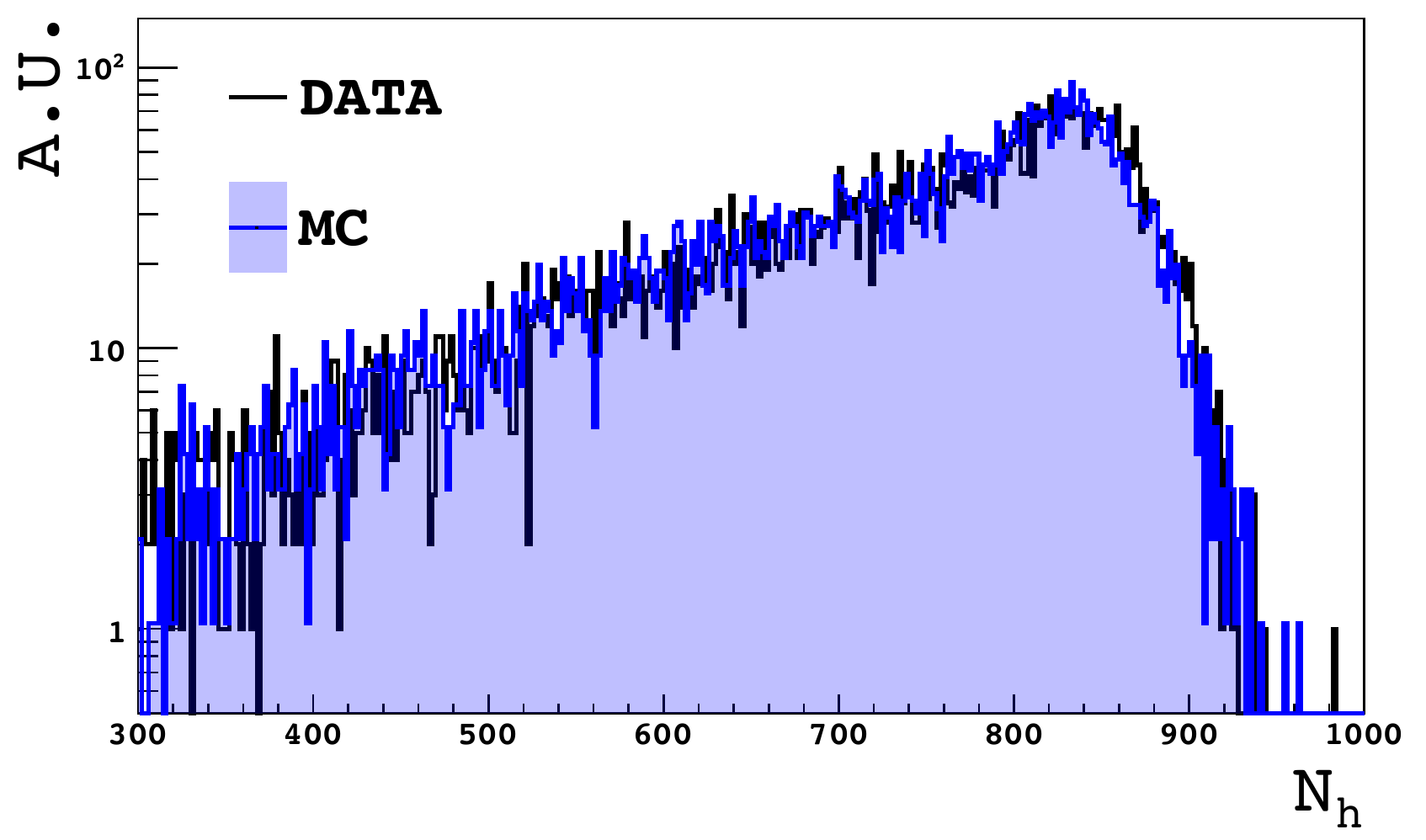}
	\includegraphics[width = 1\columnwidth]{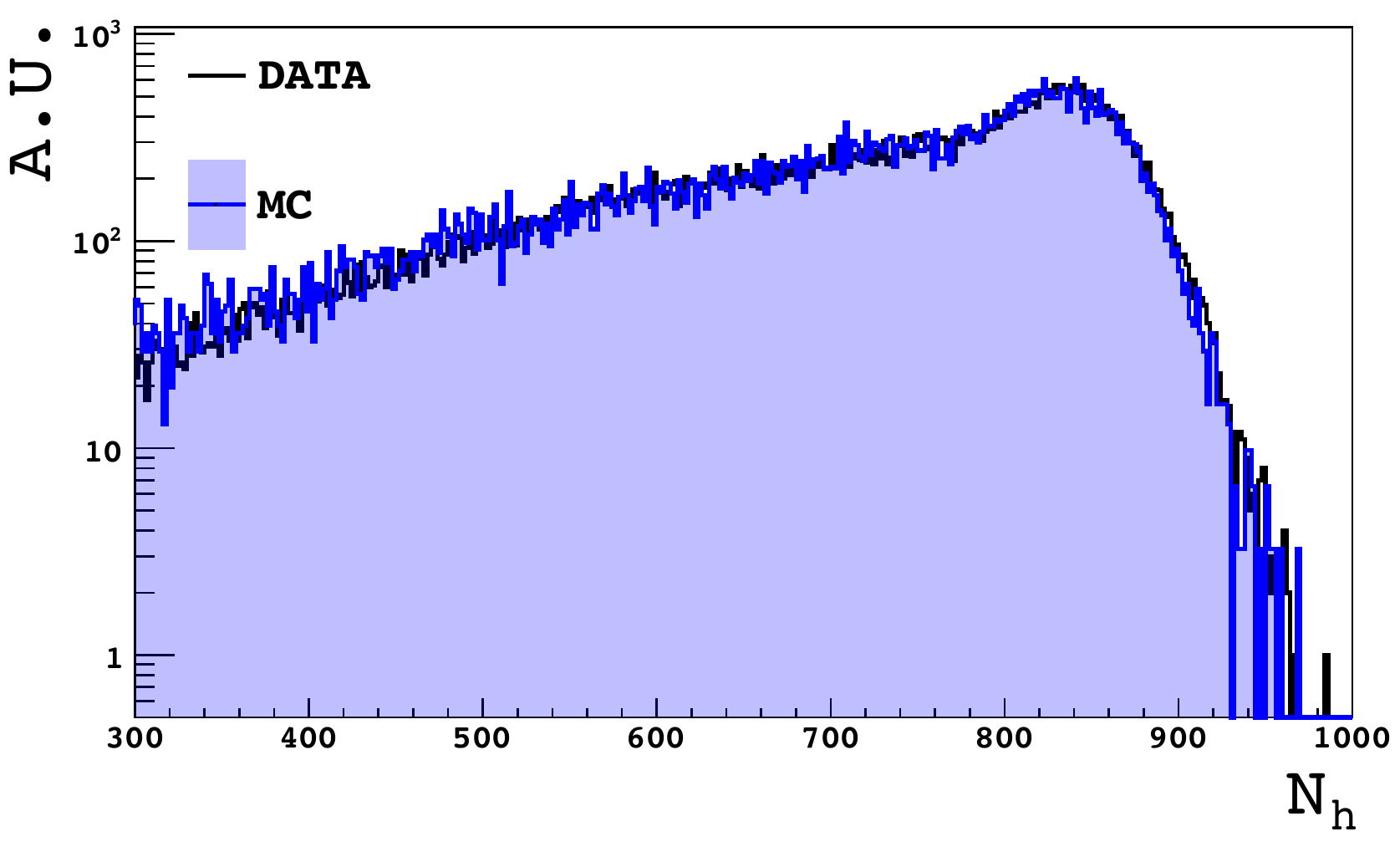}
	\caption{\emph{Left Panel:} Comparison of the external \gr\ energy spectrum ($N_h$) for \Theight\ source-induced events from the N3 position. The fiducial volume is defined as a sphere  
		with $3.5\,\mbox{m}$ radius.
		\emph{Right Panel:} Comparison of the external \gr\ energy spectrum ($N_h$) for source-induced events from the N5 position. The fiducial volume  cut
		is a $3\,\mbox{m}$ radius sphere.}
	\label{fig:extspectrum}
\end{figure*}

\begin{figure}
	\centering
	\includegraphics[width = 1\columnwidth]{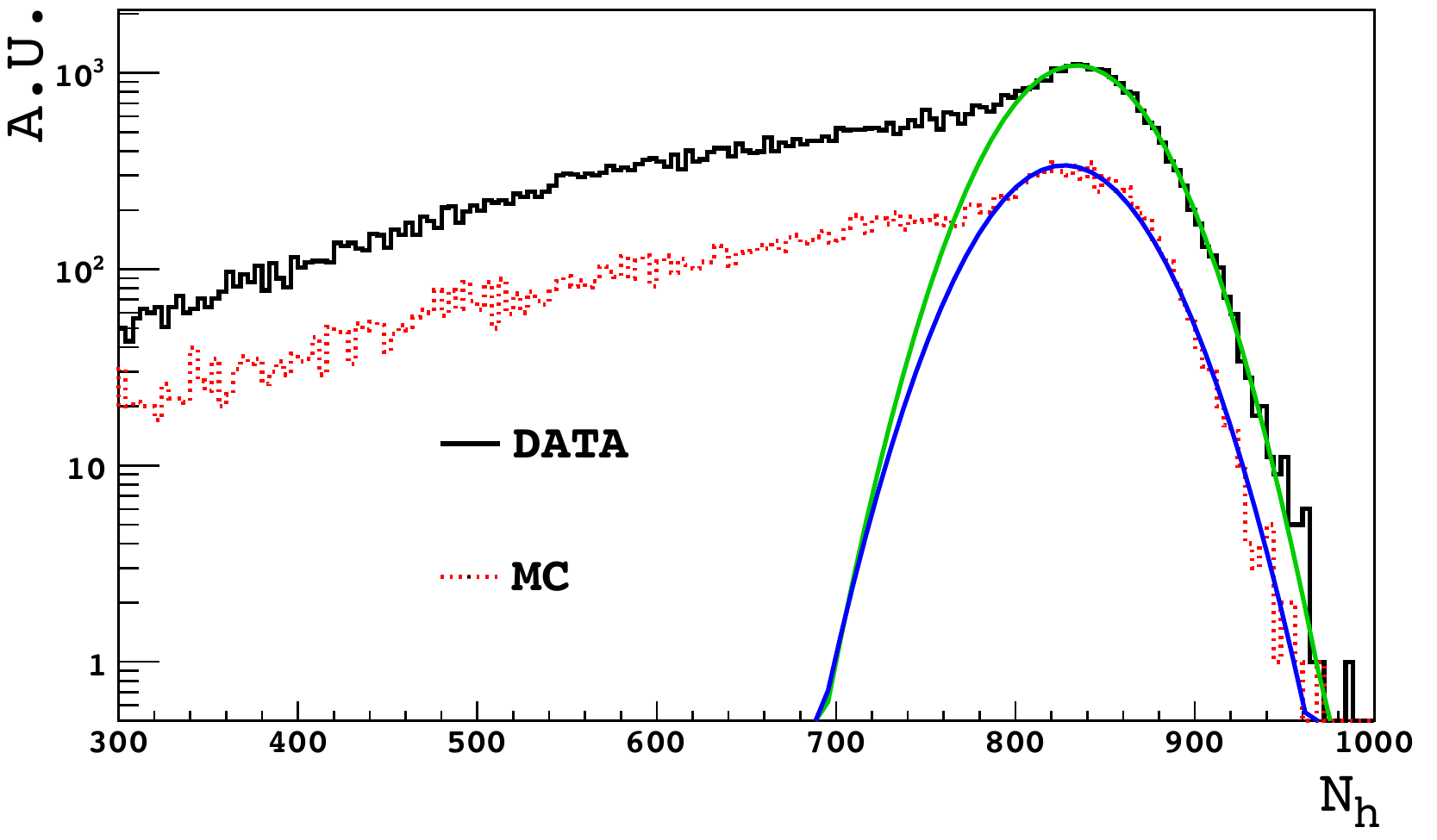}
	\caption{Gaussian fit examples of the full \gr\ absorption peak for data and MC for the \Theight\ external calibration source in the N3 position from Table~\ref{tab:extpos}. The relative normalization of the two spectra is arbitrary.}
	\label{fig:extpos}
\end{figure}
The external calibration campaign was carried out using a \Theight\ source  ($\tau=2.76$ yr), with $\sim$2.9~MBq  activity placed 
in several positions both in the top and the bottom hemispheres, as listed in Table \ref{tab:extpos}.
\Tl\ is one of \Theight's daughters and the emission probability of the 2.615\,MeV \gr\ line is $35.6\%$. 
$\alpha$ and $\beta$ particles emitted by daughter nuclides of \Theight\ are absorbed by the source encapsulation, while low energy \grs\ are absorbed in the buffer, very close to the generation point. 
Data were acquired with the source in positions on two sides of the detector (north and south,
geographically) and with different heights with respect to the ground. In Table\ \ref{tab:extpos} the ``N'' (``S'')
letter indicates the north (south) side while the numbers refer to the position's height. 
It is important to study the detector response to external \grs\ in different positions and especially at different heights to test both the effect of 
the nylon vessel shape and that of the asymmetry of the live PMTs distribution.

Using the variance reduction procedure described in Sec.~\ref{sec:EB}, \Tl\ events originating from all source positions (see Table \ref{tab:extpos}) were simulated. 
The reconstructed energy spectra and event positions were compared to the measured ones, recording an excellent agreement between simulations and real data.
Examples are given in Fig.\ \ref{fig:posrecoext}, where the event positions reconstructed in data and MC are shown, and in Fig.\ \ref{fig:extspectrum},
where the \gr\ energy spectrum built with the $N_h$ variable is compared in data and MC for different fiducial volume cuts. 
Both the event position and the energy spectrum are well reproduced.
Particularly, it is possible to note a good agreement even with the $3.5\,\mbox{m}$ fiducial volume cut (left panel of Fig.\ \ref{fig:extspectrum}), which is already well beyond the needs
of the solar neutrino analysis (i.\,e. good light collection reconstruction in the $3\,\mbox{m}$ sphere). The different shape of the energy spectrum between the two plots in Fig.\ \ref{fig:extspectrum}
is due to the different fiducial volume cut, and thus to the different ratio between the Compton tail (of events degraded in energy by the buffer) and the full absorption peak.  

A more systematic study was performed to assess the solidity of the method in the whole inner vessel volume. 
In particular, a parameter of interest is the ratio between the number of events in the full absorption peak and in the Compton tail, $\rho_{\mbox{peak/tail}}$,
which mostly depends on the vessel shape non-uniformity.
This ratio is quantified by fitting the full absorption peak with a Gaussian (as shown in Fig.\ \ref{fig:extpos}).
The integral of the Gaussian is then compared to the total number of events in the spectrum and thus to those in the Compton tail.

\begin{figure}
\centering
\includegraphics[width = 1\columnwidth]{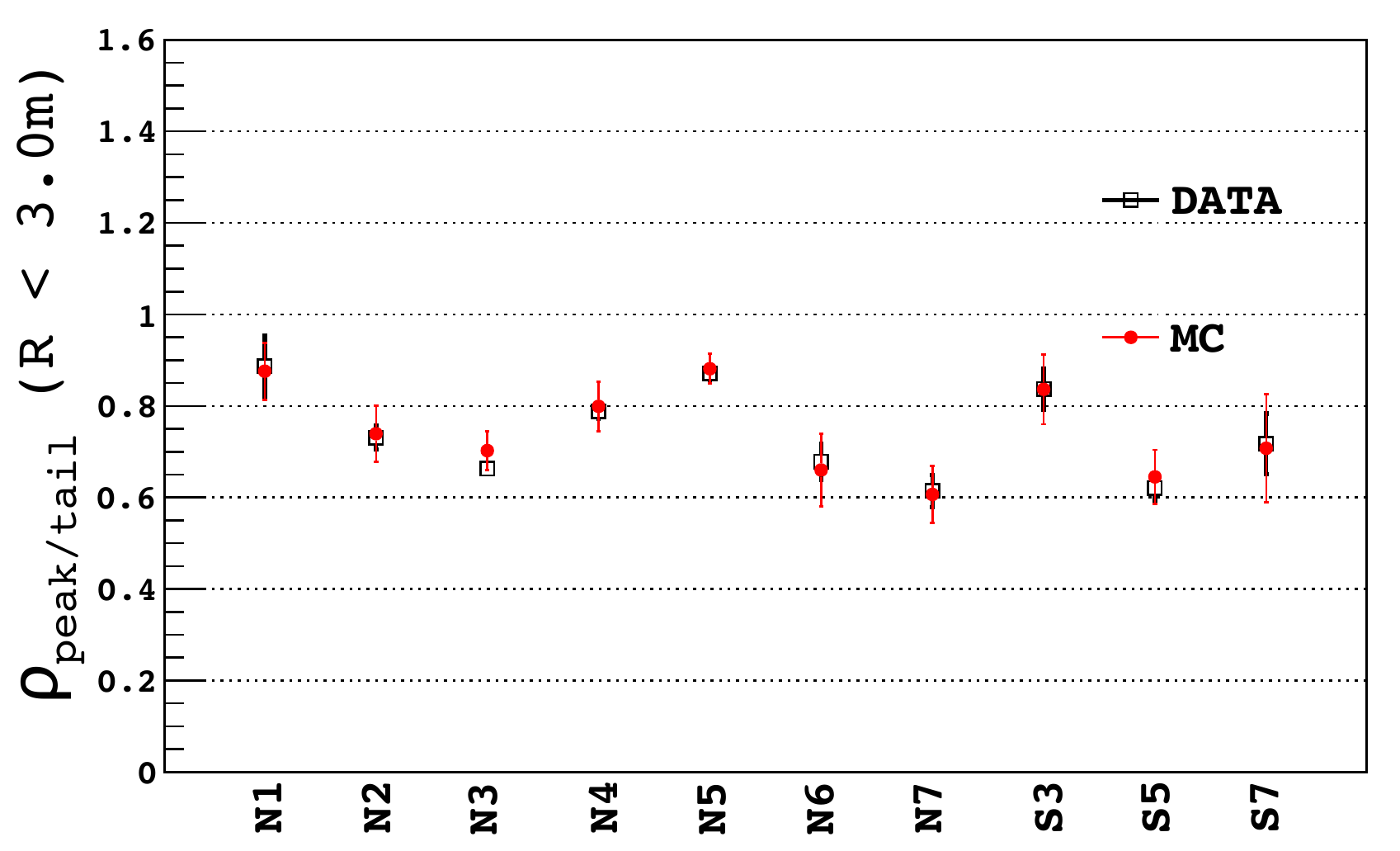}
\caption{Value of the full absorption peak/Compton tail ratio as a function of the source position both in MC and in data.}
\label{fig:compton}
\end{figure}

Figure\ \ref{fig:compton} shows the values of $\rho_{ \mbox{peak/tail}}$ as a function of the source position both for MC and data for events selected in a $3\,\mbox{m}$ fiducial sphere. 
The excellent agreement found demonstrates that the simulation is working properly. 
Moreover, it proves once more that the dynamical vessel shape reconstruction in Borexino is
solid, since $\rho_{ \mbox{peak/tail}}$ as a function of the source position depends on it.
	

\subsection{Position reconstruction}
\begin{figure*}
\centering
\includegraphics[width = 2\columnwidth]{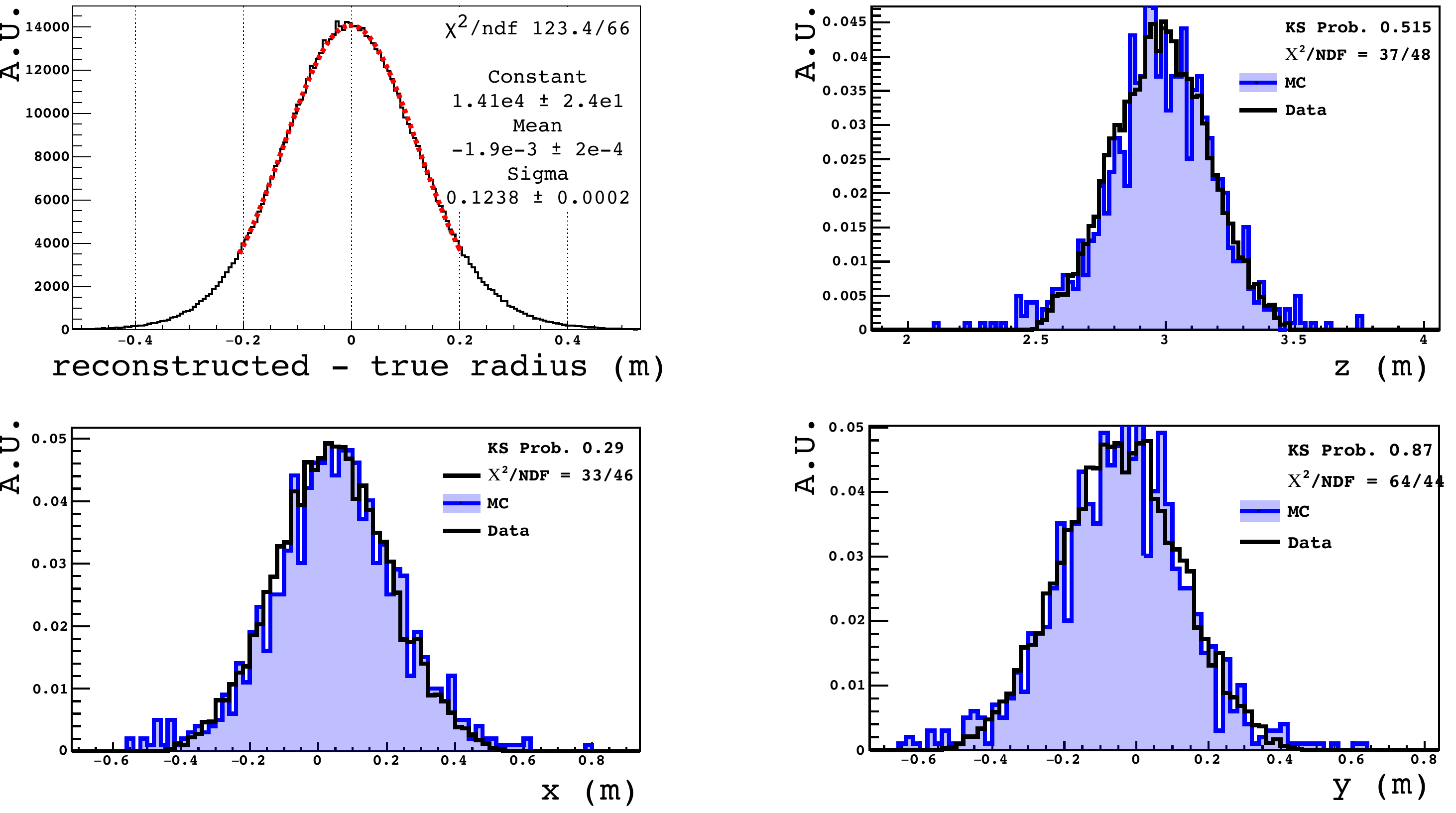}
\caption{The top left panel shows the difference between real and reconstructed radius for electron events uniformly simulated in a $3.5$ m sphere. The distribution is not Gaussian, thus the $\chi^{2}$/NDF of the fit is not very good. However, the mean of the distribution is only a couple of mm different from zero. The other panels show the comparison between data and MC of the three cartesian coordinates of the \ce{^{85}Sr} source calibration events. The first entry in the legend indicates the outcome of a Kolmogorov-Smirnov test.}
\label{fig:position}
\end{figure*}

The  ability of the MC to reproduce the position reconstruction was evaluated in two steps.
As the first step, 
a large sample of events homogeneously distributed over the 3.5\,m fiducial volume was generated to check for the presence of systematic shifts in the reconstruction.
As shown in the top left panel of Fig.~\ref{fig:position}, the difference between the real vertex position (recorded in the MC truth structure for each event) and the reconstructed position is compatible with zero
within the error. 
As second step, the reconstructed position resolution
was compared for data from calibration sources at defined positions and corresponding MC samples.
The agreement between data and MC is shown in Fig.~\ref{fig:position}. 
Since no tuning parameters are introduced to reproduce the position reconstruction accuracy, this result is an indication of a reliable modeling of light production, propagation and detection.

In conclusion, the correct reproduction of the position reconstruction resolution validates the accuracy of the MC, while the absence of systematic shifts 
demonstrates the overall quality of the position reconstruction algorithm.

\subsection{Event pileup}
Pileup events are due to two or more scintillation events happening so close in time that the clustering algorithm cannot disentangle them. It is the most critical
background for very low energy solar neutrino analyses, particularly for the measurement of the \pp\ neutrino interaction rate~\cite{bib:BxPp}.
For the analysis presented in Ref.~\cite{bib:BxPp}, the pileup energy spectrum was built using a data driven method (we call it ``synthetic pileup'') and the energy estimators defined with a fixed cluster duration
(i.\,e. $N_p^{dti}$, see Sec.\ \ref{sec:reco}). We recall that for every triggered event, we register all the hits in a time window of 16.5 $\mu$s, 
which is much longer than the cluster duration (typically around 1 $\mu$s). The algorithm for the synthetic pileup production considers the hits recorded
within a time interval of duration $dti$ in the second half of the gate and superimposes them with the primary cluster, which caused the trigger to occur and is placed at the beginning of the DAQ gate. 
These synthetic events are then processed with the standard reconstruction code. 
The pileup spectrum is defined as the set of synthetic events whose energy estimator $N_p^{dti}$ is increased by at least $E_{min}$, with respect to
the corresponding original event, after the overlapping of the additional hits. The major sources of systematics in this method are the choice of $E_{min}$
and the assumption that all pileup events contain at least one primary event which would have triggered on its own. 
Another critical point is the fact that the PMT dark noise is double counted, since it is automatically contained both in the original clustered event and in the hits which are shifted in time.

The tuned MC simulation allows to reproduce the synthetic pileup and to gain understanding on its different components. 
In addition, the MC pileup is free of the above mentioned systematic effects.
 
The MC pileup generation works as follows. 
Events of given kinds are produced with the tracking code and saved. 
Two different events are correlated in time by an external application which creates a new file, preserving the same output structure of the tracking code. 
The time correlation is typically created by sampling a random delay between the events uniformly in a selectable range (typically equal to a few $\mu$s).  
The output so created is then processed through the standard electronics simulation and reconstruction code. More details can be found in Ref.~\cite{bib:tesisimone}.
This procedure has the advantage of replicating what happens in reality: real events overlap at the level of light production in the scintillator, 
and the already piled-up pattern of hits undergoes the electronics processing and the reconstruction. This is performed in the same way in the simulation. 
The disadvantage of this method is the need of knowing the most abundant primary classes of events which contribute non negligibly to the pileup. 
The final pileup spectrum is constructed by summing up the results of the overlap of pairs of events (such as \C\-\C, see below for details). 
As a final remark, this method allows the inclusion in the analysis of those pileup events which might not trigger the detector individually, but which do if they happen close enough in time.
This is not possible with the synthetic pileup method.

Among the classes of events with the highest rate, we identified three important ingredients for the pileup: \C\ (120 Bq) , \Po\ (a few hundreds of Bq), 
and \grs\ of the external background, which interact in the buffer or in the outermost region of the inner vessel. The developments carried out on the external background simulation (see Sec.\ \ref{sec:EB})
were essential to assess that the order of magnitude of the expected external \gr\ rate is not negligible for the pileup construction.
Extrapolating the measurements of the external contaminations inside the inner vessel according to the simulated radial distribution for the outermost shells, we found that 
the expected count rate of external \grs\ interacting in the whole SSS volume is about  7\,kBq.
However, it can be easily proven (also by using the MC), that the overlap of two external \gr\ events is negligible in practice.
Despite the very high rate of external \grs, most of their pileup events are overlaps of a few hits coming from two external \gr\ depositions in the buffer. 
Given the very low visible energy of these events, almost none of them triggers or is identified as a cluster. Therefore, they can be neglected in the analysis. 
Thus, the most important components turn out to be:
\begin{itemize}
\item  \C\ overlapped with external \grs.
\item \C\ overlapped with $\rm{^{14}C}$.
\item  \Po\ overlapped with external \grs.
\end{itemize}
The pileup with \Po\ results in an effective smearing of the \Po\ peak width. Furthermore, the contribution of \Po\ overlapped with \C\ can be neglected, 
being of the second order with respect to that of \Po\ with external \grs.

\begin{figure}[htb]
\centering
\includegraphics[width = 1\columnwidth]{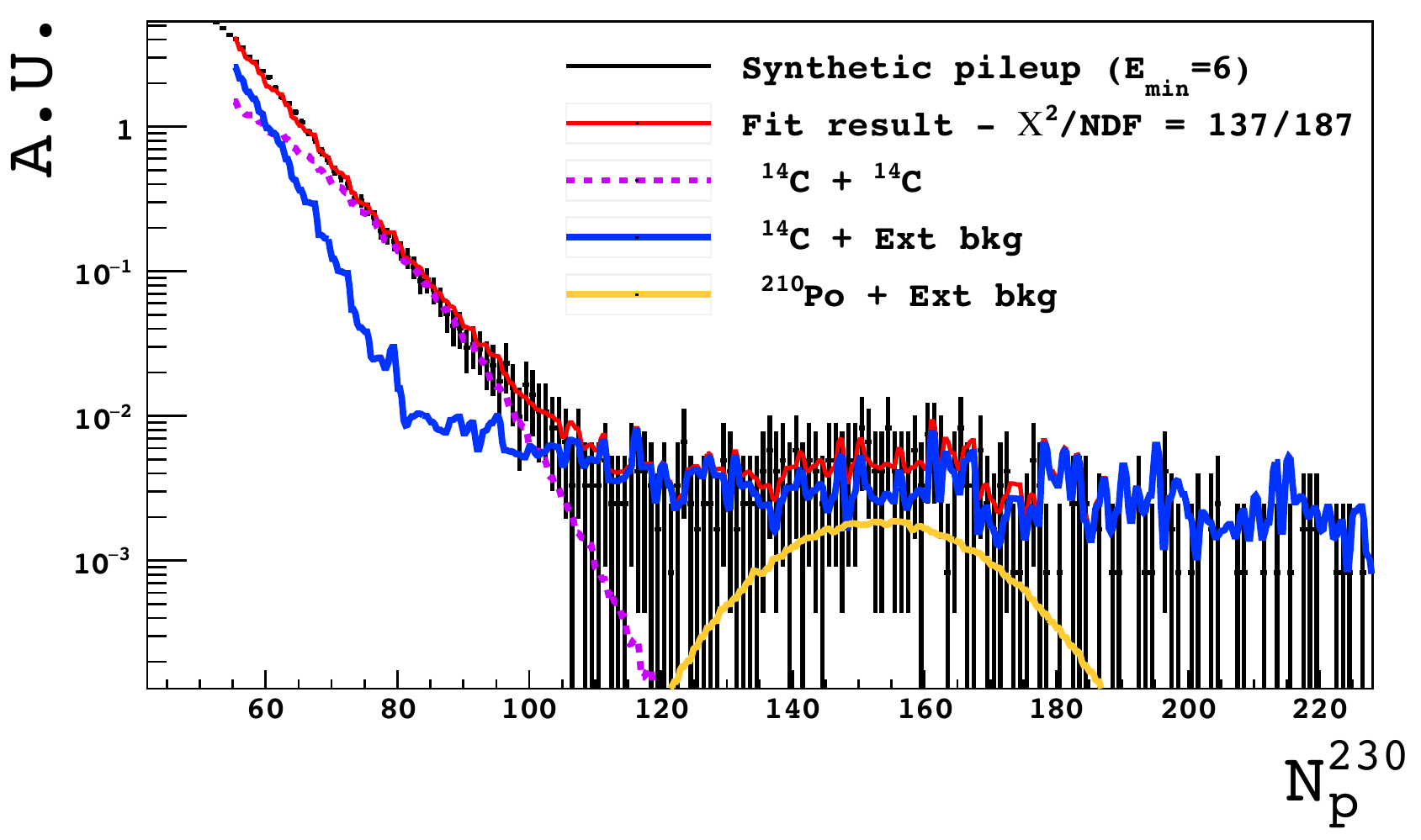}
\caption{ Fit of the synthetic pileup spectrum with the considered MC pileup spectral components for $E_{min}=6\,N_p^{230}$.
}
\label{fig:fitemin6}
\end{figure}

At this point, the MC production of these three pileup components is straightforward. However, the relative rates must be known in order to sum them up and produce 
the final pileup spectrum.
We obtained them by fitting the synthetic pileup spectrum with a fixed $E_{min}$ value with the superposition of pileup spectra produced by the MC and obtained with the application of the same $E_{min}$ cut.
The rates of the primary components can be established with $\sim$2\% accuracy.
Then, MC pileup spectra are produced with no other modifications or assumptions. Note that in the final MC pileup, there is no $E_{min}$ parameter.
The fit to the synthetic pileup spectrum produced for the analysis shown in Ref.~\cite{bib:BxPp}\footnote{In Ref.~\cite{bib:BxPp}, the synthetic pileup spectral shape with 
$E_{min}=5\,N_p^{230}$ was used to describe the pileup.} and $E_{min}=6\,N_p^{230}$ is shown in Fig.~\ref{fig:fitemin6}. 
The spectral components used for fitting are \C\ overlapped with external \grs,  \C\ overlapped with \C, and \Po\ overlapped with external \grs. 
The fit result provides the relative weights of the different components, depurated from threshold and reconstruction effects.

\begin{figure}[htb]
\centering
\includegraphics[width =0.9\columnwidth]{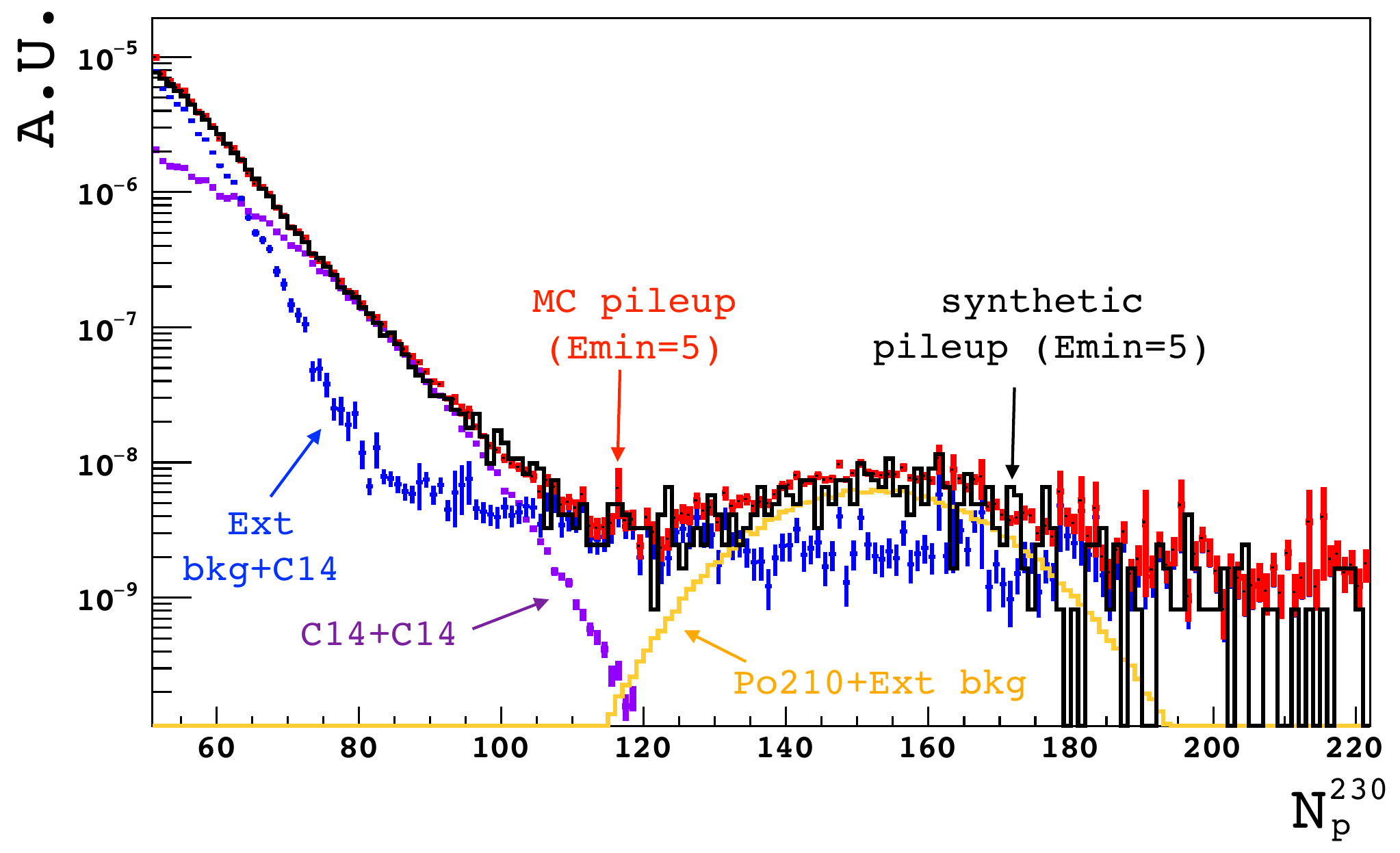} \\
\caption{Synthetic and MC pileup for $E_{min}=5\,N_p^{230}$. The different components are shown in various colors.
}
\label{fig:MCemins5}
\includegraphics[width = 0.9\columnwidth]{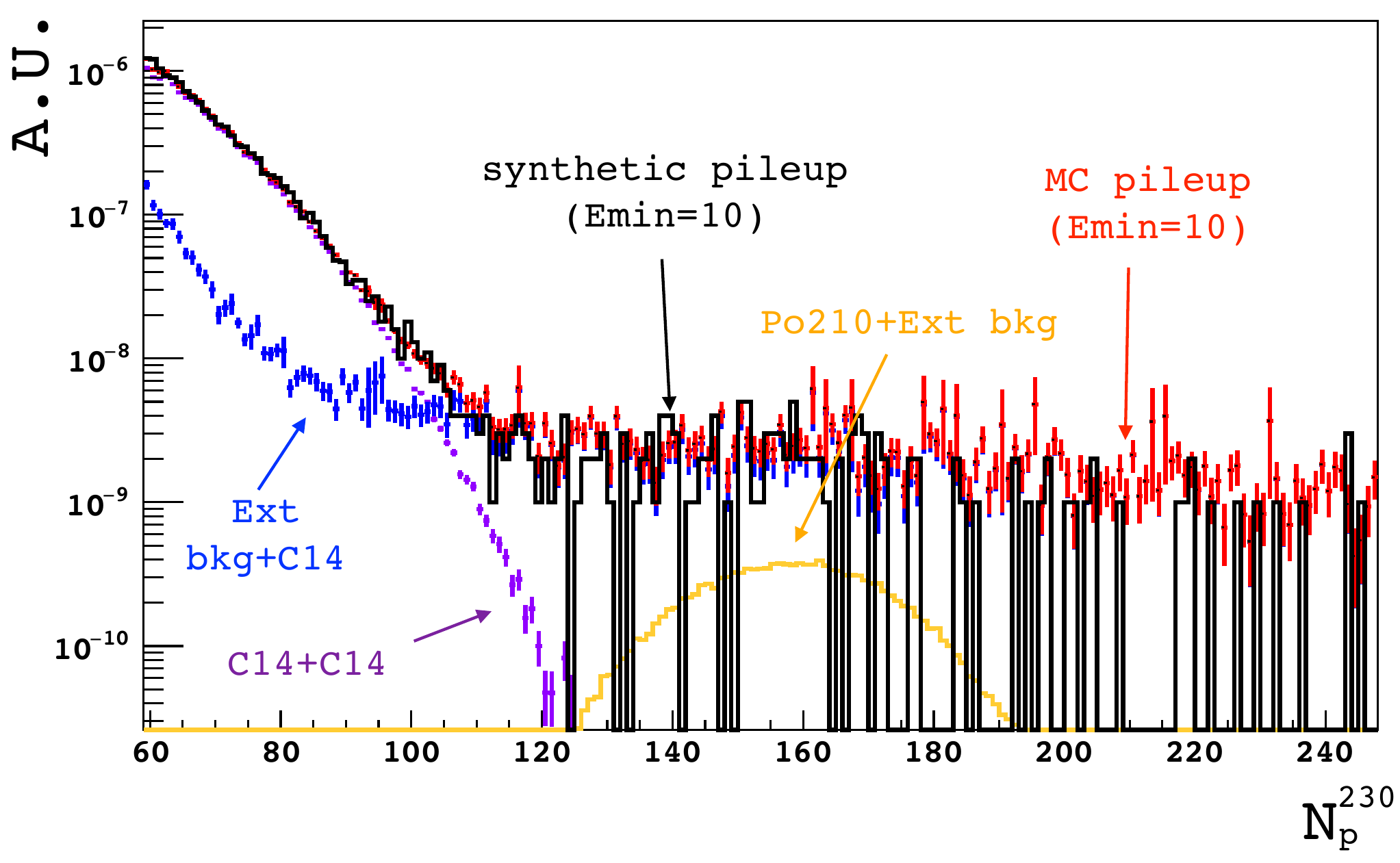} \\
\caption{Synthetic and MC pileup for $E_{min}=10\,N_p^{230}$. The different components are shown in various colors.
}
\label{fig:MCemins10}
\end{figure}

Once the various contributions were established with $E_{min}=6\,N_p^{230}$, the MC pileup was validated by comparing the predicted pileup shape with that of the synthetic one
for other values of $E_{min}$. The results are shown in Figs.\ \ref{fig:MCemins5} and \ref{fig:MCemins10}.
The pictures show that the synthetic pileup shape changes a lot as a function of $E_{min}$ and the MC pileup is able to reproduce it well. 
In particular, as $E_{min}$ grows, the contribution of the pileup with external \grs\ is reduced, since the mean visible energy of buffer events is of only a few hits.
In the case of $E_{min}=10\,N_p^{230}$, the pileup can mostly be explained by \C\ on \C\ events, with a higher energy tail of \grs\ interacting in the inner vessel and reconstructed in the
FV. The agreement of the MC pileup with the synthetic one tests not only the procedure of the pileup creation and composition, but also the quenching description,
the light collection reproduction, the light propagation modeling, and the external background production. 
Pileup events with a given visible energy are formed by events whose energy, if summed up in real single events,
would correspond to a lower quenching, and thus a higher amount of detected light. The energy scale of MC pileup matches  well with that of data, giving confidence on the solidity of the quenching description.
Furthermore, the hit time distribution is essential for reproducing the behavior of position reconstruction. Pileup events happen ``far away'' from each other
in the detector and thus the behavior of the reconstruction algorithm cannot be established \emph{a priori}, since it is based on a likelihood maximization. 
The MC pileup spectral composition after the cuts agrees with that of the synthetic pileup, further strengthening the trust in the whole simulation procedure.

\section{Conclusions}
In this paper we describe the details of the physical processes and instrumental response implemented in the MC simulation of the Borexino detector.
The ab initio simulation of the energy loss of particles interacting with the liquid scintillator target and of the generation, propagation and detection 
of scintillation photons yield better than $1\%$ accurate results for the reconstructed energy and for the time response of the detector for events in the innermost 100 tonnes of Borexino (a 3 m radius sphere).
The use of calibration sources has proven essential in providing a large statistical sample of events with well-defined energy and position, complementary to data collected in physics mode.

The Borexino simulations have steadily evolved over time and demonstrate an exquisite level of understanding of the processes governing the light propagation in large volume liquid scintillators, 
suggesting that the techniques presented in this report could easily be adapted to similar detectors. The implementation of the ionization quenching, the detailed model of the \v Cerenkov light emission, 
interaction and timing, and the treatment of the effective quantum efficiency of the PMTs are just a few examples of features that could successfully be exported to other experiments 
featuring different geometries and scintillator mixtures.

The Borexino MC code is an invaluable tool for all physics analyses and we expect it to play an equivalently crucial role in the forthcoming SOX sterile-neutrino phase of the experiment. 
The measurement of the solar neutrino interaction rates requires the highest level of accuracy from the MC, which is used to generate the energy spectra of all the components used in 
the final fit of the recorded energy spectrum.
We note that the accuracy of the energy response reported here exceeds that achieved for previous analyses (see Refs.~\cite{bib:BxLong, bib:BxBe3}) 
while covering a wider energy range (it now extends down to $\sim$100~keV). This suggests that a reduction of the systematic uncertainty on the solar 
neutrino interaction rate over previous measurements is possible.


\section{Acknowledgements}
The Borexino program is made possible by funding from INFN (Italy), NSF (USA), BMBF, DFG, HGF, and MPG (Germany), 
RFBR (Grants 16-02-01026 A, 15-02-02117 A, 16-29-13014 ofi-m, 17-02-00305 A) and RSF (Grant 17-02-01009) (Russia), 
and NCN Poland (Grant No. UMO-2013/10/E/ST2/00180). We acknowledge the generous hospitality and support of the Laboratori Nazionali del Gran Sasso (Italy).

\bibliography{MC_paper}

\end{document}